\documentclass[12pt]{article}
\newlength{\pubnumber} 

\catcode`\@=11
\@addtoreset{equation}{section}

\def\section{\@startsection{section}{1}{\z@}{3.5ex plus 1ex minus .2ex}
 {2.3ex plus .2ex}{\large\bf}}
\def\subsection{\@startsection{subsection}{2}{\z@}{2.3ex plus .2ex}
 {2.3ex plus .2ex}{\bf}}

\usepackage{latexsym}

\begin{document}

\begin{titlepage}
\samepage{
\setcounter{page}{0}
\rightline{November 2009}
\vfill
\begin{center}
     {\Large \bf On the Inconsistency of Fayet-Iliopoulos Terms in Supergravity Theories\\}
\vfill
   {\large
      Keith R. Dienes$^{1,2,3}$\footnote{
     E-mail address:  dienes@physics.arizona.edu}
        $\,$and$\,$ Brooks Thomas$^3$\footnote{
     E-mail address:  brooks@physics.arizona.edu}
    \\}
\vspace{.10in}
 {\it  $^1$  Physics Division, National Science Foundation, Arlington, VA 22230  USA\\}
 {\it  $^2$  Department of Physics, University of Maryland, College Park, MD  20742  USA\\}
 {\it  $^3$  Department of Physics, University of Arizona, Tucson, AZ  85721  USA$\footnote{Permanent address.}$\\}
\end{center}
\vfill
\begin{abstract}
  {\rm  Motivated by recent discussions, we revisit the issue of whether globally 
    supersymmetric theories with non-zero Fayet-Iliopoulos terms may be consistently coupled 
    to supergravity.  In particular, we examine claims that a fundamental inconsistency 
    arises due to the conflicting requirements which are imposed on the 
    $R$-symmetry properties of the theory by the supergravity framework.  
    We also prove that certain kinds of Fayet-Iliopoulos contributions to the supercurrent
    supermultiplets of theories with non-zero Fayet-Iliopoulos terms fail to exist.
    A key feature of our discussion is an explicit comparison
    between results from the chiral (or ``old minimal'') and linear (or ``new minimal'')
    formulations of supergravity, and the effects within each of these formalisms
    that are induced by the presence of non-zero Fayet-Iliopoulos terms.} 
\end{abstract}
\vfill
\vfill
\smallskip}
\end{titlepage}

\setcounter{footnote}{0}

\def\beq{\begin{equation}}
\def\eeq{\end{equation}}
\def\beqn{\begin{eqnarray}}
\def\eeqn{\end{eqnarray}}
\def\half{{\textstyle{1\over 2}}}
\def\quarter{{\textstyle{1\over 4}}}

\def\calO{{\cal O}}
\def\calE{{\cal E}}
\def\calT{{\cal T}}
\def\calM{{\cal M}}
\def\calF{{\cal F}}
\def\calS{{\cal S}}
\def\calY{{\cal Y}}
\def\calV{{\cal V}}
\def\calL{{\cal L}}

\def\Re#1{{{\rm Re}\,({#1})}}
\def\Im#1{{{\rm Im}\,({#1})}}
\def\ibar{{\overline{\imath}}}
\def\chibar{{\overline{\chi}}}
\def\qbar{{\overline{q}}}
\def\thetabar{{\overline{\theta}}}
\def\sigmabar{{\overline{\sigma}}}
\def\lambdabar{{\overline{\lambda}}}
\def\epsilonbar{{\overline{\epsilon}}}
\def\etabar{{\overline{\eta}}}
\def\Dbar{{\overline{D}}}
\def\Wbar{{\overline{W}}}
\def\Tbar{{\overline{T}}}

\def\ttwo{{\vartheta_2}}
\def\tthree{{\vartheta_3}}
\def\tfour{{\vartheta_4}}
\def\ttwob{{\overline{\vartheta}_2}}
\def\tthreeb{{\overline{\vartheta}_3}}
\def\tfourb{{\overline{\vartheta}_4}}

\def\wtPhi{{\widetilde{\Phi}}}
\def\Phid{{\Phi^\dagger}}
\def\wtPhid{{\widetilde{\Phi}^\dagger}}
\def\FI{{\mathrm{FI}}}
\def\SW{{\mathrm{SW}}}
\def\SigL{{\Sigma_L}}
\def\SigbL{{\overline{\Sigma}_L}}

\newcommand{\newc}{\newcommand}
\newc{\gsim}{\lower.7ex\hbox{$\;\stackrel{\textstyle>}{\sim}\;$}}
\newc{\lsim}{\lower.7ex\hbox{$\;\stackrel{\textstyle<}{\sim}\;$}}

\hyphenation{su-per-sym-met-ric non-su-per-sym-met-ric}
\hyphenation{space-time-super-sym-met-ric}
\hyphenation{mod-u-lar mod-u-lar--in-var-i-ant}


\def\inbar{\,\vrule height1.5ex width.4pt depth0pt}

\def\IC{\relax\hbox{$\inbar\kern-.3em{\rm C}$}}
\def\IQ{\relax\hbox{$\inbar\kern-.3em{\rm Q}$}}
\def\IR{\relax{\rm I\kern-.18em R}}
 \font\cmss=cmss10 \font\cmsss=cmss10 at 7pt
\def\IZ{\relax\ifmmode\mathchoice
 {\hbox{\cmss Z\kern-.4em Z}}{\hbox{\cmss Z\kern-.4em Z}}
 {\lower.9pt\hbox{\cmsss Z\kern-.4em Z}}
 {\lower1.2pt\hbox{\cmsss Z\kern-.4em Z}}\else{\cmss Z\kern-.4em Z}\fi}

\long\def\@caption#1[#2]#3{\par\addcontentsline{\csname
  ext@#1\endcsname}{#1}{\protect\numberline{\csname
  the#1\endcsname}{\ignorespaces #2}}\begingroup
    \small
    \@parboxrestore
    \@makecaption{\csname fnum@#1\endcsname}{\ignorespaces #3}\par
  \endgroup}
\catcode`@=12

\input epsf

\def\mm{{\tilde m}}
\def\nn{{\tilde n}}
\def\rep#1{{\bf {#1}}}
\def\ie{{\it i.e.}\/}
\def\eg{{\it e.g.}\/}

\def\alphadot{{\dot \alpha}}
\def\betadot{{\dot \beta}}
\def\gammadot{{\dot \gamma}}
\def\deltadot{{\dot \delta}}
\def\Noth{{Noether\ }}
\def\Phibar{{\overline{\Phi}}}
\def\ibar{{\overline{\i}}}
\def\jbar{{\overline{\j}}}

\def\Phitilde{{\widetilde{\Phi}}}
\def\Phidaggertilde{{{\widetilde{\Phi}}^\dagger}}
\def\phitilde{{\widetilde{\phi}}}
\def\Wtilde{{\widetilde{W}}}
\def\Ktilde{{\widetilde{K}}}
\def\Sigmabar{{\overline{\Sigma}}}
\def\Omegabar{{\overline{\Omega}}}

\section{Introduction and overview}
\setcounter{footnote}{0}

Within the context of supersymmetric gauge theories, the $D$-field is truly unique.
It is the only field component which is simultaneously
invariant under both gauge and supersymmetry (SUSY) transformations,
and thus the $D$-field can play roles in supersymmetric theories which are forbidden to other fields.
One such role is in the 
introduction of the Fayet-Iliopoulos (FI) term~\cite{FI}
\begin{equation}
          \xi D ~\in~ \xi  \int d^2 \theta d^2\thetabar\, V~ 
\label{FFIterm}
\end{equation}
where $\xi$ is the FI coefficient and $V$ is a real vector superfield.
As is well known, FI terms figure prominently 
in the history of supersymmetry, and constitute one of only two 
ways of inducing spontaneous supersymmetry breaking.
As a result, over the past two decades,
FI terms have appeared in attempts at supersymmetric model-building
which are too numerous to list.

Yet experience has shown that even within these contexts,
FI terms face significant limitations.
For example, they have very restrictive renormalization group flows.
It also turns out that they almost never
dominate in dynamical supersymmetry breaking.
It also turns out to be very difficult to couple FI terms to supergravity,
and progress in this area has a long and somewhat tortu(r)ous history (see, 
\eg, Refs.~[2--25]).  As a result, despite the existence of a vast
literature on this subject spanning several decades, 
the question of whether it is even consistent to couple theories
with FI terms to gravity remains largely unresolved.

The first steps in attempting to generalize FI terms to local supersymmetry were
undertaken in Ref.~[3], where it was shown that the action for theories with such terms is invariant
under a set of transformations which combines $U(1)$ gauge transformations with local 
chiral rotations of the supergravity fields.
This picture
was further refined in Refs.~\cite{StelleAndWestEToTheV,BarbieriEtAlEToTheV},
where it was shown that the presence of an FI term in supergravity
results in a mixing between the $U(1)_\FI$ gauge symmetry and $U(1)_R$ rotations, 
leaving behind only a single $U(1)$ invariance which is 
a linear combination of the two original symmetries.
Following this work, a number of investigators 
then endeavored to couple such a theory to matter
fields in a consistent, gauge-invariant manner.
Because the presence of the FI term effectively gauges the global $U(1)_R$ symmetry,
any theory involving an FI term would have to satisfy an additional set of
anomaly-cancellation conditions involving $U(1)_R$.  Unfortunately, these
conditions turn out to be quite stringent.  In
Ref.~\cite{EarlyAnomalyIssues}, for example, it was found that these conditions are
not satisfied in the MSSM or in the simplest extensions thereof.
While there do exist rather baroque models (\eg, models with flavor-dependent
$R$-charges) which satisfy these constraints, such models tend to
be riddled with other undesirable traits, such as rapid
proton-decay rates or a broken $SU(3)$ color group.
Subsequent investigations~\cite{AnomalyIssues} yielded
similar results, and while such models have not been ruled out,
phenomenologically successful ones are not easy to construct.

But this is not the only difficulty.
Even if a phenomenologically-consistent model could be realized,
a further problem arises because
the charge shifts of the sort required to preserve gauge
invariance in FI models have been shown to be inconsistent
with Dirac quantization in the presence of magnetic
monopoles~\cite{WittenDiracQuant}.
As a result, no consistent and phenomenologically viable models with FI terms 
are known to exist, either in field theory or in string theory.
Taken together,
these considerations have even led to speculations that fundamental
FI terms may be incompatible with supergravity altogether.

In a recent paper~\cite{Seiberg}, an explanation was proposed for these
difficulties:  
although the FI term is indeed invariant under supersymmetry
transformations as well as the regular abelian gauge transformations that remain after 
passing to Wess-Zumino gauge, the 
FI term breaks the invariance of the theory
with respect
to the larger, full set of abelian gauge transformations 
that are required in a fully supersymmetric context.
Specifically, 
the supercurrent  and 
energy-momentum tensor  fail
to be fully gauge-invariant when an FI term of the form in Eq.~(\ref{FFIterm})
is added to the theory.
Moreover, in the presence of an FI term, 
the associated $R$-current
also fails to be gauge invariant, even after the truncation to Wess-Zumino gauge.

In principle, these results would have enormous consequences~\cite{Seiberg}.
For example, if a supersymmetric field theory has no FI term at high scales,
then no FI term can be generated at lower scales. 
These results would also imply that FI terms cannot be 
generated either perturbatively or non-perturbatively.
Finally, these results would imply, once and for all, that 
all supergravity theories must have vanishing FI terms.
Thus, if true, the results in Ref.~\cite{Seiberg} would truly amount
to a death penalty for FI terms.  While FI terms might still arise in
certain limited contexts (\eg, string theories whose particle spectra exhibit
FI charges with non-vanishing traces), their role in most of supersymmetric
particle physics would be seriously curtailed.

This paper is devoted to developing a more complete understanding of these
results.
In particular, because the $D$-field does not carry an $R$-charge,
an FI term cannot be responsible for the breaking of the $R$-symmetry
that would otherwise exist in a superconformal theory.
Yet the results which are found in Ref.~\cite{Seiberg} explicitly
break this $R$-symmetry.
Similarly, the FI term, by itself, preserves the Fayet-Iliopoulos gauge symmetry.
Yet the results which are found in Ref.~\cite{Seiberg} explicitly break
this gauge symmetry --- indeed, this is the primary point of Ref.~\cite{Seiberg}.

The reason for these apparent inconsistencies is that the 
results of Ref.~\cite{Seiberg} were implicitly derived using what is known
as the ``chiral'' (or ``old minimal'') formulation of supergravity~\cite{ChiralFormalismSUGRA}.
Indeed, one of the subtleties associated with making supersymmetry local 
is that there exist several different formulations for off-shell supergravity.  
These differ from each other in the auxiliary-field content of the
supergravity multiplet, and as a result they also differ from each other in their
formulations of the supercurrrent supermultiplet to which the supergravity multiplet 
must couple. 
However,
even the classical symmetry properties of a given theory 
differ from one formulation to the next.  
As we will show, the gauge-non-invariance of the 
supercurrent highlighted in Ref.~\cite{Seiberg} is such a
formalism-dependent property.  Indeed, 
the chiral formalism is well known to implicitly break
not only the $U(1)_{\rm FI}$ gauge symmetry (if originally present),
but also $R$-symmetry (if originally present).  
This then explains the results found in Ref.~\cite{Seiberg}. 

There do, however, exist other formalisms in which the supercurrent
remains gauge invariant, and in which $R$-symmetry is explicitly preserved.
One such formulation is the so-called ``linear'' (or ``new minimal'') 
formalism~\cite{LinearFormalismSUGRA}.
This formalism has the advantage of manifestly preserving the symmetries of
the original theory from the outset, and thus does not lead 
to spurious broken-symmetry effects.
One natural question, then, is to determine the extent to 
which the conclusions of Ref.~\cite{Seiberg} continue to apply,
even within a gauge-invariant formalism such as the linear formalism.

In this paper, we shall therefore undertake a general analysis 
of the supercurrent supermultiplet and general symmetry structures of 
theories with non-zero FI terms.
Throughout our analysis, we shall bear in mind
the special role played by the fact that FI terms, by themselves, preserve $R$-symmetry.
Other than this, we shall make no special assumptions about the theory in question.
A key feature of our discussion will be an explicit comparison
between results from the chiral and linear 
formulations of supergravity, and the effects within each of these formalisms
that are induced by the presence of non-zero Fayet-Iliopoulos terms.
Indeed, one of our results will be to prove that certain kinds of
Fayet-Iliopoulos contributions to the supercurrent
supermultiplets of theories with non-zero Fayet-Iliopoulos terms fail to exist.

Unfortunately, the literature on supersymmetry and supergravity is notoriously 
plagued with a plethora of different notations and conventions.
Moreover, as mentioned above, there also exist a variety of different formalisms
that can be used for describing what is often ultimately the same physics,
and one of the goals of this paper is to provide an explicit comparison
between results derived using the chiral formalism and those derived
using the linear formalism.
We have therefore attempted to write this paper
so that our discussion and analysis is as streamlined 
and as completely self-contained as possible.

In Sect.~2, we review some basic facts about FI terms, 
establishing our notation and conventions along the way.
In Sect.~3, 
we then outline some basic facts about the supercurrent supermultiplet
and discuss the various forms which it may 
take in supersymmetric theories which do not respect the full set
of symmetries of the superconformal group.
In Sect.~4, we review
the chiral-compensator formalism, paying special attention to the
issues that arise in the presence of a non-zero FI term.
We show, in particular, how this formalism allows us to obtain
one version of the FI supercurrent, and we prove that certain
$R$-symmetric FI contributions to the overall supercurrent 
supermultiplet do not exist in the chiral formalism.
We also review the arguments of Ref.~\cite{Seiberg} which demonstrate 
that the breaking of $U(1)_{\rm FI}$ gauge symmetry in this 
formalism directly leads to the appearance of an extra global symmetry,
implying that such a theory cannot be consistently coupled to supergravity 
in the chiral formalism.
In Sect.~5, we then turn our attention to the linear-compensator formalism.
In marked contrast to the chiral-compensator formalism, this
formulation has the advantage of maintaining $U(1)_{\rm FI}$ gauge invariance
throughout.  
We again focus on issues that arise in the presence of a non-zero FI term,
and demonstrate once again that
certain FI contributions to the overall supercurrent 
supermultiplet do not exist.
Finally, in Sect.~6,
we discuss how the chiral and linear formalisms can 
ultimately be related to each other through a duality transformation
in theories that respect $R$-symmetry, such as those that might include non-zero FI terms.  
We also show how the supercurrent supermultiplets in the two
formalisms can be related to each other, and discuss the symmetry properties of the theories
that emerge in the linear formalism in the presence of a non-zero FI term.

\section{The Fayet-Iliopoulos term:  Basic facts}
\setcounter{footnote}{0}

We begin, for completeness, by recording some standard results\footnote{
    Here and throughout this paper, we shall follow
    the notation and conventions of Ref.~\cite{WessBagger} exactly.
    We shall also use the phrase ``$R$-symmetry'' to refer to a generic
    $\theta$-rotation, while we shall let $R_5$ denote the 
    specific chiral
    $\theta$-rotation whose generator appears in the 
    ${\cal N}=1$ supersymmetry algebra.  
       }
concerning FI terms.
This material is completely standard and can be found in almost any
textbook on elementary supersymmetry.  As such, our goal in repeating
it here is primarily to establish our notation and conventions,
as well as to have it readily available for later use.

We begin by recalling that
a general vector superfield $V$ can be expanded in the form
\begin{eqnarray}
  V &=& C + i\theta \chi -i\thetabar \chibar + {i\over 2} \theta\theta (M+iN) 
              -{i\over 2} \thetabar\thetabar(M-iN)\nonumber\\ 
    && ~~~- \theta\sigma^\mu \thetabar A_\mu 
        + i \theta\theta\thetabar \left(\lambdabar + {i\over 2} \sigmabar^\mu \partial_\mu  \chi\right)
        - i \thetabar\thetabar\theta \left(\lambda + {i\over 2} \sigma^\mu \partial_\mu  \chibar\right)
           \nonumber\\
    &&  ~~~+  {1\over 2}\theta\theta\thetabar\thetabar \left( D+{1\over 2}
           \Box C\right)~
\label{vecsup}
\end{eqnarray}
where $C$, $D$, $M$, $N$, and $A_\mu$ are all real, and where
$\Box \equiv \partial_\mu \partial^\mu$.
Under a supersymmetry transformation with parameter $\epsilon$,
these fields transform according to
\begin{eqnarray}
   \delta_\epsilon C &=& i \epsilon \chi - i \epsilonbar\chibar~\nonumber\\
   \delta_\epsilon \chi_\alpha  &=& \epsilon_\alpha (M+iN) + (\sigma^\mu \epsilonbar)_\alpha \partial_\mu C 
                 + i(\sigma^\mu \epsilonbar)_\alpha A_\mu ~\nonumber\\
   \delta_\epsilon (M+iN) &=& 
           2i \epsilonbar \sigmabar^\mu \partial_\mu \chi + 2 \lambdabar\epsilonbar~\nonumber\\
   \delta_\epsilon A^\mu &=& 
       i \epsilonbar\sigmabar^\mu\lambda  + i \epsilon \sigma^\mu \lambdabar  + \epsilon \partial^\mu \chi
              + \epsilonbar \partial^\mu \chibar~\nonumber\\
   \delta_\epsilon \lambda_\alpha &=& i \epsilon_\alpha D  + 
            \half (\sigma^\mu \sigmabar^\nu \epsilon)_\alpha F_{\mu\nu}~\nonumber\\
   \delta_\epsilon D &=& - \epsilon \sigma^\mu \partial_\mu \lambdabar + 
          \epsilonbar \sigmabar^\mu \partial_\mu \lambda~,
\label{susytransforms}
\end{eqnarray}
where $F_{\mu\nu} \equiv \partial_\mu A_\nu - \partial_\nu A_\mu$.
We observe from these results that only the $D$-field transforms into a total derivative.

Let us also recall the properties of these fields under $U(1)$ gauge transformations.
In a supersymmetric theory, a $U(1)$ gauge transformation of the vector superfield $V$
generalizes to take the form
\begin{equation}
        V ~\to~ V + \Phi + \Phi^\dagger
\label{gengaugesuperfield}
\end{equation}
where $\Phi$ is a chiral superfield.  
For the individual component fields within $V$, 
this leads to the gauge transformations 
\begin{eqnarray}
             \delta C &=& \phi + \phi^\ast~ \nonumber\\
             \delta \chi &=& -i\sqrt{2} \psi ~\nonumber\\
             \delta (M+iN) &=& -2iF~\nonumber\\
             \delta A_\mu &=& -i\partial_\mu(\phi-\phi^\ast)~\nonumber\\
             \delta \lambda &=& 0~\nonumber\\
             \delta D &=& 0~,
\label{gaugetransforms}
\end{eqnarray}
where $\phi$, $\psi$, and $F$ are the component fields within $\Phi$.
The transformation for the vector field $A_\mu$ is the usual $U(1)$ gauge
transformation, while the remaining transformations are associated with the
generalizations of $U(1)$ gauge symmetry in a supersymmetric context.
Passing to Wess-Zumino gauge amounts to fixing these remaining gauge transformations
in such a way that  
$C$, $\chi$, $M$, and $N$ 
all vanish;  this then leaves behind only the ordinary $U(1)$ gauge transformation
associated with $A_\mu$.
For complete generality, we shall avoid restricting to Wess-Zumino gauge
and retain all fields in our theory.  However, we shall continue to distinguish between
the traditional $U(1)$ gauge symmetry associated with $A_\mu$, and
the fuller, more general gauge symmetry which is additionally associated 
with the transformations of $C$, $\chi$, $M$, and $N$.
When needed, we shall refer to these gauge symmetries as ``little'' and ``big''
respectively. 

We thus see that $\lambda$ and $D$ are fully gauge invariant:  
their values are unaltered under the most general gauge transformations  
associated with {\it all}\/ of the degrees of freedom within $\Phi$.
By contrast, the fields $C$, $\chi$, $M$, and $N$ 
are invariant only under the traditional $U(1)$ gauge transformations 
associated with $A_\mu$ that survive the truncation to Wess-Zumino gauge.
Finally, as expected, the vector field $A_\mu$ is not invariant under
any of these gauge transformations.

\begin{table}[t!]
\begin{center}
\begin{tabular}{||c||c|c|c||}
         \hline
         \hline
       ~& SUSY & traditional $U(1)$  & full gauge invariance \\
       \hline
    FI Lagrangian  w/o $C$-term &  no & yes & yes \\
    FI action      w/o $C$-term &  yes & yes & yes \\
         \hline
    FI Lagrangian  w/ $C$-term &  no & yes & no \\
    FI action      w/ $C$-term &  yes & yes & yes \\
         \hline
         \hline
\end{tabular}
\end{center}
\caption{Invariances of the FI Lagrangian and FI action under
         supersymmetry transformations,
        under the traditional $U(1)$ gauge transformations that survive the truncation
          to Wess-Zumino gauge, and
           under the full gauge transformations 
         associated with Eq.~(\protect\ref{gengaugesuperfield}).
          We list results for the case in which the (total-derivative) $C$-term 
          is dropped from Eq.~(\ref{FIterm}) as well as the case in which it is retained.}
\label{FIinvariances}
\end{table}

Given these results, 
we can now analyze the transformation properties of the FI action
\begin{equation}
             {\cal S}_{\rm FI} = \int d^4 x \, {\cal L}_{\rm FI} ~~~~{\rm where}~~~~
      {\cal L}_{\rm FI} \equiv 2\xi \int d^2\theta d^2\thetabar \, V =
             \xi \left( D+{1\over 2} \Box C\right)~. 
\label{FIterm}
\end{equation}
Under supersymmetry transformations, we see that both the $D$-term and the $C$-term 
transform into total derivatives, and these vanish  under the spacetime integral $\int d^4x$.
Thus, the FI Lagrangian is not supersymmetry invariant, but the FI action is.
By contrast, the situation regarding gauge invariance is a bit more subtle.
Under the traditional $U(1)$ gauge symmetry (\ie, the remnant of the full
gauge symmetry that survives the truncation to Wess-Zumino gauge),
both the $D$-term and $C$-term are gauge invariant.
As a result, both the FI Lagrangian and FI action are invariant
under the traditional $U(1)$ gauge symmetry.
However, under the full gauge symmetry associated with Eq.~(\ref{gengaugesuperfield}),
the $D$-term is invariant while the $C$-term is not.
Yet, even in this case, the $C$-term within Eq.~(\ref{FIterm}) 
transforms into a total derivative.  Thus, the FI action is actually invariant
under the full gauge symmetry associated with Eq.~(\ref{gengaugesuperfield})
even though the FI Lagrangian is not.
These transformation properties for the FI term are summarized in Table~\ref{FIinvariances}.

\section{Three possible structures for the supercurrent supermultiplet:   A quick review} 
\setcounter{footnote}{0}

In this section, we review the emergence and constraints that govern the supercurrent
supermultiplet.  We discuss these constraints in both the chiral and linear formalisms, 
as well as the relations between them.

\subsection{The supercurrent supermultiplet}

The ${\cal N}=1$ supersymmetry algebra is the algebra 
associated with super-Poincar\'e symmetry.
Each of these charges in this algebra is associated with
a current, and these currents (through the Noether theorem) are associated with
different symmetries that together constitute the super-Poincar\'e group.

In this paper, we shall be concerned with three of these currents:
the $R_5$-symmetry current $j_\mu^{(5)}$, associated with the chiral $R_5$-variations
(phase rotations) of $\theta$ and $\thetabar$;
the supercurrent $j_{\mu\alpha}$, associated with supersymmetry transformations;
and the energy-momentum tensor $T_{\mu\nu}$, associated with spacetime translations.
Each of these may be calculated in the usual way, through the Noether theorem;
since equations of motion are used in the Noether derivation,
such currents are intrinsically on-shell objects.

Ultimately, however, these currents are not physical:  the only parts which are
physical are their divergences $\partial^\mu j^{(5)}_\mu$, $\partial^\mu j_{\mu\alpha}$,
and $\partial^\mu T_{\mu\nu}$ (indicating whether or not these symmetries are preserved),
and their associated charges (\ie, the spatial integrals of their $\mu=0$ components).
As a result, these currents themselves are not unique, and 
each may be modified in a variety of ways through the addition
of so-called ``improvement'' terms which do
do not change the physical properties of the theory  
in question.  Thus, for any current $j_\mu$, an improvement term may affect
the value of neither $\partial^\mu j_\mu$ nor $Q\equiv \int d^3 x j_0$.
It then follows that 
if a particular theory has a Noether current corresponding to an unbroken 
symmetry, \ie, if $\partial^\mu j_\mu=0$, then no improvement term can induce
a non-zero value for $\partial^\mu j_\mu$.
Improvement terms can, however, can be useful for making certain properties of a theory manifest.
For example, in the case of abelian gauge theories, it is well known that
the Noether procedure does not 
always yield an energy-momentum tensor $T_{\mu\nu}$ which is gauge invariant
or symmetric in its indices.  However, there exist improvement terms which
can fix both of these difficulties.
Likewise, in supersymmetric theories involving K\"ahler potentials,
K\"ahler transformations will also modify the currents by 
altering the forms of these improvement terms.

In Ref.~\cite{FZ}, it was shown that the
$R_5$-current $j^{(5)}_\mu$, the supercurrent $j_{\mu\alpha}$, and 
the energy-momentum tensor $T_{\mu\nu}$  --- all suitably improved --- can be related
to the lowest-lying components of a real supermultiplet, the so-called supercurrent superfield.
This fact is ultimately a consequence of the structure of the underlying super-Poincar\'e algebra
which relates the charges associated with these currents. 
In general, the supercurrent supermultiplet is a real vector superfield
which we may write in the form
\begin{eqnarray}
  J_\mu &=& C_\mu + i\theta \chi_\mu -i\thetabar \chibar_\mu + {i\over 2} \theta\theta (M_\mu+iN_\mu)
              -{i\over 2} \thetabar\thetabar(M_\mu-iN_\mu)\nonumber\\
    && ~~~- \theta\sigma^\nu \thetabar \hat T_{\nu\mu}
        + i \theta\theta\thetabar \left(\lambdabar_\mu + {i\over 2} \sigmabar^\nu \partial_\nu  \chi_\mu\right)
        - i \thetabar\thetabar\theta \left(\lambda_\mu + {i\over 2} \sigma^\nu \partial_\nu  \chibar_\mu\right)
           \nonumber\\
    &&  ~~~+  {1\over 2}\theta\theta\thetabar\thetabar \left( D_\mu+{1\over 2}\Box C_\mu \right)~
\label{vecsup2}
\end{eqnarray}
where $C_\mu$, $D_\mu$, $M_\mu$, $N_\nu$, and $\hat T_{\nu\mu}$ are all real.
Note that this form is completely analogous to that in Eq.~(\ref{vecsup}) except that all fields now
carry an additional Lorentz index.
Consequently, under supersymmetry transformations, the components
in Eq.~(\ref{vecsup2}) 
mix according to the relations in Eq.~(\ref{susytransforms}),
suitably extended with an extra Lorentz index:
\begin{eqnarray}
   \delta_\epsilon C_\mu &=& i \epsilon \chi_\mu - i \epsilonbar\chibar_\mu~ \nonumber\\
   \delta_\epsilon \chi_{\mu\alpha}  &=& \epsilon_\alpha (M_\mu+iN_\mu) + 
                      (\sigma^\nu \epsilonbar)_\alpha (\partial_\nu C_\mu + i \hat T_{\nu\mu}) ~ \nonumber\\
   \delta_\epsilon (M_\mu+iN_\mu) &=& 
           2i \epsilonbar \sigmabar^\nu \partial_\nu \chi_\mu + 2 \epsilonbar \lambdabar_\mu~ \nonumber\\
   \delta_\epsilon \hat T_{\nu\mu} &=& 
       i \epsilonbar\sigmabar_\nu\lambda_\mu  + i\epsilon \sigma_\nu \lambdabar_\mu
           + \epsilon \partial_\nu \chi_\mu + \epsilonbar \partial_\nu \chibar_\mu~ \nonumber\\
   \delta_\epsilon \lambda_{\mu\alpha} &=& i \epsilon_\alpha D_\mu  + 
            \half (\sigma^\rho  \sigmabar^\sigma \epsilon)_\alpha 
             (\partial_\rho \hat T_{\sigma \mu}
            - \partial_\sigma \hat T_{\rho \mu}) ~ \nonumber\\
   \delta_\epsilon D_\mu &=& - \epsilon \sigma^\nu \partial_\nu \lambdabar_\mu + 
          \epsilonbar \sigmabar^\nu \partial_\nu \lambda_\mu~.
\label{susytransforms2}
\end{eqnarray}

\begin{figure}[t!]
\centerline{
   \epsfxsize 6.0 truein \epsfbox {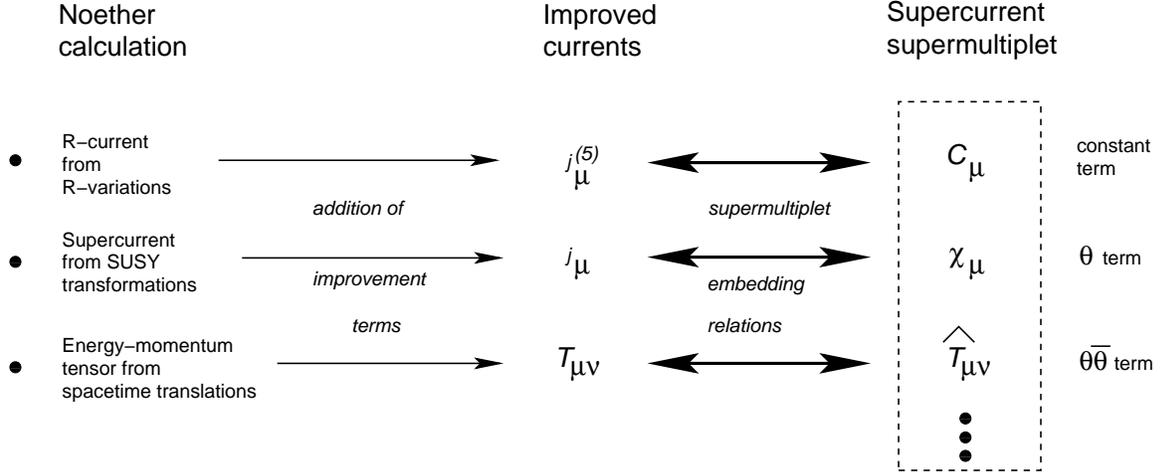} }
\caption{The general relation between the individual Noether currents and the supercurrent
   supermultiplet.  While the Noether currents follow directly from their associated symmetries,
   they can each be modified through the addition of appropriate so-called ``improvement'' terms
   which preserve their divergences and charges.
   When suitably improved, these currents can then be mapped into the lowest components of a single
   supercurrent superfield through mapping relations which depend on the particular constraints
   that the supercurrent supermultiplet is assumed to obey.} 
\label{procedurefig}
\end{figure}

The fields $C_\mu$, $\chi_{\mu\alpha}$, and $\hat T_{\nu\mu}$ are directly related to
the improved versions of $j_\mu^{(5)}$, $j_{\mu\alpha}$, and $T_{\mu\nu}$ respectively. 
In fact, we can immediately identify $C_\mu= j^{(5)}_\mu$,
but the precise mappings between $\chi_{\mu\alpha}$ and $j_{\mu\alpha}$,
and between $\hat T_{\nu\mu}$ and $T_{\mu\nu}$,
depend on the particular constraint equations satisfied by 
the supercurrent supermultiplet $J_\mu$ in Eq.~(\ref{vecsup2}). 
These in turn depend on the precise supermultiplet structure of the 
superconformal anomalies in the theory.  Several different formalisms for approaching
this issue exist, as we shall discuss in Sect.~3.2.
This situation is schematically depicted in Fig.~\ref{procedurefig}.

\subsection{Constraints on supercurrent supermultiplets in the 
   superconformal, chiral, and linear formalisms:  A review and comparison}

In this section, we review the forms that the supercurrent supermultiplet is expected to take
in three different cases.  These cases are ultimately distinguished by whether superconformal symmetry
and $R_5$-symmetry are either separately preserved or broken.
While additional cases may also exist~\cite{West,Lessons}, these three
cases represent the ``minimal'' constructions
and will be the primary focus of this paper. 

In general, there are three quantities which are relevant to determining the precise 
superconformal anomaly structure of a given theory:
$\partial^\mu j_\mu^{(5)}$, 
$\sigmabar^\mu j_{\mu\alpha}$,
and $T\equiv T_\mu^\mu$.
In a theory with unbroken superconformal invariance, 
all three of these quantities vanish, and the supercurrent supermultiplet
satisfies the constraint equation
\begin{equation}
              \Dbar^\alphadot J_{\alpha\alphadot} = 0~
\label{superconfconst}
\end{equation}
where  $D_\alpha$ and $\Dbar_\alphadot$ are the standard chiral covariant derivatives
\begin{eqnarray}
            D_\alpha &\equiv& {\partial \over \partial \theta^\alpha} 
                     + i (\sigma^\mu \thetabar)_\alpha \partial_\mu\nonumber\\ 
            \Dbar_\alphadot &\equiv&  - {\partial\over \partial \thetabar^\alphadot} - 
                    i (\theta \sigma^\mu)_\alphadot \partial_\mu~
\end{eqnarray}
and where
\begin{equation}
               J_{\alpha\alphadot} ~\equiv~ \sigma_{\alpha\alphadot}^\mu J_\mu~.
\label{alphadef}
\end{equation}
These equations therefore describe the superconformal case.
Note that these constraints do not define $J_{\alpha\alphadot}$ uniquely;  instead,
$J_{\alpha\alphadot}$ is defined only up to the addition of real superfield improvement 
terms which are annihilated by $\Dbar^\alphadot$.
This is related to the fact that the supercurrent supermultiplet 
$J_{\alpha\alphadot}$ 
is not a physical object;  indeed, only its divergence, charge, and associated anomalies
are of relevance.

By contrast, in a general theory in which the superconformal symmetry is maximally broken,
all three of the anomalies accrue non-zero values and fill out a chiral multiplet.
In such a case, the supercurrent supermultiplet satisfies a constraint equation of the form
\begin{equation}
              \Dbar^\alphadot J_{\alpha\alphadot} = D_\alpha S~
\label{chiralcase}
\end{equation}
where $S$ is a chiral multiplet.
Eq.~(\ref{chiralcase}) defines the chiral case. 
In this case, $J_{\alpha\alphadot}$ is defined up to the addition 
of real superfield improvement terms which, when acted on by $\Dbar^\alphadot$,
take the form $D_\alpha X$, where $X$ is a chiral
superfield.  Such improvement terms can be generated through
K\"ahler transformations.

However, {\it it is possible for a theory to break superconformal symmetry and yet
preserve $R_5$-symmetry}\/.
In such a case, $\partial^\mu j_\mu^{(5)}$ vanishes while
$\sigmabar^\mu j_{\mu\alpha}$ and $T$ accrue non-zero values and together form
a so-called {\it linear}\/ multiplet.\footnote{
     Recall that a linear multiplet, like a chiral multiplet, is an irreducible
     representation of the supersymmetry algebra contained within a vector multiplet $T$. 
     Specifically,
     a linear multiplet can be realized as a vector multiplet $T$
     which additionally satisfies the constraints 
     $D^\alpha D_\alpha T= \overline{D}_\alphadot \overline{D}^\alphadot T=0$.
     Equivalently, a linear multiplet can be obtained from a general vector multiplet $T$
     by setting the $\theta\theta$ and $\thetabar\thetabar$ components to zero
     and imposing the reality condition $T^\dagger=T$.  }
The supercurrent supermultiplet then satisfies a constraint equation of the form
\begin{equation}
              \Dbar^\alphadot J_{\alpha\alphadot} = L_\alpha~
\label{linearcase}
\end{equation}
where $L_\alpha$ is a {\it chiral}\/ multiplet satisfying
\begin{equation}
           \Dbar_\betadot L_\alpha=0~~~~~~ {\rm and}~~~~~~
            D^\alpha L_\alpha = \Dbar_\alphadot \overline{L}^\alphadot~.
\label{Lconstraints}
\end{equation}
In other words, despite its name, 
$L_\alpha$ is structurally identical to the gauge multiplet $W_\alpha=(\lambda,F_{\mu\nu},D)$,
where $F_{\nu\mu}= -F_{\mu\nu}$.
We will henceforth denote the components of $L_\alpha$ as $L_\alpha=(\Lambda,\Phi_{\mu\nu},\Delta)$, where
similarly $\Phi_{\nu\mu}= -\Phi_{\mu\nu}$.
Eq.~(\ref{linearcase}) then defines the linear case.
Indeed, as we shall shortly demonstrate, Eq.~(\ref{Lconstraints}) implies that $J_{\alpha\alphadot}$
is itself a linear multiplet.

In Table~\ref{constraintstable}, we indicate the explicit supercurrent supermultiplet structures 
that emerge in each of these three cases. 
In the first two columns, we provide the explicit term-by-term expansion of the superfield 
$\Dbar^\alphadot J_{\alpha\alphadot}$, where
\begin{equation}
      \Dbar^\alphadot J_{\alpha\alphadot} ~=~
        -\epsilon^{\alphadot\betadot} \sigma^\mu_{\alpha\alphadot} 
               {\partial\over \partial \thetabar^\betadot} J_\mu 
          +i (\sigma^\mu\sigmabar^\nu \theta)_\alpha \partial_\nu J_\mu
\end{equation}
with $J_\mu$ is given in Eq.~(\ref{vecsup2}).
In the remaining columns, we also provide the corresponding term-by-term expansion of the
different superfields to which $\Dbar^\alphadot J_{\alpha\alphadot}$
must be equated in each of these three cases.
In general, equating the coefficients of these expansions yields a set of constraint
equations for the fields that appear in these coefficients.  Note that in extracting
the constraint equations for
these coefficients, we have made use of two facts which apply when the coefficients
are scalars in spinor space:
\begin{eqnarray}
       \sigma^\mu A_\mu =0 ~~&\Longrightarrow&~~ A_\mu=0\nonumber\\
       \sigma^{\mu\nu} (A_{\mu\nu}+i B_{\mu\nu})=0 ~~&\Longrightarrow&~~ 
                          A^{\mu\nu} = \half \epsilon^{\mu\nu\rho\sigma} B_{\rho \sigma}~
\end{eqnarray}
where in the second line $A_{\mu\nu}$ and $B_{\mu\nu}$ are 
each taken to be real and anti-symmetric in $(\mu,\nu)$ indices.

\begin{table}
 {\footnotesize
\begin{center}
\begin{tabular}{||c||c||c|c|c||}
         \hline
         \hline
    ~Term & $\Dbar^\alphadot J_{\alpha\alphadot}$ &
          $\matrix{{\rm super-}\cr {\rm conformal}}$  
                          & $\matrix{D_\alpha S,~{\rm where}\cr 
                            S:(\phi,\psi,F)\cr}$ &
                            $\matrix{L_\alpha,~{\rm where}\cr 
                            L:(\Lambda, \Phi_{\mu\nu},\Delta)\cr}$ \\
       \hline
       \hline
       ~ & ~ & ~ & ~ & ~\\
   ~no $\theta,\thetabar$ & $-i (\sigma^\mu \chibar_\mu)_\alpha$ & 0 & $\sqrt{2} \psi_\alpha$ & $-i \Lambda_\alpha$ \\
            ~ & ~ & ~ & ~ & ~\\
       \hline
            ~ & ~ & ~ & ~ & ~\\
   ~$\theta_\alpha$: (Re)  & $-\hat T$ & 0 & $2 \Re{F}$ & $\Delta$\\
   ~$\theta_\alpha$: (Im)  & $-i\partial^\mu C_\mu$ & 0 & $2 \Im{F}$ & 0 \\
            ~ & ~ & ~ & ~ & ~\\
   ~$(\sigma^{\mu\nu} \theta)_\alpha$ 
             & 
    $\matrix{ -(\hat T_{\mu\nu} - \hat T_{\nu\mu})  \cr
           + \half \epsilon_{\mu\nu\rho\sigma} (\partial^\rho C^\sigma - \partial^\sigma C^\rho)\cr}$ 
            & 0 & 0 & $ \half \epsilon_{\mu\nu\rho\sigma} \Phi^{\rho\sigma}$\\  
            ~ & ~ & ~ & ~ & ~\\
       \hline
            ~ & ~ & ~ & ~ & ~\\
   ~$(\sigma^\mu \thetabar)_\alpha$: (Re) & $-N_\mu$ & 0 & $-2 \partial_\mu \Im{\phi}$ & 0 \\
   ~$(\sigma^\mu \thetabar)_\alpha$: (Im) & $-iM_\mu$ & 0 & $2i \partial_\mu \Re{\phi}$ & 0 \\
            ~ & ~ & ~ & ~ & ~\\
       \hline
            ~ & ~ & ~ & ~ & ~\\
   ~$\theta\theta$ & $i(\sigma^\mu \lambdabar_\mu)_\alpha$ & 0 & 0 &  $(\sigma^\mu \partial_\mu \overline{\Lambda})_\alpha $ \\ 
            ~ & ~ & ~ & ~ & ~\\
       \hline
            ~ & ~ & ~ & ~ & ~\\
   ~$\thetabar\thetabar$ & 0 & 0 & 0 &  0 \\ 
            ~ & ~ & ~ & ~ & ~\\
       \hline
            ~ & ~ & ~ & ~ & ~\\
   ~$(\sigma^\nu\thetabar)_\gamma \theta^\beta $ &
    $\matrix{ -\delta^\gamma_\beta (\sigma^\mu \partial_\nu \chibar_\mu)_\alpha \cr
       -2i \delta^\gamma_\alpha \lambda_{\nu\beta} + 2\delta^\gamma_\alpha (\sigma^\mu\partial_\mu\chibar_\nu)_\beta \cr}$ 
          & 0 &
       $\matrix{  -\sqrt{2}i \delta^\gamma_\beta \partial_\nu \psi_\alpha \cr
                   + 2\sqrt{2} i \partial^\gamma_\alpha \partial_\nu\psi_\beta \cr} $ 
          & $-\delta^\gamma_\beta \partial_\nu  \Lambda_\alpha$   \\
            ~ & ~ & ~ & ~ & ~\\
       \hline
            ~ & ~ & ~ & ~ & ~\\
   ~$(\theta\theta) (\sigma^\mu \thetabar)_\alpha$:  (Re) & 
            $D_\mu + \half\Box C_\mu + \half \epsilon_{\mu\nu\rho\sigma}
         \partial^\nu \hat T^{\rho\sigma}$ & 0 & $-\partial_\mu \Im{F}$ &  $ \half \partial^\nu \Phi_{\mu\nu}$\\  
   ~$(\theta\theta) (\sigma^\mu \thetabar)_\alpha$:  (Im) & 
       $-{i\over 2} \partial^\nu \left(
            \hat T_{\mu\nu} +
            \hat T_{\nu\mu} - g_{\mu\nu} \hat T\right)$ & 0 & 
            $i \partial_\mu \Re{F}$  &  $-{i\over 2} \partial_\mu \Delta$\\ 
            ~ & ~ & ~ & ~ & ~\\
         \hline
            ~ & ~ & ~ & ~ & ~\\
   ~$(\thetabar\thetabar)\theta_\alpha$:  (Re) & 
               $-\half \partial^\mu M_\mu$ & 0 & $\Box\Re{\phi}$ & 0 \\
   ~$(\thetabar\thetabar)\theta_\alpha$:  (Im) & 
               ${i\over 2} \partial^\mu N_\mu$ & 0 & $i\Box\Im{\phi}$ & 0 \\
            ~ & ~ & ~ & ~ & ~\\
   ~$(\thetabar\thetabar)(\sigma^{\mu\nu}\theta)_\alpha$   & 
     $\matrix{ -\half(\partial_\mu M_\nu - \partial_\nu M_\mu) \cr
               -\quarter \epsilon_{\mu\nu\rho\sigma} (\partial^\rho N^\sigma - \partial^\sigma N^\rho)\cr}$
   & 0 & 0 & 0 \\ 
            ~ & ~ & ~ & ~ & ~\\
         \hline
            ~ & ~ & ~ & ~ & ~\\
   ~$\theta\theta\thetabar\thetabar$ & 
         $-\half (\sigma^\nu\sigmabar^\mu \partial_\mu\lambda_\nu)_\alpha 
     + {i\over 4} (\sigma^\mu \Box \chibar_\mu)_\alpha$ &
         0 & $ -{1\over 2\sqrt{2}} \Box\psi_\alpha $ & 
      $-{i\over 4} \Box \Lambda_\alpha$\\
            ~ & ~ & ~ & ~ & ~\\
         \hline
         \hline
            ~ & ~ & ~ & ~ & ~\\
  ~ & ~$\matrix{{\rm superconformal}\cr
               {\rm ~~~~~anomalies:}\cr}
        ~\cases{
         \partial^\mu C_\mu & ~\cr
         (\sigmabar \chi)_\alpha & ~\cr
         \hat T & ~\cr}$
     & 
    $\matrix{ 0\cr 0\cr 0\cr }$ &
    $\matrix{ -2 \Im{F}\cr
              \sqrt{2} i \psi_\alpha\cr
                -2\Re{F} \cr}$ &
    $\matrix{0 \cr \Lambda_\alpha\cr -\Delta\cr}$\\
            ~ & ~ & ~ & ~ & ~\\
         \hline
         \hline
\end{tabular}
\end{center}
\caption{Three different structures for the supercurrent supermultiplet, corresponding to
the unbroken superconformal case, the maximally broken (chiral) case,
and the (linear) case in which superconformal symmetry is broken but $R_5$-symmetry is preserved.}
\label{constraintstable}
   }
\end{table}

Solving the resulting constraint equations in each case,
we indeed find that our anomalies take the required vanishing or non-vanishing
forms indicated at the bottom of Table~\ref{constraintstable}.
Moreover, given these solutions to the constraint equations,
we can also determine which particular combinations of $\chi_\mu$ and $\hat T_{\mu\nu}$
are conserved and can thus correspond to
the supercurrent $j_\mu$ and energy-momentum tensor $T_{\mu\nu}$, respectively.
Our results are as follows.
In the superconformal case, we find
\begin{equation}
      \cases{  j_{\mu\alpha} = \chi_{\mu\alpha} & ~\cr
               T_{\mu\nu} =  -\quarter(\hat T_{\mu\nu} + \hat T_{\nu\mu}) &~\cr}
\label{case1}
\end{equation}
while in the chiral case, we find
\begin{equation}
      \cases{  j_{\mu\alpha} = \chi_{\mu\alpha} + (\sigma_\mu \sigmabar^\nu \chi_\nu)_\alpha  & ~\cr
               T_{\mu\nu} =  -\quarter( \hat T_{\mu\nu} + \hat T_{\nu\mu}  - 2 g_{\mu\nu} \hat T) ~ &~\cr} \Longrightarrow~~
      \cases{  \chi_{\mu\alpha} = j_{\mu\alpha} + {1\over 3} (\sigma_\mu \sigmabar^\nu j_\nu)_\alpha  & ~\cr
               \hat T_{\mu\nu}|_{\rm symm}  =  -2T^{\mu\nu} + {2\over 3} g_{\mu\nu} T ~   &~\cr} 
\label{case2}
\end{equation}
where $\hat T_{\mu\nu}|_{\rm symm}$ indicates the symmetric part of $\hat T_{\mu\nu}$.
Note that Eq.~(\ref{case2}) implies the anomaly relations
$\sigmabar^\mu j_\mu = -3 \sigmabar^\mu \chi_\mu$ 
and 
$T= 3\hat T/2$.
However, in the linear case, we find
\begin{equation}
      \cases{  j_{\mu\alpha} = \chi_{\mu\alpha}  & ~\cr
               T_{\mu\nu} =  -\quarter (\hat T_{\mu\nu} + \hat T_{\nu\mu})   ~.&~\cr}
\label{case3}
\end{equation}
Thus, we see that the linear case resembles the superconformal case, even though
the superconformal symmetry is broken and $\sigmabar^\mu \chi_\mu$ and $T$ are both non-zero. 
Indeed, the critical feature that links these two cases is the fact that $R_5$-symmetry
is unbroken.

Ultimately, in passing from global supersymmetry to local supersymmetry,
these supercurrents are coupled
to the supergravity multiplet.
As a result, each of these formalisms leads to a different
self-consistent formulation of off-shell supergravity with
a unique auxiliary-field content.
In the chiral formalism~\cite{ChiralFormalismSUGRA}, 
for example, the complete supercurrent supermultiplet takes the form
$(j^{(5)}_\mu, j_{\mu\alpha}, T_{\mu\nu},P,Q)$ where $P$ and $Q$
are two additional scalar currents.
This multiplet can therefore couple to a supergravity multiplet of the
form
$(b_\mu, \psi_{\mu\alpha}, e^m_\mu,M,N)$, where 
where $b_\mu$ is the connection field associated with
local chiral rotations, 
where $\psi_{\mu\alpha}$ is the gravitino,
where $e^m_\mu$ is the vielbein,
and where $M$ and $N$ are additional real
scalar degrees of freedom.  
Indeed, this supergravity multiplet is nothing but a particular
truncation of the more general superconformal supergravity multiplet
$(b_\mu, \psi_{\mu\alpha},  e^m_\mu, a_\mu)$
where $a_\mu$ is the connection field associated with
local chiral dilatations.
Thus, we see that the form of the supercurrent supermultiplet uniquely fixes
the form of the corresponding supergravity supermultiplet. 

By contrast, in the linear formalism 
(also occasionally called the ``new'' minimal formalism)~\cite{LinearFormalismSUGRA},
the full supercurrent supermultiplet has the structure
of a linear multiplet 
$(C_\mu ,\chi_{\mu \alpha},\hat T_{\mu\nu}|_{\rm symm}, \hat T_{\mu\nu}|_{\rm anti-symm})$.
This then corresponds to a 
supergravity multiplet which can be written in the form
$(b_\mu,  \psi_{\mu\alpha}, e^m_\mu,a_{\mu\nu})$, where $a_{\mu\nu}$
is an antisymmetric tensor.
As we shall discuss in Sect.~5, this formalism can only be
applied to $R_5$-invariant theories.

Clearly, the off-shell structures of the supergravity multiplet
in the chiral and linear formalisms are quite different.
However, it has been shown in Ref.~\cite{RelateChiralLinearSUGRA}
that although the linear  and chiral formalisms  
are not  generally equivalent, 
they become so at the classical level
if the Lagrangian of the theory respects $R_5$-symmetry.
This will be discussed in more detail in Sect.~6.1.
In such a case, 
the corresponding supercurrent supermultiplets can be 
related to each other through the addition of appropriate improvement 
terms~\cite{SupercurrentAnomaliesRenorm}.
Their corresponding anomaly multiplets can also be related in a similar
way.  These relations will be discussed more fully in Sect.~6.

Each of these formalisms has its advantages
and disadvantages, as well as its range of validity in terms of the
globally supersymmetric theories to which it may be applied.
For example, the chiral formalism is applicable
in a wide variety of contexts, including models with explicit $R_5$-symmetry
breaking, though the superconformal gauge choice which renders it useful
for calculation can also obscure the symmetry properties of the underlying
theory, especially those associated with $R_5$-symmetries.
By contrast, the linear formalism can only be
applied to $R_5$-invariant theories.  However, it has the
advantage of manifesting the full $R_5$-invariance
of such theories, which is obscured by superconformal
gauge-fixing in the chiral formalism.

In the special case when the theory possesses an $R_5$-symmetry,
these two formalisms are related by a duality transformation.
This will be discussed in Sect.~6.
Nevertheless, even for $R_5$-invariant theories,
the representations of the supergravity multiplets in these 
two formalisms differ in many salient aspects.
In the chiral formalism, for example, the usual form of the FI
Lagrangian (\ie, the $D$-term within the vector superfield $\xi V$) 
is not gauge invariant.  However, this can 
be remedied~\cite{StelleAndWestEToTheV}  
by replacing the FI Lagrangian with the $D$-term associated with the 
superfield $\kappa^{-2}e^{\kappa^2\xi V}$
(with $\kappa^2\equiv M_P^{-2}$).
In this case, gauge transformations of the form
$V\rightarrow V'=V+\Phi+{\Phi^\dagger}$, where $\Phi$ denotes a
chiral superfield and $\Phi^\dagger$ its conjugate, can be
compensated for by mandating that the fields of the supergravity
multiplet transform in such a manner
as to render the action invariant~\cite{BarbieriEtAlEToTheV}.
If the theory contains matter, the matter fields must also
transform non-trivially under $U(1)$ gauge shifts in order to compensate
for the requisite shifts in the supergravity fields.  Consequently,
the $U(1)$ charges of the matter fields of the theory are modified
from their globally supersymmetric values~\cite{DasMilicharges}.  One can show
that a given action can only be made gauge invariant if the
corresponding superpotential has an $R_5$-symmetry.

By contrast,
the picture in the linear formalism is quite different.  Here, the
standard form of the FI Lagrangian --- appropriately modified to
include couplings between the fields in $V$ and the fields of
the supergravity multiplet --- is manifestly gauge
invariant~\cite{MultipletsInLinearFormalism}.  The theory is
$R_5$-invariant by assumption, and, as stated in the Introduction, the $U(1)$
invariance of the theory involves a combination of
$U(1)_{\mathrm{FI}}$ gauge shifts and $U(1)_R$ rotations.

\subsection{Theories with unbroken $R_5$-symmetry}

Because $R_5$-symmetry is preserved in both the superconformal and linear cases,
the supercurrent supermultiplet that emerges in the linear case shares many properties with
the multiplet that emerges in the superconformal case. 
In particular, both supercurrent multiplets are truncations of the general supercurrent
supermultiplet in Eq.~(\ref{vecsup2}) in which $M_\mu$ and $N_\mu$ vanish --- \ie, they 
are {\it linear}\/ multiplets.
We can therefore probe their common structure by 
examining the general properties of supermultiplets
in which these two components vanish.
Consulting Eq.~(\ref{susytransforms2}),
we find that such a truncated multiplet can indeed be 
consistent with the supersymmetry algebra
only if its remaining components satisfy the additional relations
\begin{eqnarray}
         \lambda_{\mu\alpha}  &=&  -i (\sigma^\nu \partial_\nu \chibar_\mu)_\alpha  \nonumber\\
          \partial^\nu \hat T_{\nu\mu} &=& 0~\nonumber\\
               D_\mu &=& - \Box C_\mu~.
\label{multicon}
\end{eqnarray} 
Note, in particular, that the second equation in Eq.~(\ref{multicon}) 
is  {\it not}\/ a conservation law for $\hat T_{\nu\mu}$ because 
$\hat T_{\nu\mu}$ is not generally symmetric, and because
a conservation law would be a statement about 
$\partial^\mu \hat T_{\nu\mu}$, not $\partial^\nu \hat T_{\nu\mu}$.
Given $M_\mu=N_\mu=0$ and Eq.~(\ref{multicon}), 
the supersymmetry transformations in 
Eq.~(\ref{susytransforms2}) then reduce to
\begin{eqnarray}
   \delta_\epsilon C_\mu &=& i \epsilon \chi_\mu - i \epsilonbar\chibar_\mu~,\nonumber\\
   \delta_\epsilon \chi_{\mu\alpha}  &=& 
                      (\sigma^\nu \epsilonbar)_\alpha (\partial_\nu C_\mu + i \hat T_{\nu\mu}) ~,\nonumber\\
   \delta_\epsilon \hat T_{\nu\mu} &=& 
        2 \epsilonbar \sigmabar_{\nu\rho} \partial^\rho \chibar_\mu 
       + 2 \epsilon \sigma_{\nu\rho} \partial^\rho \chi_\mu ~.
\label{reducedalgebra}
\end{eqnarray}

Several things are immediately apparent from Eq.~(\ref{reducedalgebra}).
First, acting with $\partial^\mu$ on the first equation within Eq.~(\ref{reducedalgebra}), we
see that
\begin{equation}
              (\partial^\mu C_\mu =0)   ~~~\Longrightarrow~~~  (\partial^\mu \chi_\mu=0)~.
\label{firstresult}
\end{equation}
This explains why the conservation of $R_5$-symmetry and the identification $j_\mu^{(5)}= C_\mu$
together compel the identification $j_{\mu\alpha}=\chi_{\mu\alpha}$ for such multiplets, as already seen
in Eqs.~(\ref{case1}) and (\ref{case3}).  
Moreover, taking the trace of the third equation in Eq.~(\ref{reducedalgebra}),
we find 
\begin{equation}
              (\hat T=0)  ~~~\Longrightarrow~~~  (\sigma^\mu \chibar_\mu=0)~. 
\label{secondresult}
\end{equation}
Indeed, these results are intrinsic to the structure of any multiplet 
in which $M_\mu$ and $N_\mu$ vanish. 

It is important to note that the intrinsic structure of such a multiplet  
does {\it not}\/, in and of itself, require that $\partial^\mu C_\mu=0$, nor that
$\sigma^\mu\chibar_\mu=0$ or $\hat T=0$.  
Indeed, whether or not such additional constraints hold depends on whether we demand
that such a multiplet $J_{\alpha\alphadot}$ additionally satisfy either
the superconformal condition $\Dbar^\alphadot J_{\alpha\alphadot} = 0$ or
the ``linear" condition $\Dbar^\alphadot J_{\alpha\alphadot} = L_\alpha$, where
$L_\alpha$ satisfies the constraints in Eq.~(\ref{Lconstraints}).
Consulting Table~\ref{constraintstable}, we find that 
imposing these additional conditions leads to the constraint 
$\partial^\mu C_\mu=0$ in both cases, while
$\sigma^\mu\chibar_\mu$ and $\hat T$ are either 
both zero (in the superconformal case) or non-zero (in the linear case).  
Likewise, again consulting 
Table~\ref{constraintstable}, we see that 
$\hat T_{\mu\nu}$ will be symmetric in the superconformal case,
while it will not be symmetric in the linear case.
In either case, however, the components $(C_\mu,\chi_\mu, \hat T_{\nu\mu})$
are guaranteed to satisfy Eqs.~(\ref{reducedalgebra}), (\ref{firstresult}), and (\ref{secondresult}).

Thus far, we have been discussing the behavior of $R_5$-symmetric theories
in the linear formalism.  We have been concentrating on the linear formalism
because we necessarily have $\partial^\mu C_\mu=0$ in the linear case, while
we necessarily have $\partial^\mu C_\mu\not =0$ in the chiral case.
Indeed, we cannot have $\partial^\mu C_\mu=0$ within the chiral case unless
$S=0$, and this only occurs for a superconformal theory. 
Thus, strictly speaking, an $R_5$-symmetric theory has a supercurrent
supermultiplet which can only be described through the linear formalism.

Despite this, 
there can be situations in which we might wish to use the
chiral formalism, even for theories in which 
$R_5$-symmetry is preserved but
superconformal invariance is broken.
Indeed, while the linear-case formalism 
incorporates this symmetry structure in the most
manifest way, it may still be possible to adopt the chiral-case 
formalism in which this full symmetry structure is not manifest.
Therefore, for an $R_5$-invariant theory, we have the possibility of constructing
two distinct supercurrent supermultiplets, $J_{\alpha\alphadot}^{(L)}$
and $J_{\alpha\alphadot}^{(C)}$, the first of which satisfies
$\Dbar^\alphadot J_{\alpha\alphadot}^{(L)} = L_\alpha$
and the second of which satisfies 
$\Dbar^\alphadot J_{\alpha\alphadot}^{(C)} = D_\alpha S$.
We may interpret 
$J_{\alpha\alphadot}^{(L)}$
as the supercurrent 
supermultiplet, unique up to K\"ahler transformations,
that may be formed from our original three Noether currents
(corresponding to $R_5$-chiral rotations, supersymmetry transformations,
and spacetime translations)
through the addition of a specific
set of improvement terms of each case.
By contrast, we may interpret 
$J_{\alpha\alphadot}^{(C)}$
as an 
alternative supercurrent supermultiplet
which can  be formed for such a theory
if we are willing to disregard any connection between $C_\mu$ and
$j^{(5)}_\mu$, and only focus on embedding $j_{\mu\alpha}$ and $T_{\mu\nu}$
into a supercurrent superfield.
Such an embedding is clearly different from the embedding that would occur
within $J_{\alpha\alphadot}^{(L)}$, and corresponds to different improvement
terms for $j_{\mu\alpha}$ and $T_{\mu\nu}$.  Yet, for an $R_5$-invariant theory,
 $J_{\alpha\alphadot}^{(C)}$ represents
an alternative supercurrent superfield that we may consider
if we are willing to concentrate on $j_{\mu\alpha}$ and $T_{\mu\nu}$ only,
and disregard any connection between 
$j^{(5)}_\mu$ (which is conserved) and
$C_\mu$ (which is not conserved).

Clearly, for $R_5$-invariant theories,
$J_{\alpha\alphadot}^{(C)}$ does not capture the full symmetry properties
of the theory.  
Indeed, the bottom component of 
$J_{\alpha\alphadot}^{(C)}$ is not conserved 
and therefore cannot bear any relation to   
$j_\mu^{(5)}$, regardless of the addition of any possible improvement terms.
Therefore,
strictly speaking, 
the supercurrent supermultiplet $J_{\alpha\alphadot}^{(C)}$ 
does not correspond to the original $R_5$-symmetric theory in question.
However, as we shall discuss in Sect.~4, 
$J_{\alpha\alphadot}^{(C)}$ actually corresponds to a {\it different}\/ theory which is
closely related to our original $R_5$-symmetric theory.
Thus, it is with this understanding that we shall loosely speak of an $R_5$-symmetric
theory as having two possible supercurrent supermultiplets, 
$J_{\alpha\alphadot}^{(L)}$
and  
$J_{\alpha\alphadot}^{(C)}$.

\subsection{An example, both with and without an FI term}

As an example of these results, 
let us consider the specific pure-gauge $U(1)$ theory defined by the Lagrangian
\begin{eqnarray}
 \calL &=&  \quarter \left(W^\alpha W_\alpha|_{\theta\theta} + \Wbar_\alphadot \Wbar^\alphadot|_{\thetabar\thetabar}\right)  
      \nonumber\\   
       &=& -\quarter F^2  
                   - {i\over 2} \lambda \sigma^\mu (\partial_\mu \lambdabar)
                   + {i\over 2} (\partial_\mu \lambda) \sigma^\mu \lambdabar + \half D^2 
\label{lag}
\end{eqnarray}
where $W_\alpha$ is the gauge supermultiplet and $\Box = \partial_\mu\partial^\mu $.
The equations of motion for the different fields in this theory are then given by
$\Dbar_\alphadot \Wbar^\alphadot=0$, \ie,
\begin{equation}
       \partial_\mu F^{\mu\nu}=0~,               ~~~~
        \sigmabar^\mu \partial_\mu\lambda =
        \sigma^\mu \partial_\mu\lambdabar = 0~,~~~~ D=0~,       
\label{eoms}
\end{equation}
and the Noether currents corresponding to this theory are 
given by
\begin{eqnarray}
    j^{(5)}_\mu &=&  -\lambda \sigma_\mu \lambdabar \nonumber\\
    j_{\mu\alpha} &=&   -i(F_{\mu\nu}+\tilde F_{\mu\nu})(\sigma^\nu \lambdabar)_\alpha 
                            -F_{\mu\nu}\partial^\nu \chi_\alpha   \nonumber\\  
    T_{\mu\nu} &=&  -\quarter g_{\mu\nu} F^2  + F_{\mu\rho} \partial_\nu A^\rho \nonumber\\
           && ~~~~ +{i\over 2} \lambda \sigma_\mu (\partial_\nu \lambdabar) 
                  -{i\over 2} (\partial_\nu \lambda) \sigma_\mu \lambdabar + \half g_{\mu\nu} D^2~ 
\label{Noetherresults}
\end{eqnarray}
where
$\tilde F^{\mu\nu} \equiv {\textstyle {i\over 2}} \epsilon^{\mu\nu\lambda\sigma} F_{\lambda\sigma}$.
Note that we may use the equations of motion in Eq.~(\ref{eoms}) in order 
to eliminate the $D$-term within $T_{\mu\nu}$ if we wish.
Moreover,
using the equations of motion,
it is straightforward to verify that
each of these currents is conserved, {\it i.e.}\/,
\begin{equation}
       \partial^\mu j^{(5)}_\mu =0~,~~~~~~~
       \partial^\mu j_{\mu\alpha} =0~,~~~~~~~
       \partial^\mu T_{\mu\nu} =0~,
\end{equation}
as guaranteed by the Noether construction.

At first glance, it may seem a bit surprising that 
the supercurrent and energy-momentum tensor contain terms of the form
$-F_{\mu\nu}\partial^\nu \chi_\alpha$
and 
$ F_{\mu\rho} \partial_\nu A^\rho$,
respectively, which are not gauge invariant.
However, the first of these does not affect the supercharge (as may be verified by noting
that the three-space integral of its zero component vanishes), and consequently
it can be ``improved'' away.
Likewise, the second becomes the gauge-invariant expression $F_{\mu \rho} {F_\nu}^\rho$
when it is improved via the addition of the improvement term
$-F_{\mu \rho} \partial^\rho A_\nu$.  
As required, this latter term is also conserved and has a zero-component with a vanishing
three-integral.
Thus, any failure of gauge invariance in this theory is at best only spurious.

This theory is clearly an example of the superconformal case in which the 
superconformal and $R_5$-symmetries are both unbroken.
Indeed,
each of these currents can be improved so that they together fill
out the supercurrent supermultiplet~\cite{FZ}:
\begin{equation}
          J^\mu\equiv -\half \sigmabar^{\mu\alphadot\alpha} J_{\alpha\alphadot}
  ~~~{\rm where}~~~
            J_{\alpha\alphadot}\equiv 2\, W_\alpha \overline{W}_\alphadot~.
\label{puregauge}
\end{equation}
Expanding this multiplet in components, we find
\begin{eqnarray}
         C_\mu &\equiv& -\lambda \sigma_\mu \lambdabar\nonumber\\  
      \chi_{\mu\alpha} &\equiv& - (\sigma_\mu \lambdabar)_\alpha D - i (\sigma^\nu \lambdabar)_\alpha 
                  (F_{\mu\nu} + \tilde F_{\mu\nu}) \nonumber\\
       \hat T_{\nu\mu} &\equiv& ({F_\mu}^\rho + \tilde {F_\mu}^\rho)(F_{\rho\nu} -\tilde F_{\rho\nu}) +
                    2i D\tilde F_{\mu\nu} \nonumber\\
       && ~~~~~~~~ + i(\partial_\nu \lambda)\sigma_\mu \lambdabar
                   - i\lambda\sigma_\mu (\partial_\nu \lambdabar)  + g_{\mu\nu} D^2 ~.
\label{compons}
\end{eqnarray}
Use of the relations in Eq.~(\ref{case1}), as appropriate for a superconformal theory,
then leads to the ``improved'' expressions for $j_\mu^{(5)}$, $j_\mu$, and $T_{\mu\nu}$ in
this theory.

Given these results, let us now consider the same theory with an FI term.
Indeed, this now becomes the simplest
possible theory that can be constructed with an FI term:
\begin{eqnarray}
 \calL &=&  \quarter \left(W^\alpha W_\alpha|_{\theta\theta} + \Wbar_\alphadot 
           \Wbar^\alphadot|_{\thetabar\thetabar}\right)  +  2\xi V|_{\theta\theta\thetabar\thetabar}
      \nonumber\\   
       &=& -\quarter F^2  
                   - {i\over 2} \lambda \sigma^\mu (\partial_\mu \lambdabar)
                   + {i\over 2} (\partial_\mu \lambda) \sigma^\mu \lambdabar + \half D^2 
                     + \xi (D+\half \Box C)
\label{lag2}
\end{eqnarray}
Because the $C$-term in Eq.~(\ref{lag2}) is a total derivative, it is legitimate
to drop this term completely from any subsequent analysis.\footnote{
      Note that the dropping of the $C$-term does {\it not}\/ mean that we have
      passed to Wess-Zumino gauge;  in particular, we are continuing to operate 
      in a completely general gauge.  We have 
      simply dropped the $C$-term because it 
      is a total derivative and as such contains no physics at the classical level.
      Indeed,
      if we had retained the $C$-term in the following, we would have found that the
      corresponding equations of motion would have left $C$ entirely unconstrained;
      likewise, the corresponding Noether $R_5$-current and supercurrent 
      would not have depended on $C$ in any way,
      and the Noether energy-momentum tensor would have accrued an extra $C$-dependent 
      term $\half \xi (g_{\mu\nu}\Box - \partial_\mu\partial_\nu) C$  
      which is in the form of an improvement term --- 
      \ie, a term which is conserved (in this case both on- and off-shell) 
      and which makes no contribution to the corresponding physical Noether charge.  
      It is therefore
      legitimate to drop the $C$-term from the Lagrangian, even in completely general gauge.}
We then find that the equations of motion for this theory are unchanged from those
of Eq.~(\ref{eoms}), except that we now have $D=-\xi$.
In other words, the equations of motion of this theory now take the form
\begin{equation}
            \Dbar_\alphadot \Wbar^\alphadot ~=~ 2\xi~.
\label{FIEOMs}
\end{equation}
Likewise, straightforwardly repeating the Noether procedure,
it is easy to verify that
the presence of the FI term induces the following additional $\xi$-dependent contributions 
to the Noether currents in Eq.~(\ref{Noetherresults}):
\begin{eqnarray}
          \Delta j_\mu^{(5)} &=& 0\nonumber\\
          \Delta j_{\mu\alpha} &=& \xi (\sigma_\mu \lambdabar)_\alpha\nonumber\\
          \Delta T_{\mu\nu} &=&  \xi g_{\mu\nu} D~.
\label{Noether2}
\end{eqnarray}

There are several features to note concerning the results in Eq.~(\ref{Noether2}). 
First, we see that {\it the results for these Noether currents are completely gauge invariant}\/.
This is true for the full gauge invariance associated with the shifts
in Eq.~(\ref{gengaugesuperfield})
as well as the restricted traditional $U(1)$ gauge symmetry 
which remains after truncation to Wess-Zumino gauge.
This makes sense, since we have already seen from Table~\ref{FIinvariances} that 
the FI term itself respects these symmetries as far as the action is concerned.

Second, we observe that this theory now effectively has a cosmological constant as
a result of the FI term. 
Previously, without the FI term, equations of motion could be used to eliminate
the final term $\half g_{\mu\nu} D^2$ which appeared in $T_{\mu\nu}$.
However, in the presence of an FI term, our equation of motion for $D$ becomes $D= -\xi$. 
As a result, this term now yields a constant $\half \xi^2 g_{\mu\nu}$.
Combining this with the extra contribution $\xi g_{\mu\nu} D$ coming from Eq.~(\ref{Noether2})
yields an overall  constant $-\half \xi^2 g_{\mu\nu}$.
This term is nothing but the effective cosmological constant $\Lambda = +\half \xi^2$
induced by the FI term;  indeed, $\Lambda>0$ (as required for such a broken-supersymmetry theory
in flat space) because $g^{00}= -1$.

But most importantly for our purposes, we observe from Eq.~(\ref{Noether2})
that the appearance of an FI term does not affect the $R_5$-current.
This is entirely as expected:  the FI term is composed of the $D$ and $C$ fields, and these
are both entirely neutral under the chiral $R_5$-symmetry transformations.
This, then, is  {\it an explicit Noether-derived verification of the fact
that FI terms, in and of themselves, 
do not break $R_5$-symmetries.}   
In particular, we have $\partial^\mu j_{\mu}^{(5)}=0$ both with and without the inclusion
of the FI term.  Thus, this property must remain true --- even with the possible addition
of improvement terms --- for any supercurrent supermultiplet which is to be associated
with the FI term. 

Having assembled at our disposal all the results we will need, 
we now seek to construct the corresponding supercurrent supermultiplet
for this theory in the presence of an FI term.  This should take the form
\begin{equation}
          J^\mu\equiv -\half \sigmabar^{\mu\alphadot\alpha} J_{\alpha\alphadot}
  ~~~{\rm where}~~~
            J_{\alpha\alphadot}\equiv 2\, W_\alpha \overline{W}_\alphadot~ + \Xi_{\alpha\alphadot}~
\label{puregaugeplusFI}
\end{equation}
where $\Xi_{\alpha\alphadot}$ represents the extra $\xi$-dependent contribution 
to the supercurrent supermultiplet.
As discussed at the end of Sect.~3.3, $\Xi_{\alpha\alphadot}$ can be different
depending on whether we work within the chiral or linear formalism.
But what precisely is this contribution from the FI term in each case?
That is the question to which we now turn.

\section{Analysis in the chiral formalism}
\setcounter{footnote}{0}

In this section, we address the issue of deriving $\Xi_{\alpha\alphadot}$
within the chiral formalism.
We begin by reviewing the formalism of so-called ``chiral compensators'',
as this will be our method for deriving the corresponding supercurrent.
We then discuss the implications of these results for the existence
of additional global symmetries in any theory with a non-zero FI term.
Finally, we provide a proof that there does not exist any solution
for $\Xi_{\alpha\alphadot}$ in which $R_5$-symmetry is conserved,
and discuss one possible method by which such a proof might be evaded.

\subsection{Chiral compensator formalism:  General outline}

Our chief interest in this paper concerns the manner in which
a globally supersymmetric theory can be coupled to supergravity
(\ie, be made locally supersymmetric).
Of course, for a theory which exhibits a full superconformal invariance,
the answer is simple:   since the relevant currents are all conserved, 
they can be coupled directly      
to the fields of the
the conformal supergravity multiplet $(b_\mu, \psi_{\mu\alpha}, e_\mu^m,a_\mu)$.
By contrast, a supersymmetric theory which is not superconformal cannot
be coupled directly to this multiplet.  However, it is always possible to
``promote'' such a theory to a superconformal one by adding to the
theory a set of additional fields called ``conformal compensators''
which artificially restore superconformal invariance to the theory.
These conformal compensator fields are introduced in 
such a way that salient algebraic aspects of 
the original non-conformal theory can be reproduced from the conformal
theory simply by assigning
certain fixed values to these conformal compensators;  indeed,
assigning fixed values to such fields necessarily
breaks any symmetries with respect to which these fields carry
a charge.  
In this way, the entire anomaly structure of the original non-superconformal theory 
is embedded into the structure of the compensator fields.

Given such conformal compensators, it is then straightforward to couple
any globally super-Poincar\'e invariant theory to supergravity, regardless of its
symmetry structure.  We simply promote the theory to a superconformal
theory by judiciously introducing appropriate conformal compensator fields, and then
couple the resulting superconformal theory to the conformal gravity multiplet.
As we shall see, 
setting the conformal compensator fields to fixed values then reproduces
either our original theory or a theory whose symmetry properties are the
same as those of our original theory.

As discussed in Sect.~3,
many different constructions of this sort exist, each with its own set of
conformal compensator fields.  However, it can be shown~\cite{KugoNonminimal} 
that only two such formulations may be regarded as minimal, meaning that the auxiliary-field
formulation of supergravity to which the compensated theory may be coupled
contains the minimal number of auxiliary degrees of freedom necessary to
achieve equal numbers of bosonic and fermionic degrees of freedom.  These
are the chiral (``old minimal'') and linear (``new minimal'') formulations.
In this section, we shall discuss the chiral compensator formalism,
deferring a discussion of the linear compensator formalism to Sect.~5.
In each case, we shall also emphasize how FI terms are incorporated
into the formalism, as well as the consequences that result from doing so. 

In general, the conformal (scaling) properties of a given quantity $X$ are determined by
its so-called Weyl charge (or Weyl weight) $w_X$.  
Gauge vector superfields $V$ and gauge field-strength superfields $W_\alpha$ have Weyl weights 
$(w_V,w_{W_\alpha})=(0,3/2)$, while constants (even dimensionful constants)
have vanishing Weyl weights. 
The Weyl weights $w_i$ corresponding to chiral matter superfields $\Phi_i$ 
can vary depending on the theory in question.
Since our superspace coordinates $(x,\theta)$ have Weyl weights $(w_x,w_\theta)=(-1,-1/2)$
respectively, we see that a globally supersymmetric theory will be conformally invariant
only if its K\"ahler potential $K$ and superpotential $W$ have Weyl charges
\begin{equation}
         w_K ~=~ 2~,~~~~~~~ w_W~=~ 3~.
\label{confconstraint}
\end{equation}
If the K\"ahler potential and superpotential of a given theory do not satisfy Eq.~(\ref{confconstraint}),
then the theory is not conformal.  

A similar situation exists for charges under $R_5$-transformations.
Specifically, if the K\"ahler potential and superpotential 
of the theory do not have $R_5$-charges $0$ and $2$ respectively, then the theory also breaks
$R_5$-symmetry. 
Note that while the $R_5$-charge of a given chiral superfield $\Phi_i$
is arbitrary, depending on the theory in question, a given left- (right-)chiral 
superfield $\Phi_i$ must have an $R_5$-charge $R_i^{(5)}$ and Weyl weight $w_i$ which satisfy
\beq
               R_i^{(5)} ~=~ \pm {2\over 3} \, w_i~.
\label{WeylRconstraint}
\eeq
This condition ensures that a left- (right-)chiral superfield remains left- (right-)chiral
under both $R_5$-rotations and Weyl rescalings~\cite{KugoUehara}. 
By contrast, other types of multiplets satisfy different relations.
For example, linear multiplets satisfy $R^{(5)}= 2(w-2)/3$.

The relation in Eq.~(\ref{WeylRconstraint}) is the minimum constraint needed for
a general chiral superfield.  In this context, we remark that all $R$-symmetries are 
essentially on the same footing in the absence of any coupling to supergravity;  indeed, it is only the coupling
to supergravity that selects one of these $R$-symmetries (here denoted $R_5$) to become local.
Thus the question of which symmetry can be identified as $R_5$ is ultimately a formalism-dependent question. 
In the chiral formalism, $R_5$ corresponds to the charge assignments
$R^{(5)}_i=2/3$, or $w_i=1$, for all chiral superfields $\Phi_i$.  With this convention,
$R_5$ may or may not actually be a symmetry of our original theory.
By contrast, in Sect.~5, we shall see that there is no constraint on $R_5$ in the linear formalism;
$R_5$ can be associated with whichever $R$-symmetry is preserved, and the use of the
linear formalism presupposes that there is at least one such symmetry in our theory.
However, it can be shown~\cite{SupercurrentAnomaliesRenorm}
that the linear formalism will not yield a supercurrent $J_{\alpha\alphadot}$
satisfying Eq.~(\ref{linearcase}), and likewise will not yield a symmetric energy-momentum
tensor $T_{\mu\nu}$, unless all $R_i^{(5)}$ are equal.

Conformal (Weyl) symmetry and $R_5$-symmetry are closely related.
By itself, $R_5$-symmetry is a $U(1)$ symmetry, but this is merely
an ordinary $U(1)$ symmetry, not a full $U(1)$ superfield symmetry
as in Eq.~(\ref{gaugetransforms}).
However, when joined with conformal (Weyl) symmetry and so-called ``special SUSY'' 
transformations,  $R_5$ 
fills out a full super-Weyl $U(1)_\SW$ superfield symmetry of the form 
in Eq.~(\ref{gaugetransforms}).
In other words, $R_5$-symmetry is what remains 
of the full $U(1)_\SW$ superfield symmetry in its
Wess-Zumino gauge. 
The above relations between the $R_5$-charges and conformal $w$-weights
for different kinds of superfields simply enforce the requirement that
these superfields retain their defining characteristics (chiral, vector, linear, 
 {\it etc.}\/.) under $U(1)_\SW$.
As a matter of normalization convention, a chiral superfield $\Phi_i$ will be defined
to have charge $q_\SW \equiv R_i^{(5)}$ under $U(1)_\SW$.

If a given theory breaks conformal symmetry and/or $R_5$-symmetry, 
we can restore these symmetries by introducing two conformal compensator fields:
a chiral superfield $\Sigma$, and its hermitian conjugate $\Sigmabar$.
These fields are respectively assigned $R_5$-charges $\pm 2/3$ 
and Weyl weights $w_\Sigma=w_\Sigmabar=1$.  
Given such fields, our prescription for promoting our theory to a superconformal one
is straightforward.

First, in the superpotential,
we define the new fields $\Phitilde_i$ through the
relations 
\begin{equation}
         \Phi_i \equiv \left( {\Sigma \over \sqrt{3} M_P}\right) \Phitilde_i~
\label{newfields}
\end{equation}
where $M_P$ is the Planck mass and the factor of $\sqrt{3}$ is merely conventional.
A similar definition holds for the conjugate fields $\Phi^\dagger$ and $\Sigmabar$.
Note that by construction, these new fields $\Phitilde, \Phitilde^\dagger$ have vanishing
Weyl weights and $R_5$-charges.
We then re-express our original fields $\Phi_i$ in terms of the new fields $\Phitilde_i$.
By contrast, any {\it coupling}\/ $X_n$ that appears in the superpotential with mass dimension $n$
(such as a mass $m_{ij}$, or a Yukawa coupling $y_{ijk}$) 
is algebraically replaced with a corresponding {\it superfield}\/: 
\begin{equation}
         X_n ~\to~   
         \left( {\Sigma\over \sqrt{3} M_P}\right)^{n} \widetilde X_n~.
\label{eq:DimParamSub}
\end{equation}
Here $\widetilde X_n$ can be viewed as another coupling
with the same numerical value in fixed units. 
The net effect of these operations is thus to shift
\begin{equation}
          W ~\to~  \widehat W\equiv \left( {\Sigma \over \sqrt{3} M_P}\right)^3  \Wtilde~
\label{newW}
\end{equation}
where $\Wtilde$ has vanishing $R_5$-charge and Weyl weight.
Thus, we have succeeded in constructing a new superconformal superpotential $\widehat W$.

Note that the shift from $\Phi_i$ to $\Phitilde_i$ is merely an algebraic rewriting.
By contrast, the replacement of the coupling $X_n$ with the superfield in Eq.~(\ref{eq:DimParamSub})
for $n\not=0$ fundamentally changes the theory, producing a superconformal 
superpotential from a non-superconformal one.
Of course, superpotentials $W$ without such couplings will already be superconformal.
In such cases, we find that $\widehat W=W$.
 
In general, we also must modify our K\"ahler potential $K$.  In this case,
the procedure is easy.  First, we may write $K=K(\Phi,\Phi^\dagger,V)=\sum_n K_n$ where
$K_n$ has Weyl weight $n$ and vanishing $R_5$-charge.
Given this form, we then define
a new quantity
\begin{equation}
           \Ktilde ~\equiv~ K(\Phitilde,\Phitilde^\dagger,V) ~=~ 
                \sum_n   \left( {\Sigma\Sigmabar\over 3 M_P^2}\right)^{-n/2} K_n~
\label{eq:KahlerWeylWeight}
\end{equation}
with vanishing Weyl weight and $R_5$-charge. 
In terms of $\Ktilde$, our new K\"ahler potential $\widehat K$ then takes the form
\begin{equation}
              \widehat K ~\equiv~ -\Sigmabar\Sigma  \exp\left(  -{\Ktilde\over 3M_P^2}\right)~,
\label{Kwidehatdef}
\end{equation}
and once again we see that $\widehat K$ is guaranteed to be superconformal.

Given these definitions, we can now 
easily promote an arbitrary supersymmetric
theory with broken superconformal invariance and broken $R_5$-symmetry to a theory
in which both symmetries are restored:
we simply replace
\begin{equation}
                   K\to \widehat K~~~~~{\rm and}~~~~~W\to\widehat W~.
\end{equation}
Note that the kinetic terms for gauge fields are unaffected, since they are already superconformal.
Now that we have a fully superconformal theory, we can ``couple'' this theory to conformal supergravity.
Specifically, this means that we covariantize all superspace derivatives to make them local,
and likewise replace our flat superspace integration measures with curved ones.
This is equivalent to retaining our original superspace derivatives, 
but introducing an additional Lagrangian term which couples the flat-space supercurrent superfield
$J_{\alpha\alphadot}$
to an appropriate corresponding supergravity multiplet (in this case, a superconformal one)~\cite{Lessons}.  
As a final step,
in order to return back to the symmetry structure and algebraic forms associated with our 
original theory, we simply set our compensator
fields to fixed values,
\begin{equation}
            \Sigma,~\Sigmabar ~\to~ \sqrt{3} M_P~.
\label{setfixed}
\end{equation}
Indeed, since $\Sigma$ and $\Sigmabar$ carry non-trivial $R_5$-charges and Weyl weights,
setting these fields to have fixed values as in Eq.~(\ref{setfixed}) has
the net effect of breaking $R_5$-symmetry
and superconformal invariance. 

The result of this process is then a theory with the same symmetry structure as
our original theory, but coupled to (Poincar\'e) supergravity.  The quantity $M_P$, of course,
describes the strength of this coupling, and we may compare  
the theory that results from this procedure with the original globally 
supersymmetric theory with which we started 
by taking the formal limit $M_P\to\infty$.

While it is clear that the replacements in Eq.~(\ref{setfixed}) algebraically restore
our original superpotential, reducing $\widehat W\to W$, it is perhaps less clear that 
making these replacements and taking the $M_P\to\infty$ limit 
algebraically restores our original K\"ahler term as well.
However, we note that
\begin{eqnarray}
    \int d^2\theta d^2\thetabar \, \widehat K  &=&
  \int d^2\theta d^2\thetabar \left[-\Sigmabar\Sigma  \exp\left(  -{\widetilde K\over 3M_P^2}\right)\right]        \nonumber\\
    &\to& \int d^2\theta d^2\thetabar \left[3 M_P^2   +  K + {\cal O}(M_P^{-4})\right]\nonumber\\
    &=& \int d^2\theta d^2\thetabar \,  K~,
\label{eq:CCompLag}
\end{eqnarray}
where we have used the fact that $\widetilde K\to K$ in passing to the second line.
Thus, the K\"ahler portion of our original theory is algebraically restored as well. 
Note that in this discussion, we have not explicitly shown the additional supergravity
terms which would ordinarily appear in the Lagrangian, as they are not relevant to the
present discussion.

 {\it In this context, however,
it is important to note that this ``restoration'' of the original theory
is merely an algebraic illusion.  
In truth, our original theory has not been restored at all.}\/
Although our extra factors of $(\Sigma/\sqrt{3}M_P)$ 
and $(\Sigmabar/\sqrt{3}M_P)$ 
have conveniently disappeared in this
process, leaving behind what superficially looks like our original theory, 
we must recall that these factors also carried with them certain Weyl weights and
$R_5$-charges.  Thus, although the theory that emerges at the end of the day algebraically 
resembles our original theory, all of the non-trivial Weyl weights and $R_5$-charges have been
stripped from the fields in question.  
Thus, the behavior of our final theory under conformal transformations and chiral $R_5$-rotations
will be completely different than the behavior exhibited by our original theory.  

This last point can be illustrated even more dramatically by considering a theory
in which superconformal invariance is broken but $R_5$-symmetry is preserved.
There is nothing that prevents us from using the above chiral compensator formalism
in this case as well.
However, if we use our chiral compensators to promote this theory to a fully superconformal  
theory and then attempt to return to our original theory following the above 
prescription, we find that our new theory has neither superconformal invariance
nor $R_5$-symmetry.  In other words, the $R_5$-symmetry of the original theory
has been broken in the ``hysteresis'' process of promoting and then demoting the theory.

For this reason, it is critical that we continue to distinguish between our original theory,
our promoted (compensated) theory, and the final theory that results after 
the compensator fields are ``frozen'' back to fixed values.
Indeed, these are three independent theories with entirely different properties.
Although the original theory and the final theory may algebraically resemble each other,
they behave entirely differently under Weyl rescalings and chiral $R_5$-rotations.
Thus, they ultimately cannot be identified with each other.

\subsection{Deriving the FI supercurrent supermultiplet in the chiral formalism}

We now turn to the question of deriving a supercurrent supermultiplet in the chiral
formalism.  

As discussed above, there are three distinct theories that come into play when
discussing the chiral formalism:  the original theory which lacks conformal
symmetry (and which may or may not contain an unbroken $R_5$-symmetry), the compensated 
theory which exhibits a full superconformal invariance, and the 
final theory that results when our chiral compensator fields are ``frozen'' to fixed
values.  
In principle, these are three distinct theories, and it is possible for each of them
to have a different supermultiplet of currents.  
Indeed, there is also no guarantee that the currents associated with the (compensated)
fully superconformal theory will, when subjected to the subsequent ``freezing" process,
reduce to the currents that might be calculated directly for the frozen theory.
Thus, we must distinguish precisely which theory it is for which we seek
to evaluate a supercurrent supermultiplet.

Clearly, all three theories will give rise to supercurrents and energy-momentum tensors
which are conserved.  This follows from the fact that all three theories exhibit unbroken
supersymmetry and translational invariance.
However, only the (compensated) fully superconformal theory will have a conserved $R_5$-current,   
for this is the only theory in which $R_5$-symmetry is guaranteed to be unbroken.
Therefore, our procedure will be to calculate our supercurrent supermultiplet within
the context of the fully superconformal compensated theory, and then to subject
this current to the ``freezing'' process of setting our compensators to fixed values.
We may then loosely identify this supercurrent as corresponding to our original uncompensated
theory by taking the $M_P\to\infty$ limit.

Note that regardless of the $R_5$-symmetry properties of our original theory, this 
procedure is guaranteed to yield a supercurrent supermultiplet whose bottom component
$j_\mu^{(5)}$ is not conserved.
Moreover, we shall also find that this 
is generally {\it not}\/ the same as calculating the currents directly
in the final theory that emerges after ``freezing''.
In other words, calculating our currents through the Noether procedure does not commute
with the ``freezing'' process, and it matters whether our currents are calculated before
or after the compensator fields are set to their fixed values. 

Our interest in this paper concerns theories with FI terms, such as the
pure $U(1)_{\rm FI}$ gauge theory in Eq.~(\ref{lag2}).
This theory has $W=0$, and we may identify the K\"ahler potential of this theory
as $K=\widetilde K = 2\xi V$, where $V$ is the $U(1)_{\rm FI}$ vector superfield.
We can therefore use our chiral compensator formalism to promote the $\xi$-dependent
part of the Lagrangian to the superconformal form
\begin{equation}
        \calL~\equiv~\int d^2\theta d^2\thetabar 
             \left[ -\Sigmabar \Sigma \exp\left( -{2\xi V\over 3 M_P^2}\right)\right]~.
\label{superconftheory}
\end{equation}    
We can then expand this expression in terms of the component fields of $V$,
as in Eq.~(\ref{vecsup}),
and the component fields 
$\{\phi_\Sigma,\psi_{\Sigma},F_\Sigma\}$ of $\Sigma$.
Doing this to quadratic order in $\xi/M_P^2$, calculating the $R_5$-current in
the resulting theory using the Noether procedure, and then setting $\Sigma$ and
$\Sigmabar$ to their fixed values $\sqrt{3} M_P$, we find 
the result
\begin{equation} 
         j_\mu^{(5)}~=~   -{4\over 3} \xi A_\mu - {\xi^2\over 18 M_P^2} \,\chi \sigma_\mu \chibar ~+~
              {\cal O}(M_P^{-6})~.
\label{jmu51}
\end{equation}
Note that this expression comes entirely from variations of the $\Sigma,\Sigmabar$ chiral
compensator fields, since all of the remaining matter fields have vanishing $R_5$-charges
in the chiral formalism.
Taking the $M_P\to\infty$ limit of Eq.~(\ref{jmu51}) then yields the result
\begin{equation}
         j_\mu^{(5)}~=~   -{4\over 3} \xi A_\mu~,
\label{jmu52}
\end{equation}
and this may be identified as the bottom component of the superfield
\begin{equation}
        \Xi^{(C)}_{\alpha\alphadot}~=~  {2\xi\over 3} [D_\alpha,\Dbar_\alphadot] V~.
\label{Xisolnchiral} 
\end{equation}
This, then, is the result for the FI contribution $\Xi_{\alpha\alphadot}^{(C)}$ 
to the supercurrent superfield
in the chiral formalism,
whereupon we conclude
that the total supercurrent supermultiplet for the theory in Eq.~(\ref{lag2})
is given by 
\begin{equation}
        J^{(C)}_{\alpha\alphadot} ~=~ 2 W_\alpha \Wbar_\alphadot + {2\xi \over 3}
                     [D_\alpha,\Dbar_\alphadot] V~ 
\label{badguess}
\end{equation}
in the chiral formalism.
Indeed, this is precisely the result quoted in Ref.~\cite{Seiberg}.

It is straightforward to generalize this result to a sigma model with arbitrary
K\"ahler potential $K$
and arbitrary superpotential $W$.
The corresponding supercurrent supermultiplet 
in the chiral formalism
is then given by the general expression
\begin{equation}
        J^{(C)}_{\alpha\alphadot} ~=~
        -g^{i\ibar} (D_\alpha \Phi_i) (\Dbar_\alphadot \Phibar_\ibar)
                + {1\over 3} [D_\alpha,\Dbar_\alphadot] K~,
\label{generalexp2}
\end{equation}
where
the K\"ahler metric is given by
\begin{equation}
      g^{i\ibar} ~\equiv~ {\partial^2 K\over \partial\Phi_i \partial \Phibar_\ibar}~.
\label{eq:KahlerMetric}      
\end{equation}
Indeed, the supercurrent in Eq.~(\ref{generalexp2}) is independent of the superpotential $W$
except through the equations of motion.
  
It is important to note that we derived these results by evaluating
our Noether currents in the compensated superconformal theory {\it prior}\/
to ``freezing'' our compensators to fixed values and taking the $M_P\to\infty$ limit.
By contrast, if we had analyzed the freezing properties of the Lagrangian in Eq.~(\ref{superconftheory})
directly, we would have found that only
one term ultimately survives:
\begin{equation}
        \calL~=~ {\xi\over 3 M_P^2}\, 
    \phi_\Sigma \overline{\phi}_\Sigma\, (D+\half \Box C)~ + ~...~,
\label{onlysurvivor}
\end{equation}
where $\phi_\Sigma$ and $\overline{\phi}_\Sigma$ are respectively the lowest (scalar) components
of our compensator fields $\Sigma$ and $\Sigmabar$. 
Of course, with the substitutions $\Sigma,\Sigmabar\to \sqrt{3} M_P$,
we recognize Eq.~(\ref{onlysurvivor}) as our original FI Lagrangian.
However, since the $D$- and $C$-fields have vanishing $R_5$-charges,
this term does not make any contribution to $j_\mu^{(5)}$.
This is therefore an explicit demonstration that the process of calculating a Noether current
does not commute with the process of freezing the compensators to 
fixed values and taking the $M_P\to\infty$ limit.
 
Once again, we stress that the results in Eqs.~(\ref{Xisolnchiral}) and
(\ref{badguess}) correspond to the conformally compensated theory whose K\"ahler contributions
to the Lagrangian are given in Eq.~(\ref{superconftheory}).
In particular, our expression for $j_\mu^{(5)}$ in Eqs.~(\ref{jmu51})
and (\ref{jmu52}) is nothing but the result of applying the
``freezing'' process to the Noether current associated with
Eq.~(\ref{superconftheory}).
However, as with any such results derived through the chiral formalism,
these currents do {\it not}\/ correspond to the original theory in Eq.~(\ref{lag2})
with which we started.
Indeed, our original theory in Eq.~(\ref{lag2}) has two manifest symmetries
which are crucial and which are preserved in spite of the appearance of a non-zero
FI term:  $R_5$-invariance and $U(1)_{\rm FI}$ invariance.
Both of these symmetries are explicitly broken in the results of 
Eqs.~(\ref{Xisolnchiral}) and (\ref{badguess}).
In other words, there is no way in which 
we can connect the bottom component of the superfield in Eq.~(\ref{badguess})
to the $j_\mu^{(5)}$ Noether current derived in Sect.~3.4, with or without the addition
of any possible improvement terms.
Thus, we see that the $U(1)_{\rm FI}$ gauge non-invariance of the result
in Eq.~(\ref{badguess}) is an artifact of the chiral compensator formalism,
and is not a property of the underlying physics of our original theory.
In other words, in a Noether sense, no supercurrent supermultiplet 
exists for a theory with a non-zero FI term in the chiral formalism,
independent of the compensators.

Despite this fact, the procedure we have followed does describe one of the ``minimal'' methods by
which a theory such as that in Eq.~(\ref{lag2}) might be coupled to supergravity.
Thus, the broken $U(1)_{\rm FI}$ and $R_5$-symmetries 
of the supercurrent will indeed be of relevance
insofar as this coupling to supergravity is concerned,
with far-reaching consequences that we shall now explore.

\subsection{The symmetry structure of theories with non-zero FI terms 
in the chiral formalism}

In this section, we shall explore the structure of local and global symmetries
that appear in theories with non-zero FI terms in the chiral formalism.
We shall begin by describing the general symmetry structure of such theories
and the way in which it
emerges.  We shall then provide an explicit example.

\subsubsection{General symmetry structure}

In order for the chiral compensator formalism to be sensible, 
our Lagrangian must at all times exhibit
manifest invariance under appropriate gauge transformations.
However, the gauge transformations in question 
depend on the theory in which one is working.

In our original theory prior to the introduction of chiral compensators,
we expect to have an unbroken local $U(1)$ gauge symmetry associated
with our FI term.  For reasons to become clear shortly, we shall refer to this
symmetry as $U(1)'_{\rm FI}$.  
Note that in the presence of a non-zero FI term, the K\"ahler potential
will no longer be $U(1)'_{\rm FI}$ gauge invariant:
under a $U(1)'_{\rm FI}$ gauge transformation
\begin{equation}
             V~\to~ V+ i(\Lambda - \Lambda^\dagger)~,
\label{U1gaugetransf}
\end{equation}
the K\"ahler potential transforms as
\begin{equation}
            K~\to~ K + 2i\xi (\Lambda - \Lambda^\dagger)~.
\label{Kahlergaugetransf}
\end{equation}
Another way of saying this is that gauge transformations induce
K\"ahler transformations,
a fact first noticed in Ref.~\cite{MultipletsInLinearFormalism}
and recently emphasized in Ref.~\cite{Seiberg}.
However, this does not disturb the $U(1)'_{\rm FI}$ gauge invariance
of the theory, since the corresponding $D$-field
within $V$ is gauge invariant.

This situation changes dramatically upon introduction of the chiral
compensators.
Now denoting our FI gauge-transformation group as $U(1)_{\rm FI}$ in the superconformal
chirally-compensated theory,
we once again see that gauge transformations of the form in Eq.~(\ref{U1gaugetransf})
induce transformations of the K\"ahler potential of the form in Eq.~(\ref{Kahlergaugetransf}).
Note that this is true for both $K$ and $\Ktilde$, since the relevant term $2\xi V$ within
$K$ does not experience a rescaling 
under Eq.~(\ref{eq:KahlerWeylWeight})
when passing to $\Ktilde$.
However, in the superconformal chirally-compensated
theory, our final K\"ahler potential $\widehat K$ (or more precisely, its corresponding $D$-term) 
must be neutral under $U(1)_{\rm FI}$ symmetries.
This in turn then forces us to assign a $U(1)_{\rm FI}$ charge to our chiral compensator fields,
\ie,
\begin{equation}
              Q_{\Sigma,\Sigmabar}~=~ \pm {2\xi\over 3 M_P^2}~.
\label{qsig}
\end{equation}

The fact that the chiral compensator fields carry $U(1)_{\rm FI}$ charges in the
presence of an FI term has three immediate consequences.

First, this implies that the $U(1)_{\rm FI}$ gauge symmetry --- just like the Weyl symmetry
and the $R_5$-symmetry which together form the super-Weyl $U(1)_\SW$ symmetry  --- will be  
broken when the compensator fields are set to fixed values.
 {\it This explains the lack of $U(1)_{\rm FI}$ gauge invariance exhibited by the
corresponding supercurrent supermultiplet after the compensator fields are ``frozen''.}\/
Moreover, after our chirally-compensated superconformal theory is 
coupled to supergravity, the $U(1)_\SW$-symmetry becomes local,
leading to a $U(1)_\SW$ gauge symmetry in addition to the $U(1)_{\rm FI}$ gauge symmetry.
The first symmetry is associated with the gauge boson $b_\mu$ which appears within the supergravity multiplet
and which corresponds directly to $R_5$-transformations, 
while the second symmetry is associated with the $A_\mu$ component field within $V$.
Because the $\Sigma$ field simultaneously 
carries both a non-zero $U(1)_\SW$-charge $q_\SW=2/3$ and a non-zero $U(1)_{\rm FI}$-charge
given in Eq.~(\ref{qsig}),
setting this field to a fixed value implies that 
our local gauge symmetry $U(1)_{\SW}\times U(1)_{\rm FI}$ will be
broken down to a single axial $U(1)_A$ subgroup~\cite{BarbieriEtAlEToTheV}:
\begin{equation}
  U(1)_{\SW}\times U(1)_{\rm FI} ~~\rightarrow~~ U(1)_A~\equiv~ U(1)_{\rm FI}  - {\xi\over M_P^2} U(1)_\SW~. 
\label{eq:U1FIHR}
\end{equation}
This residual $U(1)_A$ symmetry persists in the effective theory at 
energy scales below $M_P$, with a corresponding gauge boson
\begin{equation}
    A'_\mu ~\equiv~ {1\over \sqrt{1+\xi^2/M_P^4}} \left(
       A_\mu - {\xi\over M_P^2}  b_\mu\right)~.
\end{equation}
For $\xi\ll M_P^2$, we see that this symmetry is mostly $U(1)_{\rm FI}$ itself.

However, there is also a second important consequence of
the fact that our chiral compensator fields $(\Sigma,\Sigmabar)$
carry a $U(1)_{\rm FI}$ charge.  
Since our superpotential $W$ must be $U(1)_{\rm FI}$ neutral, we see that $\widetilde W$
must carry a $U(1)_{\rm FI}$ charge 
\begin{equation}
          Q_{\widetilde W}~=~  -{2\xi\over  M_P^2}~.
\end{equation}
This places important restrictions on the superpotential structure of our theory.
In general, $\widetilde W$ may contain trilinear terms of the form
$y_{ijk}\widetilde \Phi_i\widetilde \Phi_j\widetilde\Phi_k$, 
mass terms of the form $m_{ij}\widetilde \Phi_i\widetilde \Phi_j$,
and so forth.  However, if it is possible to assign $U(1)_{\rm FI}$ charges to
all of the $\widetilde \Phi$ fields 
such that each of the terms in $\widetilde W$
transforms with a uniform charge under $U(1)_{\rm FI}$ transformations,
then this tells us something additional
about the original theory that we had prior to introducing the compensators.
In particular, this tells us that our original theory must exhibit an invariance 
under not only the local $U(1)'_{\rm FI}$ symmetry, but also under
an additional global $R$-symmetry, to be denoted  $R'$.
If our original symmetry had an $R_5$-invariance as well, then this additional global 
$R'$-symmetry may or may not coincide with $R_5$.  This ultimately depends
on the structure of the theory.

\begin{figure}[b!]
\centerline{
   \epsfxsize 6.0 truein \epsfbox {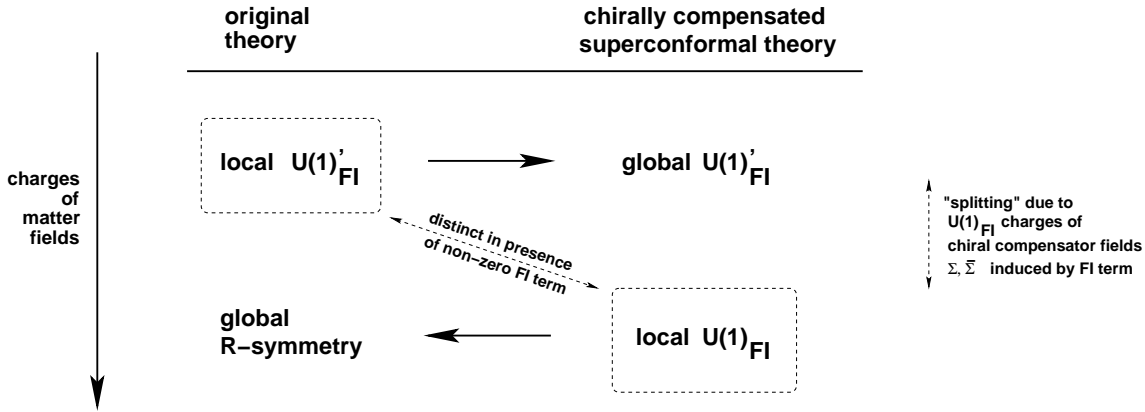} }
\caption{Comparison between the symmetries of our original theory
and those of our chirally-compensated superconformal theory.
In general, the existence of a local $U(1)_{\rm FI}$ symmetry
in either theory implies the existence of a corresponding  global symmetry in the 
other theory.  In the presence of a non-vanishing FI term, 
the local $U(1)_{\rm FI}$ symmetries in the two theories 
are distinct 
as a result of the non-vanishing $\xi$-dependent $U(1)_{\rm FI}$-charges
of the chiral compensator fields $\Sigma,\Sigmabar$. 
Thus, in the chiral formalism, the presence of a non-zero FI term requires 
the existence of additional global symmetries in both theories.  However, 
when the FI term is eliminated, the local $U(1)_{\rm FI}$ symmetries 
in both theories can be identified with
each other.  In this case, the global $U(1)_{\rm FI}$ symmetry within the 
local $U(1)_{\rm FI}$ symmetry of either theory serves as the global symmetry required by the 
local $U(1)_{\rm FI}$ symmetry of the other theory, and no
additional global symmetries are needed in either theory.}       
\label{localglobalfig}
\end{figure}

Finally,
in a similar vein,
there are also additional symmetry repercussions for the chirally-compensated theory.
Just as the local $U(1)_{\rm FI}$ invariance of the chirally-compensated theory
implies the existence of an additional global symmetry $R'$ in the original uncompensated theory,
a similar thing happens in reverse:   the local $U(1)'_{\rm FI}$ symmetry
of the original uncompensated theory implies the existence of an additional global symmetry
in the chirally-compensated theory.
To see why this is the case, let us recall that our original uncompensated theory
had a local $U(1)'_{\rm FI}$ invariance under which our matter fields $\widetilde \Phi_i$
were assigned charges leading to an invariant superpotential $W(\Phi)$.  
We may therefore promote this symmetry into the chirally-compensated theory by choosing our
chiral compensator fields $\Sigma,\Sigmabar$ to be neutral under this $U(1)'_{\rm FI}$
symmetry.  Of course, in the chirally-compensated theory, this choice can only make sense if 
this new $U(1)'_{\rm FI}$ symmetry is no longer associated with shifts in the
gauge field supermultiplet $V$ or in the K\"ahler potential $\widehat K$ --- \ie,
if this symmetry is no longer local, but global.
Indeed, in the chirally-compensated theory, it is not $U(1)'_{\rm FI}$ but $U(1)_{\rm FI}$ which
is associated with the gauge shifts in Eq.~(\ref{U1gaugetransf}).
Nevertheless, in the chirally-compensated theory, we see that our original
$U(1)'_{\rm FI}$ symmetry leaves behind a global remnant 
which is disconnected from $U(1)_{\rm FI}$ gauge transformations.
Indeed, this is a {\it bona-fide}\/ global symmetry of the chirally-compensated
theory, with charge assignments for the matter fields $\widetilde \Phi_i$
and chiral compensator fields $\Sigma,\Sigmabar$
which are distinct from those corresponding to the gauge symmetry $U(1)_{\rm FI}$.
Under the subsequent ``freezing'' process in which 
$\Sigma,\Sigmabar$ are set to fixed values, we have seen that
our local $U(1)_{\rm FI}$ is ultimately broken.
However, our global $U(1)'_{\rm FI}$ symmetry remains intact.

We emphasize that this entire structure, with matching local and global
symmetries for both the original theory and the chirally-compensated theory,
only arises in the presence of a non-zero FI term. 
Without a non-zero FI term, our K\"ahler potentials $K$ and $\widehat K$ would
both be invariant under $U(1)_{\rm FI}$ transformations.
Thus the $U(1)_{\rm FI}$ and $U(1)'_{\rm FI}$ symmetries would coincide
between our original and chirally-compensated theories,
and neither would trigger the existence of a global symmetry in the opposite
theory.  
Indeed, it is only because the non-zero FI term requires the introduction
of chiral compensator fields $\Sigma,\Sigmabar$ with non-zero $U(1)_{\rm FI}$
charges that these two symmetries become distinct, with each implying the
existence of an additional global symmetry.   
This situation is illustrated in Fig.~\ref{localglobalfig}.

Needless to say, the fact that non-zero FI terms require the existence
of additional global symmetries has far-reaching consequences.  Indeed,
it is currently believed (and indeed proven in the context of string theory~\cite{banksdixon,GlobalSymsInSUGRA,cliff})
that such additional global symmetries are inconsistent for any theory
which is ultimately coupled to gravity.  Thus, assuming this ``folk'' theorem 
to be true for all classes of gravity-coupled theories,
we then conclude that theories with non-zero FI terms cannot be consistently coupled
to supergravity in the chiral formalism.

\subsubsection{An explicit example}

For clarity, we shall now illustrate these ideas with a concrete example.
Let us consider a
simple $R_5$-invariant toy model which involves a single $U(1)'_{\rm FI}$ gauge group 
with a  non-trivial FI term, 
and matter content  
comprising three chiral superfields $\Phi_i$, $i=1,2,3$.
We shall imagine that these chiral superfields 
carry charges under 
 $U(1)'_\FI$,  as well as under a global  
$R_5$-symmetry, in the
manner indicated in Table~\ref{tab:chgs2before}.  
We have also shown the charges for the  $U(1)'_\FI$ gaugino
field $\lambda_\alpha$.
Given these charge assignments, the most general renormalizable
superpotential invariant under these symmetries is given by
\begin{equation}
    W ~=~ y_{1} \Phi_1 \Phi_2 \Phi_3 + y_2 \Phi_3^3~.
\label{eq:Wex}
\end{equation}
We will also assume that the matter field contribution $K'$ to the
K\"ahler potential takes the minimal form
\begin{equation}
   K' ~=~ \Phi_1 e^{-V} \Phi^\dagger_1 + \Phi_2 e^{V} \Phi^\dagger_2 + \Phi_3\Phi^\dagger_3~,
\label{eq:Kex}
\end{equation}
and that the full K\"ahler potential is given by $K = K' + 2\xi V$,
where $V$ denotes the vector superfield associated with $U(1)'_\FI$.   
Note that
this toy theory is completely $R_5$-invariant, with dilatation  
invariance spoiled
only by the presence of the FI term.
This model therefore contains only two symmetries:  $U(1)'_{\rm FI}$
and $R_5$, the first local and the second global.

\begin{table}[t!]
\begin{center}
\begin{tabular}{||c||c|c||}
   \hline 
   \hline 
   ~~ Field ~~ & ~ $U(1)_{\FI}'$ ~  &~ $R_5$~\\
     \hline
   \hline 
$\Phi_1$ & $+1$ & 2/3 \\
$\Phi_2$ & $-1$ & 2/3 \\
$\Phi_3$ & $0$  & ~2/3~ \\
     \hline
$\lambda_\alpha$    &  0  & 1 \\
   \hline 
   \hline 
\end{tabular}
\end{center}
\caption{$U(1)'_{\rm FI}$- and $R_5$-charges of the fields in our toy model, as discussed in the text.}
\label{tab:chgs2before}
\end{table}

When this theory is promoted to a superconformal theory in the chiral  
formalism
through the introduction of the chiral compensators $\Sigma$ and $\Sigmabar$, 
several changes occur.
First, the
matter fields of the theory are rescaled according to  
Eq.~(\ref{newfields}).  Since the chiral compensators are forced to carry $U(1)_\FI$ charge
according to Eq.~(\ref{qsig}),
we see that  
the resulting charges of the rescaled matter fields $\wtPhi_i$ are shifted
relative to their original $U(1)'_{\rm FI}$ charges.  In this way we determine
that $U(1)_\FI$ and $U(1)'_\FI$ are now distinct symmetries.
Likewise, the shift in the K\"ahler potential from $K$ to $\widehat K$ 
defined in Eq.~(\ref{Kwidehatdef}) restores conformal (Weyl) symmetry to the theory, 
even in the presence of an FI term, thereby generating a full $U(1)_\SW$ super-Weyl symmetry
as well.
Table~\ref{tab:chgs2} lists the charges of the 
matter fields $\Phi_i$,
the compensator field $\Sigma$,
the $U(1)_\FI$ gaugino field $\lambda_\alpha$, and the gravitino field $\psi_{\mu \alpha}$ 
under these three $U(1)$ symmetries, along with their $R_5$-charges and $w$-weights.

\begin{table}[b!]
\begin{center}
\begin{tabular}{||c||c|c|ccc||}
   \hline 
   \hline 
    ~~ Field ~~ & ~ $U(1)_{\FI}'$ ~ & ~$U(1)_{\FI}$~ & ~$U(1)_\SW$:~ &~ $R_5$~ &~ Weyl ~\\
     \hline
      \hline
$\wtPhi_1$ & +1 & ~$\phantom{-}1-2\xi/3M_P^2$ ~&0& 0 & 0   \\
$\wtPhi_2$ & $-1$ &$-1-2\xi/3M_P^2$ &0& 0 & 0  \\
$\wtPhi_3$ &  0 &  $-2\xi/3M_P^2$ &0& 0 & 0   \\
\hline
$\Sigma$            &  0  & $\phantom{-}2\xi/3M_P^2$ & 2/3& 2/3 & 1     \\ \hline
$\lambda_\alpha$    &  0  &     0         & $\ast$ & 1   & 3/2 \\
$\psi_{\mu \alpha}$ &  0  &     0         & $\ast$ & 1   & 3/2 \\
     \hline
     \hline
\end{tabular}
\end{center}
\caption{The symmetry structure of the chirally-compensated superconformal version
of the toy model originally defined in Table~\ref{tab:chgs2before}.
We list the unbroken symmetries that exist in this model, along with the corresponding
charges of the matter fields $\widetilde{\Phi}_i$, the chiral
compensator field $\Sigma$, the $U(1)_\FI$ gaugino $\lambda_\alpha$,
and the gravitino $\psi_{\mu\alpha}$. 
    An entry `$\ast$' indicates that the corresponding field is not a chiral
    superfield, and therefore its $R_5$- and Weyl-charges cannot 
    be packaged as a chiral charge under $U(1)_\SW$. 
\label{tab:chgs2}}
\end{table}

Note that this theory contains two distinct FI-related symmetries,  $U(1)_\FI$ and $U(1)'_\FI$,
yet only one FI vector superfield $V$.
In the fully superconformal compensated theory, $V$ is associated 
with $U(1)_\FI$.  As a result, $U(1)_\FI$ is a fully local supersymmetric $U(1)$ gauge symmetry in this theory.
By contrast, because $U(1)'_\FI$ lacks its own vector gauge superfield,
we see that the $U(1)'_\FI$ symmetry is neither local nor fully supersymmetric;
rather, it is global, and it is only a Wess-Zumino remnant corresponding to an ordinary global $U(1)$ symmetry.
Thus, borrowing the terminology introduced below Eq.~(\ref{gaugetransforms}),
we see that $U(1)_\FI$ is both local and ``big'', while 
$U(1)'_\FI$ is global and ``little''.

The full symmetry content of the chirally-compensated superconformal theory
thus consists of three $U(1)$ symmetries:  $U(1)'_\FI$, $U(1)_\FI$, and $U(1)_\SW$.
Note that this last symmetry is an $R$-type symmetry --- \ie, a symmetry under which
the superspace $\theta$-coordinate is charged, or equivalently a symmetry under which
the superpotential $\widehat W$ is charged.  By contrast, the two FI symmetries are not $R$-type.
However, writing our three $U(1)$ symmetries in this way is only a basis choice,
and we can express the symmetries of this model in terms of any linear combinations 
of these symmetries that we wish.
One particularly important linear combination that we may define is
\beq
     R_G ~\equiv~ \left({\xi\over M_P^2}\right)^{-1} \biggl[ U(1)'_\FI - U(1)_\FI\biggr] 
              + U(1)_\SW~.
\label{eq:RG}
\eeq
As a result of its definition, we see that $R_G$ is a 
global, ``little'', $R$-type symmetry.
Under this symmetry
the matter fields $\wtPhi_i$ each have $R_G$-charge $2/3$, while
the compensator fields $\Sigma$, $\Sigmabar$ are uncharged.
However, we emphasize that $R_G$ is not an 
additional symmetry of
the theory, but merely a recasting of our global $U(1)'_\FI$ symmetry  
into a new basis.
Indeed, in all bases, our chirally-compensated superconformal
theory contains only one independent global symmetry.

Let us now examine how the symmetry properties of this theory are altered by
the freezing of the compensators.  Since $\Sigma$ and $\Sigmabar$ transform
non-trivially under both $U(1)_\FI$ and $U(1)_\SW$, 
these symmetries will not be preserved individually in the
frozen theory.  However, as discussed above, a linear combination of these
two symmetries, namely the gauged $R$-symmetry $U(1)_A$ of Eq.~(\ref{eq:U1FIHR}),
is preserved in this theory.  In addition, there is also a global $U(1)$ symmetry
which survives:  
this may be alternatively interpreted as $U(1)'_\FI$, as in  
Table~\ref{tab:chgs2after},
or as the additional $R_G$ symmetry defined in Eq.~(\ref{eq:RG}).  Thus,
after the compensator fields are frozen, the symmetry structure of our theory
reduces to that shown in Table~\ref{tab:chgs2after}.

\begin{table}[t!]
\begin{center}
\begin{tabular}{||c||c|c||}
   \hline 
   \hline 
    ~~ Field ~~ & ~ $U(1)'_{\FI}$ ~ & ~$U(1)_{A}$ ~ \\
     \hline
   \hline 
$\wtPhi_1$ & +1   & ~$\phantom{-}1-2\xi/3 M_P^2$~  \\
$\wtPhi_2$ & $-1$ & $          -1-2\xi/3 M_P^2$ \\
$\wtPhi_3$ &  0 & $-2\xi/3 M_P^2$   \\
     \hline
$\lambda_\alpha$    &  0 & $-\xi/3 M_P^2$  \\
$\psi_{\mu \alpha}$ &  0 & $-\xi/3 M_P^2$  \\
     \hline
     \hline
\end{tabular}
\end{center}
\caption{The symmetry structure of the final version of our toy model in Table~\ref{tab:chgs2},
     after the compensator fields are ``frozen'' to their fixed values in the chiral formalism.
     We list the charges of the matter fields $\wtPhi_i$, 
          the $U(1)_\FI$ gaugino $\lambda_\alpha$,
   and the gravitino $\psi_{\mu\alpha}$.
   Note that the $R_5$-symmetry of our original model in Table~\ref{tab:chgs2before} is broken;
   likewise, the local $U(1)'_{\rm FI}$ gauge symmetry of our original
   model has also been broken, leaving behind only a global remnant, while
   the $U(1)_{\rm FI}$ gauge symmetry of the superconformal version
   of our theory has been broken entirely.}
\label{tab:chgs2after}
\end{table}

As we explicitly see from this example,
the symmetry structure that survives after the chiral compensators are frozen
contains not only a single $U(1)_A$ gauge symmetry, but also an exact global symmetry $U(1)'_\FI$. 
Indeed, this is the entire symmetry structure that survives, even  
in cases such as this in which the original theory is $R_5$-symmetric.
Moreover, we observe that this structure remains intact for all $\xi \not= 0$.
However, for $\xi=0$, we observe that $U(1)'_{\rm FI}$ and $U(1)_A$ become
redundant.
In this case, the local $U(1)_A=U(1)'_{\rm FI}$ symmetry survives,
but no additional independent global symmetry remains in the theory.


\subsection{Proof of the non-existence of an FI supercurrent supermultiplet 
   with conserved $R_5$-symmetry in the chiral formalism}

As we have discussed, the supercurrent supermultiplet in the chiral formalism 
must always satisfy the constraint in Eq.~(\ref{chiralcase}), where $S$ is a chiral supermultiplet.
However, even though the chiral formalism can accommodate the special case in which $R_5$-symmetry
is preserved (such as for FI terms), this feature is not enforced by the formalism itself.  
Indeed, in the chiral formalism,
the bottom component of the supercurrent supermultiplet is not conserved.

This result makes sense, since the supercurrent supermultiplet in this formalism
actually corresponds not to our original global theory (in which $R_5$-symmetry is preserved),
but to the chirally-compensated version of this theory (in which $R_5$-symmetry is broken).
However, strictly speaking, this feature prevents us from associating the resulting
supercurrent supermultiplet with our original global theory.
Indeed, no possible improvement term can allow us to connect the resulting
value of $C_\mu$ with an improved Noether current $j_\mu^{(5)}$.
A natural question, therefore, is whether it might be possible to construct 
a supercurrent supermultiplet in the chiral formalism which {\it does}\/ exhibit
$R_5$-symmetry conservation, perhaps as a special case.

We shall now show that this cannot be done.
Specifically, we shall assume the constraint in Eq.~(\ref{chiralcase}), and then
attempt to impose $R_5$-symmetry conservation by hand as an additional constraint.  
By imposing both constraints simultaneously,
we shall derive
a condition on the most general supercurrent supermultiplet $J_{\alpha\alphadot}^{(C)}$ 
in the chiral formalism that can be consistent with $R_5$-symmetry conservation, \ie,
consistent with the requirement that $\partial^\mu j^{(5)}_\mu=0$.

Since $j^{(5)}_\mu$ is by definition the lowest component of the supercurrent
supermultiplet $J_\mu$, the $R_5$-symmetry constraint can be expressed as the superfield constraint
\begin{equation}
      \partial^\mu J_\mu ~=~ -{i\over 4} \lbrace D^\alpha, \Dbar^\alphadot\rbrace J_{\alpha\alphadot}~=~0~,
\end{equation}
where $J_{\alpha\alphadot}$ is defined in Eq.~(\ref{alphadef}).
However, we now use the chiral-case constraint in Eq.~(\ref{chiralcase}), along with the fact
that an arbitrary chiral superfield $S$ can be written, without loss of generality,
in terms of an unconstrained vector superfield $T_S$
as
\begin{equation}
               S ~=~ \Dbar^2 T_S~.
\end{equation}
This enables us to write
\begin{equation}
      \lbrace D^\alpha, \Dbar^\alphadot\rbrace J_{\alpha\alphadot}~=~ D^2 \Dbar^2 T_S - 
                \Dbar^2 D^2 \Tbar_S~,
\end{equation}
whereupon we find that any supercurrent supermultiplet $J_{\alpha\alphadot}$ 
in the chiral formalism will
be consistent with unbroken $R_5$-symmetry if and only if
\begin{equation}
       D^2 \Dbar^2 T_S~=~ \Dbar^2 D^2 \Tbar_S~.
\label{prechiralconstraint}
\end{equation}
Thus, in cases where superconformal invariance is broken while
$R_5$-symmetry is preserved, Eq.~(\ref{prechiralconstraint})
should follow as a direct consequence of the equations of motion of the theory.
Note that in cases for which $T_S$ turns out to be real, 
Eq.~(\ref{prechiralconstraint}) 
reduces to 
\begin{equation}
       [D^2, \Dbar^2] T_S~=~ 0~,
\label{chiralconstraint} 
\end{equation}
or equivalently
\begin{equation}
          \Box T_S~=~ -{i\over 2} (D\sigma^\mu \Dbar) \partial_\mu T_S ~.  
\label{chiralconstraint2}
\end{equation}

So what goes wrong in the case of an FI term?
Given the above results, the answer is quite easy to see.

First, we observe that in the case of an FI term,
the corresponding contribution $\Xi^{(C)}_{\alpha\alphadot}$ to the total supercurrent
supermultiplet would have to correspond to a case
in which $T_S$ is proportional to the real vector superfield $V$ itself
(up to the addition of harmless terms annihilated by $D^2 \Dbar^2$). 
This result follows immediately from dimensional analysis and Lorentz symmetry.
Moreover, as we have already seen in Sect.~4.2,
this result is also natural from the supposition that the FI term should
follow the expectations associated with a general sigma model, leading to the result
in Eq.~(\ref{badguess}).
Indeed, given the result in Eq.~(\ref{badguess}),
we find that $S= -(\xi/3) \Dbar^2 V$, whereupon we have
\begin{equation}
             T_S ~=~ -{\xi\over 3} V~.
\end{equation}
Thus, once again, we find that $T_S$ would be proportional to $V$. 

Unfortunately, the problem with having $T_S$ proportional to $V$ is that
the constraint equation~(\ref{chiralconstraint2}) on $T_S$ does not hold as a result of the
equations of motion for $V$.
Instead, Eq.~(\ref{chiralconstraint2}) now has the devastating effect of 
imposing a structural truncation directly on the multiplet $V$ which goes beyond 
its equations of motion.
Specifically, with $V$ expanded as in Eq.~(\ref{vecsup}) and with $T_S\sim V$,
we find that Eq.~(\ref{chiralconstraint2}) becomes
\begin{eqnarray}
          \partial_\mu A^\mu &=& 0~\nonumber\\
          \partial_\mu D &=&  -\partial_\mu \Box C~\nonumber\\
        \Box M &=&  \Box N ~=~ 0~\nonumber\\
        \Box \chi_\alpha &=&  -i(\sigma^\mu \partial_\mu \lambdabar)_\alpha~.
\label{consequences}
\end{eqnarray}
These constraints go far beyond the equations of motion for $V$:
they imply that $V$ must be a {\it linear}\/ multiplet, up to possible K\"ahler transformations
which do not change the physics.
However, this represents a severe and unjustified truncation of the unconstrained real vector 
multiplet $V$ with which we started.

Thus, we reach a contradiction:  if we wish to impose $R_5$-conservation on $\Xi^{(C)}_{\alpha\alphadot}$ in
the chiral formalism, we find that $V$ cannot be the unconstrained real gauge field 
with which we started, and with which we constructed
our FI term.  Instead, we see that $V$ must actually be another beast 
entirely --- a linear multiplet, up to K\"ahler transformations  
--- if Eq.~(\ref{badguess}) is to remain valid for this theory.
But if $V$ is presumed constrained according to Eq.~(\ref{consequences}), 
then our supposed FI term with which we began was not an FI term at all, 
but something entirely different.
Thus, we conclude that we cannot 
self-consistently impose $R_5$-conservation on the supercurrent supermultiplet for the FI term
in the chiral formalism.

This, then, is the fundamental impasse that emerges 
upon attempting to construct 
an FI supercurrent supermultiplet in the chiral formalism
while simultaneously demanding manifest $R_5$-invariance.  
Dimensional analysis 
indicates that any supercurrent that could possibly correspond to the FI term in the chiral
formalism must have $T_S\sim V$.
However,
self-consistency then requires that
the multiplet $V$ be truncated in a way that transcends 
its general equations of motion.  This in turn  
prohibits from $V$ from corresponding to
the real gauge field with which we started, and in terms of which we constructed our FI term.
Consequently, even in the most general possible case, 
we conclude that there is no self-consistent supercurrent superfield 
in the chiral formalism
which can correspond directly to the FI term 
and thereby exhibit manifest $R_5$-invariance. 

These results do, however, illustrate one important theme:
the breaking of $U(1)_{\rm FI}$ gauge invariance 
in the chiral formalism is directly related to the 
breaking of  $R_5$-symmetry.
We have already seen this connection at the level of the 
chiral compensators in Sect.~4.2:  because the chiral compensator $\Sigma$
carries both an $R_5$-charge and a $U(1)_{\rm FI}$ charge,
both symmetries are broken simultaneously when the chiral compensator is
given a VEV. 
However, this connection is also apparent from our supercurrent expressions.
Identifying Eq.~(\ref{badguess}) as our supercurrent supermultiplet
in the chiral formalism, 
we find that
$j^{(5)}_\mu \sim A_\mu$, and
on the basis of this result (and other results of a similar nature) 
we see that the FI supercurrent supermultiplet
fails to be $U(1)_{\rm FI}$ gauge invariant~\cite{Seiberg}.
This makes sense, since the chiral formalism is well known to explicitly break
$U(1)_\FI$ in the presence of a non-zero FI 
term (see, \eg, Refs.~\cite{StelleAndWestEToTheV,BarbieriEtAlEToTheV}).
However, we now see from the first equation 
in Eq.~(\ref{consequences}) that if we could also demand  
consistency with $R_5$-current conservation, we would simultaneously be imposing
the constraint $\partial^\mu A_\mu=0$ --- \ie, a compensating gauge choice.
At an algebraic level, the constraint equations in Eq.~(\ref{consequences})
would have the net effect of correctly restoring $R_5$-current conservation, as they must, while
simultaneously eliminating 
the $A_\mu$-dependence within $j^{(5)}_\mu$ which was the source
of the $U(1)_{\rm FI}$ gauge non-invariance of the theory.
Thus, we see that the issue of the gauge non-invariance of the FI supercurrent supermultiplet 
in the chiral formalism is 
a direct consequence of fact that the chiral
formalism also breaks manifest $R_5$-symmetry.
Indeed, both of these features emerge only because the 
supercurrent supermultiplet in Eq.~(\ref{badguess})
corresponds
not to our original globally supersymmetric theory in Eq.~(\ref{lag2}) (in which both
$R_5$-symmetry and $U(1)_{\rm FI}$ gauge invariance are preserved),
but to its chirally-compensated cousin (in which both symmetries are broken).

There is yet another way in which we can demonstrate our inability to
consistently demand $R_5$-current conservation in the chiral formalism, as would be required
if Eq.~(\ref{badguess}) 
were the
FI supercurrent supermultiplet which directly corresponds to the theory in Eq.~(\ref{lag2})
(as opposed to its chirally-compensated cousin).
Using the equations of motion~(\ref{FIEOMs}) for this theory, 
we find from Eq.~(\ref{badguess}) that
\begin{eqnarray}
       \Dbar^\alphadot J_{\alpha\alphadot} &=&  -{1\over 3} \xi \Dbar^2 D_\alpha V 
               + {4\over 3} \xi (\sigma^\mu \Dbar)_\alpha \partial_\mu V\nonumber\\
              &=& -{1\over 3}\xi D_\alpha \Dbar^2 V~.
\label{badguessonshell}
\end{eqnarray}
Moreover, if this supercurrent were to conserve $R_5$-symmetry, we have already seen
that $V$ would have to be truncated according to Eq.~(\ref{consequences}) --- \ie, $V$ would
have to become a linear multiplet, up to K\"ahler transformations which do not affect the physics.
However, if $V$ were to become a linear multiplet (up to K\"ahler transformations), then by definition
$\Dbar^2 V=0$ (up to K\"ahler transformations), and consequently 
we see from Eq.~(\ref{badguessonshell}) that $\Dbar^\alphadot J_{\alpha\alphadot}=0$
(up to K\"ahler transformations which do not affect the physics).
This in turn implies that our theory would actually have to exhibit not only an unbroken
$R_5$-symmetry (as required), but also a {\it full unbroken superconformal symmetry}\/.
This, of course, is inconsistent with the fact that the FI term introduces a mass scale
into the theory.

This is in fact a general phenomenon:  a supercurrent supermultiplet in the
chiral formalism can exhibit manifest $R_5$-current conservation only when the theory
itself is superconformal.  Indeed, consulting Table~\ref{constraintstable},
we see that $R_5$-current conservation in the chiral formalism requires 
that ${\rm Im}\,F=0$, where $F$ is the auxiliary field within the 
chiral multiplet $S$.
However, the irreducibility of $S$ with respect to supersymmetry transformations
implies that we cannot set   ${\rm Im}\,F=0$ without setting $S=0$.
This will only happen for a superconformal theory.

Thus, to summarize:  Eq.~(\ref{badguess}) does not represent 
the supercurrent supermultiplet corresponding to Eq.~(\ref{lag2}). 
As we have shown in Sect.~3.4,  any potential supercurrent superfield
$J_{\alpha\alphadot}$
corresponding to the FI term 
  {\it must}\/ preserve $R_5$-symmetry,
since $j^{(5)}_\mu$-conservation is guaranteed by the Noether theorem, and no possible improvement
terms beyond the Noether result can change such a critical piece of physics. 
However, Eq.~(\ref{badguess}) does represent the  
supercurrent supermultiplet of a theory which is
a close cousin to that in Eq.~(\ref{lag2}), namely its chirally-compensated counterpart.
Not surprisingly, this counterpart theory has broken $R_5$-invariance and broken
$U(1)_{\rm FI}$ invariance as a result of the assignment of a non-zero VEV to the
chiral compensator field $\Sigma$.
This, then, is the source of the $U(1)_{\rm FI}$ gauge non-invariance
of the result in Eq.~(\ref{badguess}), and the source of its ensuing implications.

\subsection{Evading the proof?}
\setcounter{footnote}{0}

In Sect.~4.4, we demonstrated that one cannot employ the chiral formalism
in order to derive a supercurrent superfield corresponding to an $R_5$-invariant theory
such as the FI theory in Eq.~(\ref{lag}).   
Indeed, we showed that imposing $R_5$-invariance on the final result
yields the
constraint equations in Eqs.~(\ref{consequences}),
and these transcend the equations of motion for $V$.
As we discussed, these constraint equations essentially imply that $V$ must be a linear multiplet,
up to K\"ahler transformations which do not change the physics.
In general, such a truncation of $V$ is unacceptable, as it does not embody the full
set of symmetries of the action.

However, if our Lagrangian also were to contain other terms
(in addition to the FI term) 
which modify the equations of motion for $V$ so that they would now
be consistent with the constraint equations in Eq.~(\ref{consequences}),
no inconsistency would result.
In such a case, 
a fully consistent, $R_5$-conserving FI supercurrent supermultiplet $\Xi^{(C)}_{\alpha \alphadot}$
could potentially be constructed. 

One major clue towards a possible choice for such extra Lagrangian terms comes from the
fact that the resulting modified equations of motion for $V$, along with their supersymmetric extensions,
would have to include the gauge non-invariant constraint that $\partial_\mu A^\mu=0$.
Thus, any suitable extra Lagrangian term must 
break the $U(1)_{\rm FI}$ gauge invariance of the theory.

The obvious choice is to introduce a supersymmetric mass term $m^2 V^2$ into the Lagrangian.
In other words, let us now take our theory to be given by
\begin{equation}
 \calL ~=~  \quarter \left(W^\alpha W_\alpha|_{\theta\theta} + \Wbar_\alphadot 
           \Wbar^\alphadot|_{\thetabar\thetabar}\right)  +  m^2 V^2 |_{\theta\theta\thetabar\thetabar}
    + 2\xi V|_{\theta\theta\thetabar\thetabar}~.
  \label{lag3}
\end{equation}
We then find that the new equations of motion take the general form
\begin{eqnarray}
          \partial_\mu F^{\mu\nu} &=& m^2 A^\nu\nonumber\\
             m^2 M &=& m^2 N ~=~0\nonumber\\
    m^2 \lambda_\alpha &=& -i m^2 (\sigma^\mu \partial_\mu \chibar)_\alpha\nonumber\\
    m^2 D &=& - m^2 \Box C \nonumber\\
    D &=& -m^2 C -\xi\nonumber\\
    m^2 \chi_\alpha &=& -i (\sigma^\mu \partial_\mu \lambdabar)_\alpha~,
\end{eqnarray}
along with the Bianchi identity $\partial_\mu \tilde F^{\mu \nu}=0$, where 
$\tilde F^{\mu\nu}\equiv {i\over 2} \epsilon^{\mu\nu\lambda\sigma}F_{\lambda\sigma}$. 
For $m=0$, of course,
these equations reduce to the equations of motion of the usual gauge-invariant $U(1)_{\rm FI}$
theory with an FI term.
However, for $m\not=0$, these equations become
\begin{eqnarray}
          \partial_\mu F^{\mu\nu} &=& m^2 A^\nu\nonumber\\
             M &=& N ~=~0\nonumber\\
     \lambda_\alpha &=& -i (\sigma^\mu \partial_\mu \chibar)_\alpha\nonumber\\
     D &=& -  \Box C \nonumber\\
    D &=& -m^2 C -\xi\nonumber\\
    m^2 \chi_\alpha &=& -i (\sigma^\mu \partial_\mu \lambdabar)_\alpha~.
\label{neweqs}
\end{eqnarray}
Indeed, combining these equations, we now have 
\begin{equation}
    \Box' C -\xi =
    \Box' \chi_\alpha =
    \Box' A_\mu = 
    \Box' \lambda_\alpha = 
    \Box' D = M=N=0~,
\end{equation}
where $\Box'\equiv \Box - m^2$.
Thus, for $m\not=0$, we see that the equations of motion themselves reduce $V$ to a linear
multiplet, whereupon we also have the constraint $\partial_\mu A^\mu=0$
as a consequence of the supersymmetry algebra.

It is clear that these equations are a subset of those in Eq.~(\ref{consequences}).
Thus, for $m\not=0$, Eq.~(\ref{consequences}) is automatically satisfied and there
is no inconsistency in taking $T_S\sim V$.
In other words, in the presence of a supersymmetric mass for the $U(1)_{\rm FI}$ gauge field,
there is no fundamental obstruction to constructing 
an $R_5$-conserving FI contribution $\Xi^{(C)}_{\alpha\alphadot}$ 
to the supercurrent supermultiplet, even within the chiral formalism.

However, this observation begs the question:   to what extent can we claim that such 
a broken-$U(1)_{\rm FI}$ theory really has an FI term?
After all, 
the K\"ahler potential for this theory can be written in the form
\begin{equation}
               K ~=~ m^2 V^2 + 2 \xi V~,
\label{eeqn1}
\end{equation}
but thanks to the mass $m$, 
we are always free to define a shifted vector superfield $V'$, 
\begin{equation}
                 V'~\equiv~ V + {\xi\over m^2}~,
\label{eeqn2}
\end{equation}
in terms of which the K\"ahler potential now takes the form
\begin{equation}
               K ~=~ m^2 (V')^2 - {\xi^2\over m^2}~.
\label{eeqn3}
\end{equation}
Note that both the shift in Eq.~(\ref{eeqn2}) and the overall constant in Eq.~(\ref{eeqn3})
have no physical effects on a theory with global supersymmetry.
Even in a theory with local supersymmetry, 
these implications of taking $m\not=0$ still hold, 
even though these shifts will have other physical effects. 
As a consequence, it makes perfect sense that this is the one case in which 
there is no obstruction to building an FI supercurrent supermultiplet $\Xi_{\alpha\alphadot}$:
indeed, this is the one case in which our theory really has no FI term at all.

Thus, our central result still stands:  
true FI terms
do not lead to self-consistent $R_5$-preserving contributions $\Xi^{(C)}_{\alpha\alphadot}$ 
for supercurrent supermultiplets.
Only when the theory has a ``fake'' FI term, as discussed above, does such a corresponding
supercurrent supermultiplet exist.
  
It is important to note that
even in this case, this still does not yield a total supercurrent supermultiplet
$J^{(C)}_{\alpha\alphadot}$
which exhibits an unbroken $R_5$-symmetry in the chiral formalism.  The reason is that 
the addition of the $m^2 V^2$ term into the Lagrangian induces a further, $m$-dependent
contribution to the supercurrent supermultiplet, and this
further contribution will necessarily break $R_5$-symmetry again.
Thus, while it is possible to achieve a partial success in which $\Xi^{(C)}_{\alpha\alphadot}$
exhibits an unbroken $R_5$-symmetry, use of the chiral formalism guarantees that 
this can never be a property of the total supercurrent
supermultiplet as a whole.

\section{Analysis in the linear formalism}
\setcounter{footnote}{0}

As we have seen in Sect.~4, the difficulties that arise in theories
with FI terms in the chiral formalism arise essentially because the 
chirally-compensated $D$-term
action in Eq.~(\ref{superconftheory}) is not gauge invariant under $U(1)_{\mathrm{FI}}$
transformations unless the chiral compensators $\Sigma$ and $\overline{\Sigma}$ are charged under
$U(1)_{\mathrm{FI}}$.  Furthermore, since the chiral compensators also carry 
non-zero $R_5$-charges, we see that $R_5$-invariance is always broken in the chiral
formalism, regardless of
whether it was preserved in the globally-supersymmetric version of the theory.
As a consequence of this artificial breaking of the $R_5$ symmetry, we found that
the FI supercurrent supermultiplet $\Xi^{(C)}_{\alpha\alphadot}$ in the chiral
formalism is not only 
non-vanishing in theories with FI terms, but also fails to be gauge invariant.

Unfortunately, many of these conclusions hinged on the 
structure of the chiral formalism itself.
It is therefore 
unclear to what extent these inconsistencies are general
truths about FI terms in supergravity, or merely 
artifacts of the conformal-compensator formalism employed.  
Indeed,     as discussed in the Introduction, there exist other, 
alternative formulations of supergravity with different sets of
conformal compensators in which the action of the 
conformally-compensated theory remains invariant under 
$U(1)_{\mathrm{FI}}$ transformations.  
It is therefore important to understand whether the primary conclusion 
of Sect.~4 --- namely that theories with non-zero FI terms must exhibit 
additional global symmetries ---  holds in such alternative formalisms as well.
If not, there would then be no problem in coupling any theory with an
FI term to supergravity, so long as this theory admitted a description in such a 
formalism.   

Our primary goal, then, is to 
understand the extent to which the conclusions of Sect.~4 and their 
implications for FI terms are modified by a change of framework. 
In this section, we will therefore re-examine the issues involved with coupling FI terms
to supergravity using an alternate framework: the so-called ``linear-compensator'' or 
``new minimal'' formalism of Ref.~\cite{LinearFormalismSUGRA}.  By its very nature, 
this formalism manifestly preserves both $R_5$-symmetry and $U(1)_{\rm FI}$ gauge 
invariance;  hence the gauge-invariance issues that arise in the chiral formalism
due to the spurious breaking of $R_5$-invariance by the compensator fields 
will not arise here. 

We begin this section with a brief review of the linear formalism.  We then 
give a proof that in this formalism, no additional FI 
contribution $\Xi^{(L)}_{\alpha\alphadot}$ to the supercurrent supermultiplet can possibly 
exist.
We shall then demonstrate how this same result can be understood through the linear
compensator formalism, and finally discuss one possible case in which this result
might seem to be altered.

\subsection{Linear compensator formalism:  General outline}

In the linear formalism, just as in the chiral formalism, one introduces a set
of conformal compensator fields, the role of which is to restore superconformal
invariance to the theory in question. 
Thus, in this way, the resulting theory may be successfully
be coupled to conformal supergravity.  
The fundamental ingredients of the linear formalism are:
\begin{itemize}
\item   a linear compensator multiplet $L$, with Weyl weight
         $w_L=2$ and vanishing $R_5$-charge;
\item   a pair of chiral compensator multiplets $\Sigma_L,\Sigmabar_L$, 
            with Weyl weights $w_{\Sigma_L}=w_{\overline{\Sigma}_L}=1$ 
            and $R_5$-charges $\pm 2/3$ respectively;  and
\item  a new local $U(1)$ symmetry, henceforth denoted $U(1)_L$, under 
        which 
\begin{equation}
  \Sigma_L~\rightarrow~ \Sigma_L e^{-i\Lambda_L} ~,~~~~~~~ 
  \Sigmabar_L~\rightarrow~ \Sigmabar_L e^{ i\overline{\Lambda}_L}~,
\label{eq:ExcessGauge}
\end{equation}
where $\Lambda_L$ is a chiral superfield parameter. 
The remaining fields in the theory will be assumed neutral under $U(1)_L$
transformations.
This $U(1)_L$ symmetry plays a crucial role in determining the structure 
of the action in the linear formalism, as we shall soon see.
Note that even though this symmetry is local, we do not introduce any corresponding
gauge bosons.  This is therefore a local $U(1)$ symmetry of the action, but not
a fully dynamical gauge symmetry.
We will discuss this symmetry further below.
\end{itemize}

We begin by discussing how an arbitrary K\"ahler potential may
be made superconformal in this context.
Just as in the case of the chiral formalism, let us assume
that we may write our K\"ahler potential in the form
 $K=K(\Phi,\Phi^\dagger,V)=\sum_n K_n$
where $K_n$ has Weyl weight $n$ and vanishing $R_5$-charge.
We can then restore scale invariance to the theory 
by multiplying each term in $K$ by an appropriate power of $L$ so as 
to define a new quantity with vanishing Weyl weight:  
\begin{equation}
  \widetilde{K}_L(\Phi_i,\Phi^\dagger_i,V) ~\equiv~ 
      \sum_n \left(\frac{L}{3M_P^2}\right)^{-n/2} K_n~.
\label{eq:KtildeL}
\end{equation}
In terms of this new quantity $\widetilde K_L$, the K\"ahler potential
$\widehat K$
of our conformally compensated theory is then given by
\begin{eqnarray}
    \widehat K	&\equiv& 
        L\ln\left[ {L \over \Sigma_L \Sigmabar_L}
     \exp \left( {\widetilde{K}_L(\Phi_{i},\Phi^\dagger_{i}) \over 3M_P^2}\right)
    \right]~ \nonumber\\
  &=& 
        L\ln\left( {L \over \Sigma_L \Sigmabar_L}\right) +
        L \, {\widetilde{K}_L(\Phi_{i},\Phi^\dagger_{i}) \over 3M_P^2}~,
\label{KhatL}
\end{eqnarray}
whereupon the corresponding 
$D$-term Lagrangian of the compensated theory is
\begin{equation}
  \mathcal{L} ~=~ \int d^4 \theta\, 
     L\ln\left[ {L \over \Sigma_L \Sigmabar_L}
     \exp \left( {\widetilde{K}_L(\Phi_{i},\Phi^\dagger_{i}) \over 3M_P^2}\right)
    \right]~.
\label{eq:LCompLag}
\end{equation}
As in the chiral formalism, we can then reproduce our original Lagrangian
by setting our compensator fields to fixed values, \ie, 
\begin{equation}
  L~\rightarrow~ 3M_P^2 ~,~~~~~~ \Sigma_L,\Sigmabar_L~\rightarrow~ \sqrt{3}M_P~,
\label{eq:FrozenLinearComps}
\end{equation}
and then taking the $M_P\rightarrow\infty$ limit.  Indeed,
following this procedure, we see that $\widehat K\to K$ directly.  
 
Several comments are in order.  
First, it should be noted that this is already quite different from
the analogous situation in the chiral formalism.  In the chiral formalism,
we did not find $\widehat K\to K$ upon ``freezing'' our compensator fields
and taking $M_P\to \infty$;  indeed, this only occurred upon
integrating over $d^4\theta$.  Moreover, even after doing this integration,
we still did not precisely reproduce our original theory in the chiral
formalism;  we only reproduced its overall algebraic form.  Indeed, in the
final version, all of the non-trivial Weyl weights and $R_5$-charges were ultimately
stripped from the fields in question.  
By contrast, in the linear formalism, we have not tampered with our fundamental
matter fields $\Phi$ at all.  Thus, upon fixing our compensator fields
to fixed values and taking the $M_P\to\infty$ limit, our original 
theory is reproduced exactly.
In other words, there is no ``hysteresis'' effect that emerges 
upon introducing our compensators
and then freezing them to fixed values.

Second,
we note that the K\"ahler portion of the compensated Lagrangian 
is invariant not only under
Weyl rescalings and $R_5$-symmetries, but also under local $U(1)_L$ transformations. 
Under the $U(1)_L$ transformation in Eq.~(\ref{eq:ExcessGauge}),
we see from Eq.~(\ref{KhatL}) that
\begin{equation}
        \widehat K ~\rightarrow~ \widehat K + i L(\Lambda_L - \overline{\Lambda}_L)~.
\end{equation}
However, it may easily be shown as a mathematical identity that
for any linear superfield $L$ and chiral superfield $\Lambda_L$, 
the quantity
\begin{equation}
        \int d^4\theta  \, L \Lambda_L
\end{equation}
is a total derivative. 
Specifically, in component form, we find
\begin{eqnarray}
  \delta \calL &=& \int d^4\theta \, L \Lambda_L\nonumber\\ 
     & = & 
  -\frac{1}{4}(\phi_L\Box C - C\Box\phi_L)-\frac{i}{2}A^\mu\partial_\mu\phi_L 
   -\frac{1}{\sqrt{2}}\left(\chibar\sigmabar^\mu\partial_\mu\psi_L- 
   \psi_L\sigma^\mu\partial_\mu\overline{\chi}\right) \nonumber \\
  & = &  \partial_\mu \left[
  \frac{1}{4}(C\partial^\mu\phi_L-\phi_L\partial^\mu C)
    +\frac{1}{\sqrt{2}}(\psi\sigma^\mu\overline{\chi})-\frac{i}{2}A^\mu\phi_L
   \right]  
\end{eqnarray}
where $\{C,\chi_\alpha,\overline{\chi}_{\dot{\alpha}},A^\mu\}$ and
$\{\phi_L,\psi_L,F_L\}$ are the component fields within 
$L$ and $\Lambda_L$ respectively.  
Thus, the action of our theory is invariant under $U(1)_L$ transformations.
Note that $\Lambda_L$ and $\overline{\Lambda}_L$ have been treated
as fields in this analysis rather than as constant parameters;
consequently the 
full symmetry of the action is local rather than global.   
This remains true even though no gauge multiplet corresponding to the $U(1)_L$ symmetry
has been introduced.  

Finally, 
we note that the K\"ahler portion of our action is also invariant
under $U(1)_{\mathrm{FI}}$ gauge 
transformations.
The logic is similar to the previous case.
Under $U(1)_{\rm FI}$ transformations of the form
in Eq.~(\ref{U1gaugetransf}),
we see that
\begin{equation}
          \widehat K ~\to~ \widehat K +  {2i \xi \over 3 M_P^2} L(\Lambda - \overline{\Lambda})~.
\label{u1fishift}
\end{equation}
Under $\int d^4\theta$ integration, this too yields a total divergence.

We now consider the superpotential $W(\Phi)$ in the linear
formalism.
It turns out that the form that the superpotential 
can take in the linear formalism is far more restricted
than it was in the chiral formalism.  
This is  due to the presence
of the additional $U(1)_L$ symmetry, which we are demanding be a symmetry
of our superconformal compensated theory.  
Because the chiral compensator fields $\Sigma_L$ and $\Sigmabar_L$ carry
non-zero $U(1)_L$ charges in addition to their Weyl weights and $R_5$-charges, 
these fields can no longer compensate 
for explicit breakings of Weyl or $R_5$-invariance in the superpotential
in the same way that $\Sigma$ and ${\Sigmabar}$ were previously 
able to do in the chiral formalism.  
Indeed, while $\Sigma$ and $\Sigmabar$ in the chiral
formalism were able to ``soak up'' unwanted Weyl weights and $R_5$-charges
from the different terms of our superpotential,
we see that $\Sigma_L$ and $\Sigmabar_L$ can no longer do the same thing
without simultaneously breaking $U(1)_L$.

As a result, we see that our superpotential
$W(\Phi)$ cannot be appropriately compensated
in the linear formalism
in order to build a fully superconformal theory
unless it was already $R_5$-invariant and 
Weyl-invariant to begin with.
Thus, we conclude that the linear compensator formalism can be only used
to produce a superconformal, compensated theory 
from an original theory which is already $R_5$-invariant and whose
superpotential is also already conformally (Weyl) invariant~\cite{LinearFormalismSUGRA}.      
However, if a theory satisfies these criteria, we may still 
define an alternative set of rescaled matter fields
$\widetilde{\Phi}_{Li}$, 
$\widetilde{\Phi}_{Li}^\dagger$
through the relations
\begin{equation}
    \Phi_i~ =~\left({\Sigma_L\over \sqrt{3}M_P}\right)^{w_i} \widetilde{\Phi}_{Li}
     ~,~~~~~~
    \Phi_i^\dagger~ =~  \left({\Sigmabar_L\over \sqrt{3}M_P}\right)^{w_i}
             \widetilde{\Phi}_{Li}^\dagger 
\label{eq:RedefinePhiLinear}
\end{equation}
Like $\widetilde{\Phi}_i$ and $\widetilde{\Phi}_i^\dagger$ in the chiral formalism,
these new fields will have
vanishing Weyl weight.  Furthermore, as we have seen,
the Weyl weight and $R_5$-charge of any chiral supermultiplet
must be related through Eq.~(\ref{WeylRconstraint}). 
Thus $\widetilde{\Phi}_{Li}$ and 
$\widetilde{\Phi}_{Li}^\dagger$ also have vanishing $R_5$-charge.  They
are, however, charged under $U(1)_L$.

Note that although $\Sigma_L$ and $\Sigmabar_L$ have non-trivial 
$R_5$-charges, the act of setting these compensator multiplets to the
constant values 
in Eq.~(\ref{eq:FrozenLinearComps})
will not affect the $R_5$-invariance of the
theory.  This is because these compensator fields appear in 
Eq.~(\ref{KhatL}) only in the $R_5$-invariant combination
$\Sigma_L \Sigmabar_L$.
Thus,
once we fix our compensators to the fixed values in Eq.~(\ref{eq:FrozenLinearComps}),
we see that ``frozen'' theory will continue to preserve 
$R_5$-invariance, in sharp contrast to what happens in the chiral formalism.  This, in
turn, implies that $\partial^\mu j_\mu^{(5)}=0$, and that the conservation
equation for the supercurrent supermultiplet takes the form 
$\overline{D}^{\alphadot}J_{\alpha\alphadot}^{(L)}=L_\alpha$, as discussed in
Sect.~3.    
Indeed, as we shall show explicitly in Sect.~5.3,
the variations of $\Sigma_L$ and $\Sigmabar_L$ --- unlike those of $\Sigma$ and $\Sigmabar$ 
in the chiral formalism ---  do not contribute to $j_\mu^{(5)}$.


\subsection{Deriving the FI supercurrent supermultiplet in the linear formalism:
    A proof of the non-existence of an FI contribution $\Xi_{\alpha\alphadot}$
    to the supercurrent supermultiplet}

We now turn to address the FI contribution $\Xi^{(L)}_{\alpha\alphadot}$ to the 
supercurrent supermultiplet within the linear formalism.  As we shall prove,
the only possible solution consistent with the symmetries of the theory 
is $\Xi^{(L)}_{\alpha\alphadot}=0$.  Clearly, this result differs from the corresponding,
non-zero result for $\Xi^{(C)}_{\alpha\alphadot}$ in Eq.~(\ref{Xisolnchiral}), which
was derived using the chiral formalism.  
This then provides graphic illustration that the form of the supercurrent supermultiplet
in theories with non-zero FI terms is highly formalism-dependent.
 
In order to show that any additional contribution $\Xi^{(L)}_{\alpha\alphadot}$ 
to the supercurrent supermultiplet must vanish in the linear formalism, we begin
by noting that amongst all of the possible terms that may appear in a supersymmetric
Lagrangian with a $U(1)$ gauge symmetry, the FI term is unique in that it simultaneously
exhibits three properties:
\begin{itemize}
\item   First, because the FI coefficient has a mass dimension, the FI term introduces a 
         mass scale (or equivalently a cosmological constant) into the 
          theory and consequently breaks superconformal invariance. 
\item   Second, because both $D$ and $C$ are neutral under $R_5$-symmetries, 
           the FI term preserves $R_5$-symmetry.  This was explicitly verified 
           in Sect.~3.4 using the Noether procedure,
             and this must remain true --- regardless of the addition
          of any possible improvement terms --- for any supercurrent supermultiplet which is 
            to be directly associated with the FI term.
          This $R_5$-invariance should be manifest for any supermultiplet constructed 
            within the linear formalism.
\item   Finally, unlike kinetic terms 
          or mass terms which are quadratic in the fundamental fields of the theory ---
        and likewise unlike superpotential terms which are often 
        cubic or higher in the fundamental fields ---
        the FI term is {\it linear}\/ in the fundamental fields.
\end{itemize}
As we shall now prove, no term which has these three properties simultaneously 
can yield a non-vanishing contribution $\Xi^{(L)}_{\alpha\alphadot}$
to the supercurrent supermultiplet of the theory.  

Our proof proceeds as follows.
As we discussed in Sect.~3, 
any theory which breaks superconformal invariance while preserving $R_5$-symmetry must
give rise to a supercurrent supermultiplet within the linear formalism 
whose lowest components 
$(C_\mu,\chi_{\mu\alpha},\hat T_{\nu\mu})$ 
satisfy the reduced supersymmetry algebra in Eq.~(\ref{reducedalgebra}).
However, given this reduced supersymmetry algebra,
it is possible to consider two successive supersymmetry transformations
of magnitudes $\eta$ and $\epsilon$ respectively
and thereby derive a self-consistency constraint on the single field $\chi_{\mu}$:
\begin{eqnarray}
 \delta_\epsilon\delta_\eta \chi_{\mu\alpha} &=& 
        (\sigma^\nu \etabar)_\alpha 
         (\partial_\nu \delta_\epsilon C_\mu + i \delta_\epsilon \hat T_{\nu\mu})\nonumber\\
  &=& i (\sigma^\nu \etabar)_\alpha [ -\epsilon \sigma^\rho \sigmabar_\nu (\partial_\rho \chi_\mu)
            + \epsilonbar \sigmabar_\nu \sigma^\rho (\partial_\rho \chibar_\mu)] \nonumber\\
  &=& -2i (\epsilon \sigma^\nu \etabar) (\partial_\nu \chi_{\mu\alpha}) 
      +2i (\epsilonbar \etabar) (\sigma^\nu \partial_\nu \chibar_{\mu})_\alpha~.
\label{mastereq}
\end{eqnarray}
This single constraint equation then governs what possible solutions for $\chi_\mu$  
might exist:
any $\chi_\mu$ which fails to satisfy this constraint equation cannot possibly be a component of
the appropriate supercurrent supermultiplet.  By contrast, any $\chi_\mu$ which satisfies
this equation {\it might or might not}\/ lead an appropriate self-consistent
supercurrent supermultiplet;  in such cases, it would still remain to verify 
that appropriate solutions for $C_\mu$ and $\hat T_{\nu\mu}$ in Eq.~(\ref{reducedalgebra}) also exist.

One critical feature of the constraint equation in Eq.~(\ref{mastereq}) 
is that its right side is independent of $\eta$, depending only on $\etabar$.
As we see from Eq.~(\ref{reducedalgebra}),
this is a direct consequence of the fact that $M_\mu=N_\mu=0$.  As discussed
in Sect.~3.3, this in turn
is a general feature of supercurrent supermultiplets in theories with
unbroken $R_5$-symmetry. 

In general, there are potentially many forms for $\chi_{\mu\alpha}$ 
which will satisfy Eq.~(\ref{mastereq}).
However, because we are focusing on the specific case of an FI term $\Xi^{(L)}_{\alpha\alphadot}$, 
we see that $\chi_{\mu\alpha}$ must 
be {\it linear}\/ in the fields that appear as components of our original
supermultiplet $V$ in Eq.~(\ref{vecsup}).
The Lorentz vectorial/spinorial nature of $\chi_{\mu\alpha}$, 
along with elementary dimensional analysis,
then require that $\chi_{\mu\alpha}$ can at most take the form
\begin{equation}
          \chi^\mu_\alpha ~=~  X\,  (\sigma^\mu \lambdabar)_\alpha + Y\, \partial^\mu \chi_\alpha
                      + Z\, (\sigma^{\mu\nu} \partial_\nu\chi)_\alpha~,          
\label{bigansatz}
\end{equation}
where $X$, $Y$, and $Z$ are unknown (generally complex) coefficients.
Inserting this {\it ansatz}\/ into Eq.~(\ref{mastereq}),
it is then possible to obtain self-consistency constraints 
on the coefficients $X$, $Y$, and $Z$.

As a first step, it is fairly easy to show that we must have $Y=Z=0$.
Indeed, taking Eq.~(\ref{bigansatz}) for $\chi^\mu_\alpha$ and evaluating the double-variation
$\delta_\epsilon \delta_\eta \chi^\mu_\alpha$, we obtain an expression whose $\eta$-dependent terms
are given by
\begin{equation}
   \delta_\epsilon \delta_\eta \chi^\mu_\alpha\biggl|_\eta ~=~  
        \left[  (Y g^{\mu\nu} + Z \sigma^{\mu\nu}) \eta\right]_\alpha
        \,\left(  2i \epsilonbar \sigmabar^\rho \partial_\nu\partial_\rho \chi + 2\epsilonbar \partial_\nu
    \lambdabar\right)~. 
\label{etaterm}
\end{equation}
However, we see from Eq.~(\ref{mastereq}) that all $\eta$-dependent terms must cancel.
Since $g^{\mu\nu}$ and $\sigma^{\mu\nu}$ are respectively even and odd under exchange of their
Lorentz indices, this requires that each term vanish separately in the coefficient of Eq.~(\ref{etaterm}).
We thus have $Y=Z=0$. 
 
Given this, 
we can now proceed to test whether the sole remaining possible term
\begin{equation}
          \chi^\mu_\alpha ~\stackrel{?}{=}~ X (\sigma^\mu \lambdabar)_\alpha ~=~ 
         X \sigma^\mu_{\alpha\alphadot}\lambdabar^\alphadot~
\label{ansatz}
\end{equation}
satisfies the constraint equation in Eq.~(\ref{mastereq}).
Evaluating the double-variation of this expression directly in terms of the underlying
fields in the theory and 
focusing first on the terms which are holomorphic in $\epsilon$, we find 
\begin{eqnarray}
      \delta_\epsilon\delta_\eta \lambdabar_\alphadot\biggl|_\epsilon &=& 
            i\etabar_\alphadot (\epsilon \sigma^\mu \partial_\mu \lambdabar) 
 + {i\over 2} (\etabar \sigmabar^\nu \sigma^\mu)_\alphadot (\epsilon \sigma_\nu \partial_\mu \lambdabar) 
 - {i\over 2} (\etabar \sigmabar^\nu \sigma^\mu)_\alphadot (\epsilon \sigma_\mu \partial_\nu \lambdabar) 
              \nonumber\\
      &=&
            i\left[ \etabar_\alphadot (\epsilon \sigma^\mu \partial_\mu \lambdabar) 
    + (\epsilon \sigma^\mu)_\alphadot (\etabar \partial_\mu \lambdabar) 
    - (\partial_\mu \lambdabar_\alphadot) (\epsilon \sigma^\mu \etabar)\right]  \nonumber\\
    &=& -2i \partial_\mu \lambdabar_\alphadot (\epsilon \sigma^\mu \etabar)~.
\label{testcase}
\end{eqnarray}
In passing to the third line of Eq.~(\ref{testcase}), we have
used the hermitian conjugate of the cyclic identity
$A_\alpha (BC) + B_\alpha (CA) + C_\alpha (AB) =0$, where $A$, $B$, and $C$ are all spinors.
From Eq.~(\ref{testcase}), it then follows that
\begin{equation}
      \delta_\epsilon \delta_\eta \chi^\mu_\alpha\biggl|_\epsilon~=~ 
    -2i X(\sigma^\mu \partial_\nu \lambdabar)_\alpha (\epsilon \sigma^\nu \etabar)~=~
     -2i X(\partial_\nu \chi^\mu_\alpha) (\epsilon \sigma^\nu \etabar)~, 
\label{testcase1}
\end{equation}
in complete accordance with Eq.~(\ref{mastereq}).
Thus, the terms which are holomorphic in $\epsilon$ succeed in satisfying 
the constraint equation~(\ref{mastereq})
for any $X$.

However, if we examine the terms which depend on $\epsilonbar$, we find that 
\begin{eqnarray}
      \delta_\epsilon\delta_\eta \lambdabar^\alphadot\biggl|_\epsilonbar &=& 
      -i\etabar^\alphadot (\epsilonbar \sigmabar^\mu \partial_\mu\lambda) +
           {i\over 2} \epsilon^{\alphadot\betadot} (\etabar \sigmabar^\nu\sigma^\mu)_\betadot \left(
  \epsilonbar \sigmabar_\nu \partial_\mu\lambda - \epsilonbar \sigmabar_\mu \partial_\nu\lambda
          \right)\nonumber\\
    &=&  -i\left[ 
        \etabar^\alphadot (\epsilonbar \sigmabar^\mu \partial_\mu \lambda)
          - (\sigmabar^\mu \partial_\mu \lambda)^\alphadot (\epsilonbar\etabar)
         + \epsilonbar^\alphadot (\etabar \sigmabar^\mu \partial_\mu \lambda)\right]\nonumber\\
    &=&   2i (\sigmabar^\mu \partial_\mu \lambda)^\alphadot (\epsilonbar\etabar)~,  
\end{eqnarray}
whereupon we find that 
the double-variation of $\chi^\mu_\alpha$ in Eq.~(\ref{ansatz}) is given by
\begin{eqnarray}
      \delta_\epsilon\delta_\eta \chi^\mu_\alpha\biggl|_\epsilonbar  ~=~
     2 i X (\sigma^\mu \sigmabar^\nu \partial_\nu \lambda)_\alpha (\epsilonbar\etabar)~.
\label{lhs}
\end{eqnarray}
By contrast, the $\epsilonbar$-dependent terms on the right side of Eq.~(\ref{mastereq})
yield
\begin{equation}
      2i X^\ast  (\sigma^\nu \partial_\nu \chibar^{\mu})_\alpha (\epsilonbar\etabar)~=~ 
              -2 i X^\ast (\sigma^\nu \sigmabar^\mu \partial_\nu \lambda)_\alpha (\epsilonbar\etabar)~,
\label{rhs}
\end{equation}
and we see that the expressions in Eqs.~(\ref{lhs}) and (\ref{rhs}) are unequal.  
Indeed, 
we find that Eqs.~(\ref{lhs}) and (\ref{rhs}) differ by the non-zero quantity
\begin{equation}
      -4 
      \left[   ({\rm Re}\, X) (i\partial^\mu \lambda_\alpha) 
                 +2 ({\rm Im}\, X) (\sigma^{\mu\nu} \partial_\nu \lambda)_\alpha \right]\,
     (\epsilonbar\etabar)~,
\end{equation}
which does not even vanish on-shell.
We thus conclude that the ansatz
in Eq.~(\ref{ansatz}) fails to satisfy the constraint equation in Eq.~(\ref{mastereq}),
or equivalently that there is only one self-consistent solution for $\chi^\mu_\alpha$ in Eq.~(\ref{bigansatz}):
\begin{equation}
               X~=~Y~=~Z~=~0~.
\end{equation}
Thus, we conclude that
\begin{equation}
          \Xi_{\alpha\alphadot}^{(L)}~=~0~.
\label{proofresult}
\end{equation}

We stress again that this result does {\it not}\/
imply that no supercurrent supermultiplet can ever be constructed for theories
in which superconformal symmetry is broken while $R_5$-symmetry is preserved.
Rather, what we have shown is that this cannot be done using expressions 
which are {\it linear}\/ in the component fields that appear 
within the vector multiplet $V$ in Eq.~(\ref{vecsup}).
Such expressions would be required for a potential FI contribution $\Xi^{(L)}_{\alpha\alphadot}$
to the total supercurrent supermultiplet of any theory containing an FI term.

It is worth emphasizing that our results were derived without use of the equations of motion
of the theory.  Equations of motion, of course, are the one feature of a theory which 
connect the different terms in its Lagrangian and thereby allow the presence of one term in the
Lagrangian to affect
the on-shell appearance of the supercurrent contributions from another term.
Therefore, it might seem that use of equations of motion could potentially invalidate the term-by-term approach
to calculating the supercurrent supermultiplet which is inherent in our implicit separation of
the supercurrent into a non-FI piece and an FI piece.
However, our proof of the non-existence of an FI
supercurrent supermultiplet $\Xi^{(L)}_{\alpha\alphadot}$ is 
intrinsically an {\it off}\/-shell proof:
it asserts that there exist no self-consistent solutions
for $(C_\mu,\chi_{\mu\alpha},\hat T_{\nu\mu})$ within $\Xi^{(L)}_{\alpha\alphadot}$ 
 {\it regardless}\/ of the equations of motion.
This permits our proof to hold in all generality.

Similarly, our proof
is also independent
of the precise mapping relations between
the supercurrent superfield components $(C_\mu,\chi_{\mu\alpha},\hat T_{\nu\mu})$
and the supercurrents $(j^{(5)}_\mu,j_{\mu\alpha},T_{\nu\mu})$.
Thus our proof should hold even in formulations in which these
mapping relations are modified, as long as the underlying supersymmetry algebra is consistent
with the constraint $M_\mu=N_\mu=0$ which characterizes $R_5$-invariant theories.

Finally, we remark that the supercurrent superfield is not a physical object;  
it can be modified
by K\"{a}hler transformations and other sorts of unphysical improvement terms.
However, since our proof has shown that there exist no consistent 
solutions for the components
of the FI contribution $\Xi^{(L)}_{\alpha\alphadot}$ to the total supercurrent superfield, 
this remains true despite the possibility of performing
K\"ahler transformations or adding improvement terms.
Indeed, all that was assumed in our proof was the fact that our supercurrent superfield is consistent
with unbroken $R_5$-symmetry, and this in turn led directly to our constraint equation
in Eq.~(\ref{mastereq}).
Consequently, our proof holds despite the possibility of K\"ahler transformations and other
improvement terms.

Thus, to summarize:  in the linear formalism,
theories that contain FI terms do not yield corresponding FI
contributions $\Xi^{(L)}_{\alpha\alphadot}$ to their supercurrent
supermultiplets.  Of course, such theories continue to have $R_5$-currents, supercurrents,
and energy-momentum tensors (each of which can be derived through the Noether procedure and
improved in various ways), and in general the FI term contributes to both the supercurrent
and the energy-momentum tensor.
However, what we have proven is that
there is no self-consistent $\xi$-dependent supermultiplet structure that can
be associated with these FI contributions.
It is this feature which stands in stark contrast to the analogous result 
in the chiral formalism.

It is important to note that the results of this proof do not necessarily lead
to any inconsistency insofar as the total supercurrent supermultiplet for the entire theory
in question is concerned, even in the presence of a non-zero FI term. 
Indeed, we note that applying the 
mapping relations in Eq.~(\ref{case3})
to the superfield components in Eq.~(\ref{compons})
yields results for the Noether currents 
which encapsulate not only those in Eq.~(\ref{Noetherresults}), 
 {\it but also those in Eq.~(\ref{Noether2})
when the equations of motion that are used are those
that exist in the presence of an FI term}\/.
Indeed, within the linear formalism, this is 
a general phenomenon for any theory with an FI term:
regardless of what superpotential terms might be added to the theory,
the presence of a non-zero FI term with coefficient $\xi$
has the net effect of shifting $D$ to $D+\xi$ 
in the equations of motion, and this shift,
when applied to the current contributions
frm the $U(1)_{\rm FI}$ kinetic terms,
always automatically generates the extra current contributions
given in Eq.~(\ref{Noether2}).
Thus, we see that within the linear formalism, the supercurrent superfield in Eq.~(\ref{puregauge})
encapsulates the correct individual currents
regardless of whether or not an FI term is introduced:
the appearance of an FI term simply shifts the equations of motion 
for the $D$-field in such a way that the required extra $\xi$-dependent current
contributions are automatically incorporated.


\subsection{Deriving the supercurrent supermultiplet in the linear formalism: 
Noether calculation}

In Sect.~5.2,
we gave an algebraic proof that no additional contribution 
$\Xi^{(L)}_{\alpha\alphadot}$ to the supercurrent supermultiplet exists in the 
linear formalism.  
However, it is also possible to understand this result by performing an explicit
Noether calculation in the fully compensated superconformal theory, 
taking into account the variations of
the compensator fields, in much the same manner as we did in the 
chiral formalism in Sect.~4.2.  In what follows, we perform such a 
calculation and show that these compensator fields yield no additional contribution
to $j_\mu^{(5)}$ in the $M_P\rightarrow\infty$ limit, 
\ie, that $\Xi_{\alpha\alphadot}^{(L)}=0$.
We therefore confirm, this time using the linear compensator formalism,
that there is
no additional FI contribution  $\Xi^{(L)}_{\alpha\alphadot}$ to the 
supercurrent supermultiplet.

Let us begin by considering how such an FI supercurrent
contribution $\Xi^{(L)}_{\alpha\alphadot}$ 
could possibly have arisen.
Clearly, since the linear compensator field $L$ carries no $R_5$-charge,
its variations will yield no Noether contribution to $j_\mu^{(5)}$.  However, the variations of
$\Sigma_L$ and $\Sigmabar_L$ could potentially contribute to $j_\mu^{(5)}$,
as these fields carry non-trivial $R_5$-charges. 

We  now demonstrate that no such contribution arises.  Although we could
do this through a brute-force calculation as in Sect.~4.2,
it is possible to take a useful shortcut.
First, 
we observe that the variation
of the Lagrangian of the compensated theory in Eq.~(\ref{eq:LCompLag}) 
with respect to any component field 
$\zeta$ (or derivative of such a component field) within $\Sigma_L$ or $\Sigmabar_L$ becomes   
\begin{equation}
  \frac{\partial\mathcal{L}}{\partial \zeta} ~=~ 
     - \int d^4\theta{L\over \Sigma_L\Sigmabar_L} 
        {\partial (\Sigma_L\Sigmabar_L)\over \partial \zeta}
    ~ \longrightarrow ~
     -\int d^4\theta {\partial (\Sigma_L\Sigmabar_L)\over \partial \zeta}
\end{equation}
where we have set $L$, $\Sigma_L$, and $\Sigmabar_L$
to their corresponding fixed values 
in passing to the final expression.  
However, we recognize that this is nothing but the form
that we would be dealing with if we were calculating
the Noether contributions to $j_\mu^{(5)}$ 
coming from the K\"ahler term in the trivial Wess-Zumino model
with $K=\Phi\Phi^\dagger$.
Therefore, up to an overall prefactor,
we can borrow the well-known Noether result for the Wess-Zumino model to write
\begin{equation}
     j_\mu^{(5)}|_{\Sigma,\Sigmabar} ~=~
     {2i\over 3}\,\left(
       \phi_\Sigma^\ast \partial_\mu \phi_\Sigma - 
       \phi_\Sigma \partial_\mu \phi_\Sigma^\ast
     \right)
   + {1\over 3} \,\psi_\Sigma \sigma_\mu \overline{\psi}_\Sigma~,
\label{WZjmu5}
\end{equation}
where $\lbrace \phi_\Sigma, \psi_\Sigma,F_\Sigma\rbrace$
are the component fields of $\Sigma_L$.  

The next step is to 
set the component fields in $\Sigma_L$ and 
$\Sigmabar_L$ to their fixed values
\begin{equation}
   \phi_\Sigma \rightarrow M_P ~,~~~~~~
   \psi_\Sigma, F_\Sigma \rightarrow 0 ~.
\end{equation}
However, when we do this, we find that
their contribution to $j_\mu^{(5)}$ in Eq.~(\ref{WZjmu5})
vanishes:
\begin{equation}
    j_\mu^{(5)}|_{\Sigma,\Sigmabar} ~\rightarrow~ 0~.
\label{eq:LinearCurrentNoContrib}
\end{equation}
Identifying this as the bottom component of 
the supercurrent superfield $\Xi_{\alpha\alphadot}^{(L)}$, we thus once again see that
\begin{equation}
         \Xi_{\alpha\alphadot}^{(L)}~=~ 0~,
\end{equation}
in complete agreement with the results of our proof in Eq.~(\ref{proofresult}).

We therefore conclude that $\Sigma_L$ and $\Sigmabar_L$ do not contribute to $j_\mu^{(5)}$
in the way that $\Sigma$ and $\Sigmabar$ do in the chiral formalism.
Indeed, $\Xi^{(L)}_{\alpha\alphadot}=0$, and the supercurrent supermultiplet 
in the linear formalism remains exactly what it was in the uncompensated
theory.
As a result, our supercurrent supermultiplet is manifestly $U(1)_{\rm FI}$ 
gauge invariant.


\subsection{Evading the proof?}

In Sect.~4.4, within the context of the chiral formalism,
we proved that an FI supercurrent
supermultiplet $\Xi^{(C)}_{\alpha\alphadot}$ with conserved $R_5$-symmetry does not exist.
Indeed, as we discussed in Sect.~4.5, there is only one 
exception to this result:  this occurs if
the $U(1)_{\rm FI}$ gauge boson has a supersymmetric
mass $m$, corresponding to the introduction of a mass
term $\int d^4\theta m^2 V^2$ into the Lagrangian.
In this section, we shall explain how the introduction
of a mass term also allows us to evade the linear-formalism proof 
in Sect.~5.2.
We shall also explicitly construct the $\xi$-dependent
supercurrent supermultiplet $\Xi^{(L)}_{\alpha\alphadot}$ that results. 

Recall that within the context of the linear formalism, 
we showed that our ansatz for $\chi_{\mu\alpha}$
in Eq.~(\ref{bigansatz})
can satisfy the constraint equation in Eq.~(\ref{mastereq}) only
if $X=Y=Z=0$.
Indeed, this is the result that emerged when we
applied the generic equations of motion
corresponding to a massless $U(1)_{\rm FI}$ gauge boson.
However, in evaluating quantities such as $\delta_\epsilon \delta_\eta \chi_{\mu\alpha}$ in the
presence of a mass term, we may now make use of the equations of motion in Eq.~(\ref{neweqs}).
It is easy to see that this can have a profound effect on our conclusions.
For example, we previously found that the
$\eta$-dependent terms in Eq.~(\ref{etaterm}) do not vanish, as they must for consistency,
unless $Y=Z=0$.  However, we now see that when the equations of motion in Eq.~(\ref{neweqs})
are applied, the final parenthesized factor in
Eq.~(\ref{etaterm}) vanishes all by itself.
We are therefore no longer forced to conclude that $Y=Z=0$.

Thus, we shall now quickly repeat
our analysis of the possible solutions to the constraint equation
in Eq.~(\ref{mastereq}),
bearing in mind the new equations of motion in Eq.~(\ref{neweqs}).
First, we observe that as a result of the equations of motion, we have
\begin{equation}
  (\sigma^{\mu\nu} \partial_\nu \chi)_\alpha~=~ 
        {i\over 2}  (\sigma^\mu\lambdabar)_\alpha +\half \partial^\mu\chi_\alpha  ~.
\end{equation}
Thus, even our ansatz in Eq.~(\ref{bigansatz}) simplifies from three possible terms down to two.
We shall therefore take our new ansatz to be
\begin{equation}
          \chi^\mu_\alpha ~=~  X\,  (\sigma^\mu \lambdabar)_\alpha + Y\, \partial^\mu \chi_\alpha~
\label{bigansatz2}
\end{equation}
where $X$ and $Y$ are unknown (generally complex) coefficients.
        
We now must evaluate $\delta_\epsilon\delta_\eta \chi^\mu_\alpha$, and compare this with the
right side of Eq.~(\ref{mastereq}).
We have already seen above that the $\eta$-dependent terms cancel as a result of the new
equations of motion, as they must, so this does not provide any constraint on $X$ or $Y$.
Likewise, we may easily verify that the $\epsilon$-dependent terms also match without providing 
any constraint on $X$ or $Y$.  However, rotating our overall phase for $\chi_{\mu\alpha}$
so that $X$ is real (without
loss of generality), we find that the  $\epsilonbar$-dependent terms do not 
match unless 
\begin{equation}
          {\rm Re}\, X ~=~- {\rm Im}\,Y~,
\end{equation}
with ${\rm Re}\, Y$ unconstrained.
Likewise, demanding that $\partial_\mu \chi^\mu_\alpha =0$ (as appropriate for a supercurrent
supermultiplet in the linear formalism), we find upon use of the equations of motion
that ${\rm Re}\, Y=0$.
Taking $X=\xi$, we therefore 
find that our solution for $\chi^\mu_\alpha$ is given by
\begin{equation}
    \chi_{\mu\alpha} ~=~ \xi (\sigma_\mu \lambdabar)_\alpha - i\xi \partial_\mu \chi_\alpha
                     ~=~ -2i\xi \sigma_{\mu\nu} \partial^\nu \chi_\alpha~,
\label{res1}
\end{equation}
whereupon it is straightforward to verify through supersymmetry transformations
that the remaining two components are given by
\begin{equation}  
      C_\mu = \xi A_\mu~~~~~~~ {\rm and}~~~~~~~~
    \hat T_{\nu\mu}  = \xi (g_{\mu\nu} \Box - \partial_\mu\partial_\nu) C + i \xi \tilde F_{\nu\mu}~.
\label{res2}
\end{equation}
Indeed, with the identifications in Eqs.~(\ref{res1}) and (\ref{res2}), it is straightforward
to check that the algebra in Eq.~(\ref{reducedalgebra}) 
closes through the equations of motion in Eq.~(\ref{neweqs}).

We conclude, then, that in the presence of a non-zero mass
for the $U(1)_{\rm FI}$ gauge boson,
an FI supercurrent supermultiplet $\Xi^{(L)}_{\alpha\alphadot}$ can indeed exist, 
and has components
\beqn
               C_\mu &=& \xi A_\mu\nonumber\\
    \chi_{\mu\alpha} &=& \xi (\sigma_\mu \lambdabar)_\alpha - i\xi \partial_\mu \chi_\alpha
                     ~=~ -2i\xi \sigma_{\mu\nu} \partial^\nu \chi_\alpha\nonumber\\
    \hat T_{\nu\mu}  &=& \xi (g_{\mu\nu} \Box - \partial_\mu\partial_\nu) C + i \xi \tilde F_{\nu\mu}~.
\label{FIsucurrentcomps}
\eeqn
Remarkably, these are nothing but the components of the superfield
\beq
    \Xi_\mu = -\half \sigmabar_\mu^{\alphadot\alpha} \Xi_{\alpha\alphadot}
     ~~~~~{\rm where}~~~~~
     \Xi_{\alpha\alphadot}= -{\xi\over 2}\, [D_\alpha, \Dbar_\alphadot] V~,
\eeq
which has the same structural form as Eq.~(\ref{Xisolnchiral}). 
The key difference here, of course, is the fact that this is valid only in
the presence of a non-zero mass for the $U(1)_{\rm FI}$ gauge boson.

\section{Connecting the chiral and linear formalisms}
\setcounter{footnote}{0}

In Sects.~4 and 5 respectively, we examined the properties of both 
the action and the supercurrent superfield for theories with FI terms,
first in the chiral formalism, then in the linear formalism.  We 
demonstrated that in the former construction, the additional FI contribution
$\Xi^{(C)}_{\alpha\alphadot}$ 
to the supercurrent superfield $J_{\alpha\alphadot}^{(C)}$ 
is given by Eq.~(\ref{Xisolnchiral}), while
in the latter, $\Xi^{(L)}_{\alpha\alphadot}=0$.  This result 
is of critical importance, for it illustrates that the 
gauge non-invariance of the supercurrent supermultiplet emphasized 
in Ref.~\cite{Seiberg} is a by-product of the particular formalism that
was used.  However, 
as we have seen, 
the breaking of $U(1)_{\rm FI}$ gauge invariance in the chiral formalism
leads to a rich local and global symmetry structure for theories with non-zero FI terms,
including   
an inconsistency when attempting to couple such theories to supergravity.
It is therefore important to understand
the local and global symmetry structure that emerges in cases 
in which the linear formalism can also be employed --- \ie,  
in theories with an unbroken $R_5$-symmetry. 
This is particularly relevant in the case of theories with FI terms,
since FI terms in and of themselves preserve $R_5$-symmetry.

We shall begin by outlining a duality
relation, first developed in Refs.~\cite{RelateChiralLinearSUGRA,VanProeyenPolandTalk}, 
between the chiral- and linear-compensator formalisms. 
We shall then use this duality relation to derive a general connection between 
the two supercurrent supermultiplets 
$J_{\alpha\alphadot}^{(C)}$ and $J_{\alpha\alphadot}^{(L)}$ 
(or between their FI contributions $\Xi_{\alpha\alphadot}^{(C)}$ and $\Xi_{\alpha\alphadot}^{(L)}$) 
that respectively emerge in the two formalisms.
This analysis will result in a general condition that a theory
must satisfy in order for these two supercurrent supermultiplets to differ.
As we will see, most theories with canonical renormalizable K\"ahler potentials
will not satisfy this condition, but theories with non-zero FI terms do.
Finally, we shall then turn our attention to the local and global symmetry structure
of theories with non-zero FI terms, this time using the linear formalism.
As we shall see, many of their properties mirror those of their chiral-formalism counterparts, 
yet there are some crucial differences.


\subsection{Duality relation between chiral- and linear-compensator formalisms}

It is clear from the results of Sects.~4 and 5 that the chiral- and 
linear-formalism descriptions of the same uncompensated theory are, in general,
distinct theories with distinct supercurrent supermultiplets 
$J^{(C)}_{\alpha\alphadot}$ and $J^{(L)}_{\alpha\alphadot}$.  However, it can
also be shown that these descriptions are related by a duality 
transformation~\cite{RelateChiralLinearSUGRA} in cases in which the original,
uncompensated theory is $R_5$-invariant.
Indeed, this is a general result, valid for any $R_5$-invariant theory.

Our method of demonstrating this duality will be quite simple:
we shall introduce a new ``intermediate'' Lagrangian which is in neither
the chiral nor linear formalisms, and then demonstrate
that from this single intermediate Lagrangian we may obtain
either Eq.~(\ref{eq:CCompLag}) or Eq.~(\ref{eq:LCompLag}) 
by substituting in for the appropriate fields (or sets of fields)
using the equations of motion.
Whether we obtain
Eq.~(\ref{eq:CCompLag}) or Eq.~(\ref{eq:LCompLag}) 
depends on which
fields are placed on shell in our intermediate Lagrangian.
This is indeed a standard technique for demonstrating a duality between
two different theories.

We shall begin with a discussion of the K\"ahler contribution
to our ``intermediate'' Lagrangian, which takes the form
\beq
  \mathcal{L}_D  ~=~ \int d^4\theta 
   \left[U\ln\left(\frac{Ue^{\widetilde{K}_U(\Phi_i,
       \Phi_i^\dagger)/3M_P^2}}{\Sigma_L  \Sigmabar_L}\right) 
     - iU(\Omega-\overline{\Omega}) \right]~.
\label{eq:UrAction}
\eeq
Here $\Sigma_L$ and ${\Sigmabar}_L$ are the same compensator 
fields introduced in the linear formalism, carrying non-trivial $U(1)_L$ charges;
$U$ is a real vector superfield, neutral under all $U(1)$ symmetries; 
$\Omega,\Omegabar$ are a pair of left- and right-chiral 
hermitian-conjugate superfields;
and 
\beq
  \widetilde{K}_U ~\equiv~ -3M_P^2 + \sum_n \left(\frac{U}{3M_P^2}\right)^{-n/2} K_n 
\eeq
is the analogue of the rescaled K\"{a}hler potential $\widetilde K_L$ defined in 
Eq.~(\ref{eq:KtildeL}), but with $U$ in place of $L$ and with an additional 
constant term $-3M_P^2$.  
Here, as in Eq.~(\ref{eq:KahlerWeylWeight}), the subscript on $K_n$
corresponds to its Weyl weight.  

Just as in the linear formalism, we would like the
Lagrangian in Eq.~(\ref{eq:UrAction}) to exhibit
a full $U(1)_{\rm FI} \times U(1)_L$ invariance.
Unfortunately, 
under $U(1)_{\rm FI} \times U(1)_L$ transformations,
we find that the first term in Eq.~(\ref{eq:UrAction}) leads
to the variations
\beq
  \delta_{\mathrm{FI}}\mathcal{L}_D  ~\ni~
     \frac{2i\xi}{3M_P^2}\int d^2 \theta \, U(\Lambda-\overline{\Lambda})
   ~,~~~~~~
 \delta_{L}\mathcal{L}_D  ~\ni~
     i\int d^2\theta \, U(\Lambda_{L}-\overline{\Lambda}_{L})~.
\label{eq:UrActionVariation}
\eeq
These variations are no longer total derivatives, as they were in the linear formalism,
because $U$ is a vector multiplet rather than a linear multiplet.
Consequently, in order to maintain $U(1)_{\mathrm{FI}}\times U(1)_L$
invariance, we require that $\Omega$ and $\overline{\Omega}$ in
the second term of Eq.~(\ref{eq:UrAction}) 
transform linearly under both of these symmetries so as to cancel Eq.~(\ref{eq:UrActionVariation}):
\beq
  \Omega\rightarrow \Omega+\left( {2\xi\over 3 M_P^2}\, \Lambda +\Lambda_L\right) ~,~~~~~~ 
  \overline{\Omega}\rightarrow \overline{\Omega}+\left( {2\xi\over 3 M_P^2}\, \overline{\Lambda} 
           +\overline{\Lambda}_L\right) ~.
\label{eq:HowOmegaShifts}
\eeq
With this transformation, the Lagrangian in Eq.~(\ref{eq:UrAction}) 
is completely $U(1)_{L}\times U(1)_{\mathrm{FI}}$ invariant.

Our aim will be to demonstrate that we may obtain either
Eq.~(\ref{eq:CCompLag}) or Eq.~(\ref{eq:LCompLag}) from this Lagrangian by
substituting in for $U$ in the former case,
and for $\Omega$ and $\overline{\Omega}$ in the latter case,
using the equations of motion.  

Let us focus first on reproducing the linear case~(\ref{eq:LCompLag}).
In superfield language, the equations of motion for the $\Omega$ and $\overline{\Omega}$
superfields are
\beq
  \frac{\partial \mathcal{L}_D}{\partial \Omega} ~=~ 
  \frac{\partial \mathcal{L}_D}{\partial \overline{\Omega}} ~=~ 0.
\eeq
Given the Lagrangian in Eq.~(\ref{eq:UrAction}), these equations 
require not only that $U$ be real, but also that it
have a $\theta^2\thetabar^2$ component which is a total
derivative.
There is only one way in which this can happen:
$U$ must take the form of a linear superfield, so that we may write
\beq
                U~=~L~.
\eeq
Substituting this into Eq.~(\ref{eq:UrAction}) then yields
\beqn
  \mathcal{L}_D &=&  \int d^4\theta 
     \left[ L \ln\left(\frac{L e^{\widetilde{K}_L(\Phi_,\Phi_i^\dagger)/3 M_P^2}}
               {\Sigma_L\Sigmabar_L}\right) - L\right]~.
\label{eq:FinalFormLDAct}
\eeqn
Since $\int d^4\theta L  $ is a total divergence (as can be verified from the
final column of Table~\ref{constraintstable}),
we see that Eq.~(\ref{eq:FinalFormLDAct}) implies the same physics as Eq.~(\ref{eq:LCompLag}).   
In other words, we have successfully reproduced the linear-compensator formalism.
Indeed, from this perspective, 
we see that $\Omega$ and $\overline{\Omega}$ are nothing but superfield Lagrange multipliers 
which enforce the linearity constraint on the superfield $L$.

We shall now demonstrate that Eq.~(\ref{eq:UrAction}) 
also leads to the chiral-compensator formalism.  
To do this, 
we shall return to Eq.~(\ref{eq:UrAction}) but now consider
the equation of motion for $U$:
\beq
    \sum_n\left(1-\frac{n}{2}\right)\left(\frac{U}{3M_P^2}\right)^{-n/2}
     \frac{K_n}{3M_P^2} + 
     \ln\left(\frac{U}{\Sigma_L{\Sigmabar_L}}\right)
            ~=~i(\Omega-\overline{\Omega})~. 
\label{eq:UgenEOM}
\eeq
For arbitrary K\"ahler potentials, it is impossible to solve this equation for $U$ exactly.
However, we may make a simplifying (and ultimately temporary) assumption that  
our K\"{a}hler potential takes the form
\beq
  K(\Phi_i,\Phi,V) ~=~ K_0(\Phi_i,\Phi^\dagger_i,V) + K_2(\Phi_i,\Phi^\dagger_i,V)~. 
\label{eq:KahlerPotSpecialCase}  
\eeq
Note that this is not a particularly restrictive assumption,
as many theories of theoretical and phenomenological interest frequently take this 
simplified form.  For example, any theory comprising a set of matter fields $\Phi_i$ with canonical
kinetic terms which are charged under a $U(1)$ gauge group with a non-zero FI term 
will have $K_2=\Phi_i e^{-Q_i V} \Phi_i^\dagger$ and $K_0=2\xi V$.
Later, we shall generalize our duality argument to any K\"ahler potential of arbitrary form.

With the assumption in 
Eq.~(\ref{eq:KahlerPotSpecialCase}),  
we find that Eq.~(\ref{eq:UgenEOM})
reduces to the form
\beq
     \frac{K_0(\Phi_{i},\Phi_{i}^\dagger)}{3M_P^2} + 
     \ln\left(\frac{U}{\Sigma_L{\Sigmabar}_L}\right)
   ~=~ i(\Omega-\overline{\Omega})~.
\label{eq:UEOM}
\eeq
Solving this for $U$ then yields 
\beq
    U ~=~ \Sigma_L\Sigmabar_L \exp\left[ i(\Omega-\overline{\Omega})\right]
             \exp\left[ -{ K_0(\Phi_i,\overline{\Phi}_i)\over 3M_P^2}\right]~. 
\label{specialUsoln}
\eeq
Defining
\beq
   \Sigma ~\equiv~ \Sigma_L e^{i\Omega} ~,~~~~~~ 
   \Sigmabar ~\equiv~ \Sigmabar_L e^{-i\overline\Omega} ~,~
\label{eq:SShifts}
\eeq
we now see that $\Sigma$ and ${\Sigmabar}$ appearing in these 
relations can be identified with the compensators 
$\Sigma$ and ${\Sigmabar}$ of the chiral formalism.  
In particular, as a result of Eq.~(\ref{eq:HowOmegaShifts}),
we see that these $\Sigma, \Sigmabar$ fields carry exactly the same non-trivial 
$U(1)_{\rm FI}$ charges as they do in the chiral formalism, even though the previous 
$\Sigma_L,\Sigmabar_L$ fields were $U(1)_{\rm FI}$-neutral.   
On the other hand,
we see that the 
the new fields $\Sigma,\Sigmabar$ are $U(1)_L$-neutral, while
the previous 
$\Sigma_L,\Sigmabar_L$ fields were $U(1)_L$-charged.   
Thus, we see that the passage from $\Sigma_L,\Sigmabar_L$ to $\Sigma,\Sigmabar$ 
essentially trades $U(1)_L$ charges for
$U(1)_{\rm FI}$ charges.

Substituting Eq.~(\ref{specialUsoln}) into Eq.~(\ref{eq:UrAction}),
we then obtain
\beqn
   \calL_D &=&  \int d^4\theta\, (-U + K_2)\nonumber\\
           &=&  \int d^4\theta\, \left[
         -\Sigma\Sigmabar \exp\left( -{K_0\over 3M_P^2}\right)  + K_2 \right]\nonumber\\
           &=&  \int d^4\theta\, \left[
               -\Sigma \Sigmabar \left( 
                1- {K_0\over 3 M_P^2} - {3 M_P^2 \over \Sigma \Sigmabar} {K_2\over 3 M_P^2} \right)+ 
                {\cal O}\left( {K_n^2 \over 9 M_P^4}\right) \right]\nonumber\\
           &=&  \int d^4\theta\, \left\lbrack
               -\Sigma\Sigmabar \left( 1- {\widetilde K\over 3 M_P^2}\right) 
                + {\cal O}\left( {K_n^2 \over 9 M_P^4}\right) \right]\nonumber\\
           &=&  \int d^4\theta\, \left\lbrack
               -\Sigma\Sigmabar \exp \left( - {\widetilde K\over 3 M_P^2}\right) 
                + {\cal O}\left( {K_n^2 \over 9 M_P^4}\right) \right\rbrack~
\eeqn
where we have 
assumed $K_n\ll 3 M_P^2$ 
and identified $\widetilde K\equiv K_0 + 3 M_P^2 K_2/(\Sigma\Sigmabar)$
in passing to the fourth line, where $\widetilde K$ is defined in Eq.~(\ref{eq:KahlerWeylWeight}).
Thus, up to higher-order terms which vanish in the $M_P\rightarrow\infty$ limit,
we successfully recover the chiral-formalism expression 
in the first line of Eq.~(\ref{eq:CCompLag}).

This result holds for K\"ahler potentials of the form in Eq.~(\ref{eq:KahlerPotSpecialCase}).  
However, it actually holds more generally.
Recall that for general K\"ahler potentials,
the equation of motion for $U$ is given in Eq.~(\ref{eq:UgenEOM}), or equivalently
\beq
    U ~ = ~ \Sigma\Sigmabar\,
    \exp\left[-\sum_n\left(1-\frac{n}{2}\right)
     \left(\frac{U}{3M_P^2}\right)^{-n/2}
     \frac{K_n}{3M_P^2}\right]~. 
\label{Usolln}
\eeq
In general, we cannot solve for $U$ in this relation explicitly, as was
possible for the simplified case above.  
However,
we can insert this solution into Eq.~(\ref{eq:UrAction}) to obtain
\beqn
  \calL_D &=& 
   \int d^4\theta\left\lbrack
      U  \ln\left({U\over \Sigma\Sigmabar}\right) - U + 
            \sum_n \left( {U\over 3M_P^2}\right)^{1-n/2}K_n\right\rbrack\nonumber\\
   &=& 
   \int d^4\theta\left\lbrack
        -\sum_n U \left( 1-{n\over 2}\right) 
      \left({U\over 3M_P^2}\right)^{-n/2}  {K_n\over 3M_P^2}   
       - U + 
            \sum_n \left( {U\over 3M_P^2}\right)^{1-n/2}K_n\right\rbrack \nonumber\\
   &=& 
   \int d^4\theta\left\lbrack
   \sum_n \left({n\over 2}\right) \left({U\over 3 M_P^2}\right)^{1-n/2} K_n ~-~ U\right\rbrack~.
\label{stepo}
\eeqn
Thus far, we have made no approximations.
However, once again 
taking $K_n\ll 3M_P^2$
and using Eq.~(\ref{Usolln}) to iteratively substitute $U\sim \Sigma\Sigmabar$ at 
higher orders,
we find that Eq.~(\ref{stepo}) yields
\beqn
    \calL_D &=&  
      \int d^4\theta\Biggl\lbrack
   \sum_n \left({n\over 2}\right) \left( {\Sigma\Sigmabar\over 3 M_P^2}\right)^{1-n/2} K_n
         ~-~\Sigma\Sigmabar \nonumber\\
     && ~~~~~~~~~~~~~~~~~~~~
       +\sum_n \left(1-{n\over 2}\right) \left( {\Sigma\Sigmabar\over 3 M_P^2}\right)^{1-n/2} K_n
          + {\cal O}\left( {K_n^2\over 9 M_P^4}\right) \Biggr\rbrack \nonumber\\
    &=&  
      \int d^4\theta\left\lbrack
        \sum_n \left( {\Sigma\Sigmabar\over 3 M_P^2}\right)^{1-n/2} K_n - \Sigma\Sigmabar
         + {\cal O}\left( {K_n^2\over 9 M_P^4}\right)
          \right\rbrack
        \nonumber\\
    &=&  
      \int d^4\theta\left\lbrack
        -\Sigma\Sigmabar \exp\left( - {\widetilde K\over 3M_P^2}\right)  
           + {\cal O}\left( {K_n^2\over 9 M_P^4}\right) \right\rbrack~
\eeqn
where we have identified $\widetilde K = \sum_n (\Sigma \Sigmabar/3 M_P^2)^{-n/2} K_n$.
Once again, we see that 
this is identical to the chiral-formalism 
Lagrangian in the $M_P\rightarrow\infty$ limit. 

Our discussion thus far has focused on the contributions to the action
coming from the K\"ahler potential, where we have shown that
the chiral- and linear-compensator formalisms are essentially related
through a duality.  Indeed, as we have seen,
this ``duality'' is nothing but a superfield-level 
Legendre transformation~\cite{RelateChiralLinearSUGRA,VanProeyenPolandTalk}.
However, it remains to verify that we can likewise relate the 
$F$-term contributions coming from a superpotential $W(\Phi_i)$
between the two formalisms.
Fortunately, this is quite simple to do, and we shall find that
the two contributions are related for all 
cases in which 
the superpotential of 
the original, uncompensated theory has an $R$-symmetry~\cite{RelateChiralLinearSUGRA}.

In order to understand this, let us consider a theory written in the chiral formalism, 
with an arbitrary superpotential $\widehat{W}\sim \Sigma^3 \widetilde W $.
To obtain the corresponding expression in the linear formalism, we may rewrite this 
expression in terms of $\Sigma_L$ by using Eq.~(\ref{eq:SShifts}):
\beq
     \widehat W  ~=~ \Sigma^3 \, \widetilde W ~=~ 
            e^{-3i\Omega}  \, \Sigma_L^3 \, \widetilde W~.
\label{eq:ItsAPhase}
\eeq
In order to cancel the $\Omega$-dependent phase in this expression, we must
be able to absorb this phase through the redefinitions of the chiral matter fields $\Phi_i$
given in Eq.~(\ref{newfields}).  
However, this will only happen
if the sum of the $R$-charges of the
fields $\Phi_i$ appearing in each term in the superpotential $W(\Phi_i)$ of
the linear-formalism theory is equal to 2.
This requirement is, of course, nothing but the statement that the 
superpotential of the original, uncompensated theory must be 
$R$-invariant.  Since this is also a necessary condition 
for the linear-compensator formalism to be valid,  
this result proves that any theory which
may be described using the linear formalism 
also has a dual description in the chiral formalism.
Of course, the converse is not true in general~\cite{RelateChiralLinearSUGRA}.

Likewise, in order to establish the relationship 
between $U(1)_{\rm FI}$ gauge transformations in the
two dual pictures, we recall from Eq.~(\ref{eq:HowOmegaShifts})
that the Lagrange-multiplier
superfields $\Omega$ and $\overline{\Omega}$ are required to transform non-trivially
under $U(1)_{\mathrm{FI}}$ gauge transformations in order to keep
Eq.~(\ref{eq:UrAction}) gauge invariant.  As a consequence, the field
redefinitions in Eq.~(\ref{eq:SShifts}) result in $\Sigma$ and ${\Sigmabar}$
acquiring $U(1)_{\mathrm{FI}}$ charges equal to $\pm 2\xi / 3M_P^2$, just as they
do in the chiral formalism. 

Given this discussion, it is straightforward to see how all of this 
relates to theories containing non-zero FI terms.
We have already shown in Sect.~4, through an analysis involving the chiral formalism, 
that any theory involving an
FI term must possess a global $R$-symmetry. 
Moreover, in order to admit a description in the linear formalism,
it is necessary that a theory possess an unbroken $R_5$-symmetry.
Consequently, any theory that can be described in the linear formalism
satisfies the criteria for the duality to hold.
Thus,  we conclude that any 
globally-supersymmetric theory which satisfies the consistency conditions 
for having FI term can be conformally compensated using the linear formalism as well.


\subsection{Duality relation between supercurrent supermultiplets}

We have seen that any theory which admits a description in the
linear formalism (and which therefore has an unbroken $R_5$-invariance)
has a  dual description in the chiral 
formalism.  Given this, we now discuss the general relationship between 
the supercurrent superfields $J^{(L)}_{\alpha\alphadot}$ and 
$J^{(C)}_{\alpha\alphadot}$ in these two dual theories. 
We also derive a
relationship between the quantities 
$L_\alpha$ 
and
$D_\alpha S$ 
in their respective conservation laws.  
In this way, we shall essentially be demonstrating that
the supercurrent-supermultiplet conservation 
law~(\ref{linearcase}) for a theory with a linear multiplet of anomalies
can be ``traded'' for a conservation law~(\ref{chiralcase}) involving
a chiral multiplet of anomalies through a modification of the 
supercurrent superfield.
While this result is already known~\cite{ClarkAndLove,SupercurrentAnomaliesRenorm},   
our demonstration of this 
relationship will be derived directly from
a strict Noether calculation, using the duality between the chiral and 
linear formalisms outlined in Sect.~6.1.
We shall also discuss the direct implications of these results for theories
with non-vanishing FI terms, and show that theories with non-vanishing FI terms
are essentially unique amongst theories with renormalizable K\"ahler potentials
in having supercurrent  superfields
$J^{(L)}_{\alpha\alphadot}$ and 
$J^{(C)}_{\alpha\alphadot}$ which actually differ.

Rather than perform direct superfield calculations, our procedure will
be to focus on deriving contributions to the $R_5$-current $j_\mu^{(5)}$  
in both formalisms.  As discussed in Sect.~4.2, 
this current may be calculated through the Noether procedure, even in the chiral formalism,
because we are working within the framework of the compensated theories in which
superconformal invariance is preserved.  We then ``freeze'' our compensator fields
in order to derive our final expressions for $j_\mu^{(5)}$ in each theory.
As a final step, we then recognize these results as the bottom components of 
supercurrent superfields, and thereby promote these results to full superfield expressions.

First, we recall that within the frameworks of our two different formalisms,
there are only certain terms within the K\"ahler portions of the corresponding Lagrangians
which contribute to non-zero $R_5$-currents after the relevant compensator fields are frozen. 
These Lagrangian terms are
\beqn 
            {\rm chiral:}&&~~~~~ \calL ~=~ \int d^4 \theta\, \left[ {\Sigma\Sigmabar\over 3 M_P^2}
              \, K(\Phitilde_i,{\widetilde \Phi}_i^\dagger,V) \right] \nonumber\\
            {\rm linear:}&&~~~~~ \calL ~=~ \int d^4 \theta\, \left[ {L \over 3 M_P^2}
              \, \widetilde K_L(\Phi_i,\Phi_i^\dagger,V) \right]~
\eeqn
where the superfields $\Phitilde_i$ in the chiral formalism are defined in Eq.~(\ref{newfields}) 
and where $\widetilde K_L$ in the linear formalism is defined in Eq.~(\ref{eq:KtildeL}). 
Note that only the chiral compensator fields $\Sigma,\Sigmabar$ and the original superfields $\Phi_i,
\Phi_i^\dagger$ carry non-vanishing $R_5$-charges;
by contrast, the linear compensator field $L$ and the rescaled superfields $\Phitilde_i,\Phitilde^\dagger_i$ do not carry $R_5$-charge. 
This implies that there are two independent reasons why the corresponding $R_5$-currents
$j_\mu^{(5)}$ of these two theories might differ:
\begin{itemize}
\item  There will be  
      contributions from the $R_5$-variations of the compensator fields $\Sigma,\Sigmabar$ in the
      chiral formalism which are not present in the linear formalism.  
     In complete analogy with Eq.~(\ref{jmu52}), which was derived for $K=2\xi V$, we find that these
       contributions take the form
\beq
  j_\mu^{(5)} ~=~ -\frac{2}{3} K_\mu(\widetilde{\Phi}_i,\widetilde{\Phi}_i^\dagger,V)
\label{analogg}
\eeq
       where $K_\mu$ is the coefficient of $-\theta\sigma^\mu \thetabar$ within the vector superfield  
       $K(\Phitilde_i,{\widetilde \Phi}_i^\dagger,V)$.
Promoting this contribution to a superfield expression, we have
\begin{equation}
    J_{\alpha\dot{\alpha}} ~=~ \frac{1}{3}
    \left[D_{\alpha},{\Dbar}_{\dot{\alpha}}\right] 
    K(\widetilde{\Phi}_i,\widetilde{\Phi}_i^\dagger)~.
\label{eq:SigSigAlteredJ}
\end{equation}
\item  There will also be differences in the 
          variations of the matter fields $\Phi_i$ in the linear formalism
        versus $\Phitilde_i$ in the chiral formalism. 
         Specifically, the difference in the overall $R_5$-charge of these superfields 
        changes the $R_5$-charges of their individual field components, and thereby alters
          the way in which these individual field components contribute to $j^{(5)}_\mu$.
\end{itemize}

As an example of how these combined modifications affect the supercurrent
superfield, let us begin by considering a simple case:
a Wess-Zumino model in which the 
K\"{a}hler potential takes the minimal form $K = \Phi_i^\dagger e^V \Phi_i$ 
and all of the matter fields are assigned $R_5$-charge $R^{(5)}_i=2/3$.  
Given this K\"ahler potential, we find that Eq.~(\ref{analogg}) takes the form
\begin{equation}
    \frac{2i}{3}
         (\widetilde \phi^\ast_i\partial_\mu\widetilde \phi_i -
     \widetilde \phi_i\partial_\mu\widetilde \phi^\ast_i )
    -\frac{2}{3} (\widetilde \psi_i\sigma_\mu{\overline{\widetilde \psi}}_i)~
\end{equation}
where $\widetilde \phi$ and $\widetilde\psi$ 
are respectively the lowest and next-lowest components of $\Phitilde$.
By contrast, the results of varying the appropriate matter fields in the two
formalisms take the general form
\begin{equation}
     i {r_\phi}  
    (\phi^\ast_i\partial_\mu\phi_i
    -\phi_i\partial_\mu\phi^\ast_i) 
    -  r_\psi (\psi_i\sigma_\mu{\overline{\psi}}_i)~
\label{ccon}
\end{equation}
where $r_\phi$ and $r_\psi$ are the $R_5$-charges 
of the appropriate fields $\phi$ and $\psi$ respectively, with
$(r_{\widetilde \phi},r_{\widetilde\psi})=(0,-1)$ for the chiral formalism and
$(r_\phi,r_\psi)=(2/3,-1/3)$ for the linear formalism.
We therefore find that the total contributions to the $R_5$-current $j_\mu^{(5)}$ are identical 
in each formalism, differing only in whether the relevant fields 
are the original fields or the rescaled fields:
\beqn
  {\rm chiral:} &&~~~~~~~~~  j_\mu^{(5)}~=~ 
    \frac{2i}{3}
         (\widetilde \phi^\ast_i\partial_\mu\widetilde \phi_i -
     \widetilde \phi_i\partial_\mu\widetilde \phi^\ast_i )
    +\frac{1}{3} (\widetilde \psi_i\sigma_\mu{\overline{\widetilde \psi}}_i)~\nonumber\\
  {\rm linear:} &&~~~~~~~~~  j_\mu^{(5)}~=~ 
    \frac{2i}{3}
         (\phi^\ast_i\partial_\mu\phi_i - \phi_i\partial_\mu\phi^\ast_i) 
    +\frac{1}{3} (\psi_i\sigma_\mu{\overline{\psi}}_i)~.
\eeqn
We thus have
\beqn
  J_{\alpha\dot{\alpha}}^{(C)} &=&    
     -D_{\alpha}\widetilde{\Phi}_i {\Dbar}_{\dot{\alpha}}\widetilde{\Phi}_i^\dagger
     +\frac{1}{3}[D_{\alpha},{\Dbar}_{\dot{\alpha}}]
     (\widetilde{\Phi}_i\widetilde{\Phi}_i^\dagger)\nonumber\\
  J_{\alpha\dot{\alpha}}^{(L)} &=&    
     -D_{\alpha}{\Phi}_i {\Dbar}_{\dot{\alpha}}{\Phi}_i^\dagger
     +\frac{1}{3}[D_{\alpha},{\Dbar}_{\dot{\alpha}}]
     ({\Phi}_i{\Phi}_i^\dagger)~,
\label{eq:JaaWithMinimalKahler}
\eeqn
and we see that the functional forms of these two supercurrents in this special
case are identical.
Finally, after we freeze our compensator fields to their fixed values, we see that
$\widetilde \Phi_i\to \Phi$.  Thus, after freezing, we see that our two supercurrent
superfields become truly identical.

Given these results, let us now proceed to consider the general case
of an arbitrary K\"ahler potential $K(\Phi,\Phi^\dagger,V)$
built from matter fields $\Phi_i$ with $R_5$-charges $R_i^{(5)}$
and gauge fields $V$.
We shall assume, of course, that this theory preserves $R_5$-invariance, so
that a description in either the linear or chiral formalism is possible;
moreover, we observe that all fields $\Phi_i$ must have a common $R_5$-charge $R_\Phi^{(5)}$ 
if we are ultimately to obtain a symmetric energy-momentum 
tensor~\cite{SupercurrentAnomaliesRenorm}.
In the linear formalism, the supercurrent superfield $J_{\alpha\alphadot}^{(L)}$ associated 
with such a theory can then be obtained via a straightforward 
Noether calculation of $j_\mu^{(5)}$, yielding
the result
\begin{equation}
  J_{\alpha\alphadot}^{(L)} ~=~ 
    - g_{i\ibar} D_\alpha \Phi_i {\Dbar}_\alphadot \Phi_\ibar^\dagger
    +\frac{R_{\Phi}^{(5)}}{2}[D_\alpha,\Dbar_\alphadot] \left(
     \Phi_i K_i\right)~.
\label{eq:JaaLGenForm}
\end{equation}
Here $K_i\equiv \partial K/\partial \Phi_i$  and the K\"ahler metric $g_{i\ibar}$
is defined in  Eq.~(\ref{eq:KahlerMetric}).      
Note that in Eq.~(\ref{eq:JaaLGenForm}), we have not explicitly indicated the
contributions $2 W_\alpha e^V \Wbar_\alphadot$ that correspond to each of the
gauge-kinetic terms;  these terms are the same in each formalism, and will not
be discussed further.
As a special case,
we observe 
that if our theory has a minimal K\"ahler potential of the form $K=\Phi_i^\dagger e^{V} \Phi_i$,
then $\Phi_i K_i = K$.
With $R_\Phi^{(5)}=2/3$, 
the expression in Eq.~(\ref{eq:JaaLGenForm}) then reduces to that given in
Eq.~(\ref{eq:JaaWithMinimalKahler}).  

Eq.~(\ref{eq:JaaLGenForm}) gives the result for $J_{\alpha\alphadot}^{(L)}$ in the
linear formalism.
However, as discussed above, the corresponding supermultiplet $J_{\alpha\alphadot}^{(C)}$ 
in the chiral formalism differs from $J_{\alpha\alphadot}^{(L)}$ in two ways:
through a shift in the charges of the redefined fields $\widetilde{\Phi}_i$
relative to those of the original fields $\Phi_i$, and through the addition
of the extra compensator-induced term in Eq.~(\ref{eq:SigSigAlteredJ}).
We therefore have
\beq
  J_{\alpha\alphadot}^{(C)}  ~=~ 
    - g_{i\ibar} D_\alpha \Phitilde_i {\Dbar}_\alphadot \Phitilde_\ibar^\dagger
    +\frac{R_{\Phitilde}^{(5)}}{2}[D_\alpha,\Dbar_\alphadot] \left(
     \Phitilde_i \Ktilde_i\right)
         + {1\over 3} [D_\alpha,\Dbar_\alphadot] \Ktilde~
\label{eq:JaaCGenForm}
\eeq
where $\tilde K\equiv K(\Phitilde,\Phitilde^\dagger,V)$.
Since $R_{\Phitilde}^{(5)}=0$ by construction,
and since $\Phitilde_i\to\Phi_i$ after our compensator fields are frozen,
we therefore find $J_{\alpha\alphadot}^{(C)}$ and 
$J_{\alpha\alphadot}^{(L)}$ are connected through the relation 
\begin{equation}
     J_{\alpha\alphadot}^{(C)} ~=~ J_{\alpha\alphadot}^{(L)} +
        \frac{1}{3}[D_\alpha,\Dbar_\alphadot]
        \left(K - \frac{3R_{\Phi}^{(5)}}{2}\Phi_iK_i\right)~.
\label{eq:JaaCJaaLdif}
\end{equation}

This result enables us to determine the general conditions under which 
$J_{\alpha\alphadot}^{(C)}$ and $J_{\alpha\alphadot}^{(L)}$ can differ:
this can happen only if there are terms in the K\"ahler
potential for which
\beq
     \kappa ~\equiv~    K - {3 R_\Phi^{(5)}\over 2}\, \Phi_i K_i ~\not=~0~.
\label{kappadef}
\eeq
Clearly, this does not occur for K\"ahler potentials of the canonical form
with $R_\Phi^{(5)}=2/3$.
By contrast, this only happens for non-renormalizable terms involving the matter fields,
or situations with
$R_\Phi^{(5)}\not = 2/3$.
 {\it However, we observe that this also occurs --- even for a canonical K\"ahler potential
and even with canonical $R_5$-charges --- in the presence of a non-zero FI term.}\/
Indeed, in such a case, we find that $\kappa=   2\xi V \not=0 $.
This is then consistent with the results in Eqs.~(\ref{Xisolnchiral}) and
(\ref{proofresult}). 

This, then, explains why our results for $\Xi_{\alpha\alphadot}^{(C)}$ in 
Sect.~4.2 and $\Xi_{\alpha\alphadot}^{(L)}$ in Sect.~5.3
are different.  Moreover, this also explains why it was
possible for the supercurrent supermultiplet in the chiral formalism to break $U(1)_{\rm FI}$
gauge invariance.  Clearly, by construction, the linear formalism preserves  
$U(1)_{\rm FI}$ gauge invariance.
Thus, we can have a gauge non-invariant $J_{\alpha\alphadot}^{(C)}$
only when the two supermultiplets are allowed to differ.

Given the result in Eq.~(\ref{eq:JaaCJaaLdif}) relating 
$J_{\alpha\alphadot}^{(C)}$ and
$J_{\alpha\alphadot}^{(L)}$,
it is also possible to derive a relation 
between the anomalies $L_\alpha$ and $D_\alpha S$ to which
$\Dbar^\alphadot J_{\alpha\alphadot}^{(C)}$ and
$\Dbar^\alphadot J_{\alpha\alphadot}^{(L)}$
are respectively equated.
Indeed, writing the linear multiplet in unconstrained form
as $L_\alpha\equiv \Dbar^2 D_\alpha T_L$ where $T_L$ is a vector superfield,
it is even possible to solve for $T_L$.
Starting with the relation
\beq
          J_{\alpha\alphadot}^{(C)}~=~ 
          J_{\alpha\alphadot}^{(L)} + {1\over 3} [D_\alpha,\Dbar_\alphadot] \kappa~
\eeq
where $\kappa$ is defined in Eq.~(\ref{kappadef}),
we then find
\beqn
      \Dbar^\alphadot J_{\alpha\alphadot}^{(C)} &=& 
                \Dbar^\alphadot J_{\alpha\alphadot}^{(L)} + 
           {1\over 3} \Dbar^\alphadot [D_\alpha, \Dbar_\alphadot] \kappa \nonumber\\
    &=&  \Dbar^2 D_\alpha \left(T_L+\kappa\right) - 
          {1\over 3} \Dbar_\alphadot D_\alpha \Dbar^\alphadot \kappa
                      \nonumber\\
    &=&  D_\alpha \Dbar^2 \left(T_L+ {2\over 3} \kappa\right) 
             + 4i (\sigma^\mu \partial_\mu \Dbar)_\alpha
                            \left(T_L+\half \kappa\right)~.
\eeqn
However, we know that 
     $\Dbar^\alphadot J_{\alpha\alphadot}^{(C)}$ should take the form $D_\alpha S$, where 
$S$ is a left-chiral superfield.  We therefore deduce that
we must have $T_L= -\half \kappa$,
whereupon we conclude that
\beqn
         L_\alpha &=& \Dbar^2 D_\alpha T_L ~=~ -\half \Dbar^2 D_\alpha \kappa~\nonumber\\
         D_\alpha S  &=&  D_\alpha \Dbar^2 \left(T_L+{2\over 3} \kappa\right) ~=~ 
                    -{1\over 3} D_\alpha \Dbar^2 T_L ~=~ {1\over 6} D_\alpha \Dbar^2 \kappa~.
\eeqn
 
\medskip


\subsection{The symmetry structure of theories with non-zero FI terms 
in the linear formalism}

In this section, we shall explore the structure of local and global symmetries
that appear in theories with non-zero FI terms in the linear formalism.
This section therefore serves as the linear-formalism counterpart to Sect.~4.3.~ 
We shall begin by describing the general symmetry structure of such theories
and the way in which it
emerges.  We shall then provide an explicit example.

\subsubsection{General symmetry structure}

Thus far in this section, we have demonstrated that a duality exists  
between the chiral and
linear formulations of supergravity.  However, while this implies a  
physical equivalence
between the compensated, superconformal theories in these two  
formulations, it is not yet
clear to what extent this equivalence persists after the compensator  
fields of each
formalism are ``frozen'' in order to break the symmetries of the  
superconformal group
down to those of Poincar\'e supergravity.
Indeed, as we have already discussed in Sects.~4 and~5, we know
that certain differences between the formalisms are inevitable
after the compensators are frozen.
For example, $R_5$-symmetry remains a symmetry in the linear
formalism, but is necessarily broken in the chiral formalism. 
Moreover, some of these differences are unique to theories with
non-zero FI terms.
For example, $U(1)_\FI$ gauge symmetry remains a good symmetry  
within the linear formalism, but is broken in the chiral formalism
if a non-zero FI term is present.
As a result,
the  supercurrent supermultiplet
$J_{\alpha\alphadot}^{(L)}$ in theories with FI terms is 
$U(1)_\FI$ gauge-invariant in the linear formalism,
whereas the  corresponding supermultiplet
$J_{\alpha\alphadot}^{(C)}$ in the chiral formalism is not.

It is therefore important to explore the extent to which
the symmetry structures that survive in the linear formalism
can be matched to those in the chiral formalism, and to determine
whether the issues that  
arise in coupling theories with non-zero FI terms 
to supergravity using the chiral formalism
also arise in the linear formalism.  
In this way, we will be testing the extent to which
the difficulties in coupling theories with non-zero FI terms
to supergravity in the chiral formalism are intrinsic to the FI
terms themselves, or are instead primarily features of the formalism that  
is being used.

With this goal in mind,
we now discuss the symmetry structure of theories that are treated in the 
linear formalism. 
Of course, any
theory which admits a description in the linear formalism must already possess  
a global
$R_5$-symmetry (which will become local once the theory is coupled to  supergravity).
Moreover, by assumption, our original theory will have an FI symmetry which we will
call $U(1)'_\FI$.
While the presence of
the non-zero FI term $2\xi V$ in the K\"ahler potential breaks Weyl invariance  
explicitly,
the superpotential of the theory is Weyl-invariant by assumption.  The  
theory may
possess additional gauge symmetries, of course, but these symmetries  
will not be
affected by the compensator calculus and are therefore not relevant to the
present discussion.

The first step in coupling a theory to supergravity in the linear formalism
is to introduce the compensator fields $L$, $\SigL$, and $\SigbL$, thereby
modifying the $D$-term Lagrangian of the original theory so that it takes
the form given in Eq.~(\ref{eq:LCompLag}).  By design, the compensated  
theory is invariant under both conformal rescalings and $R_5$ rotations in addition
to  special SUSY
transformations, 
implying that it is invariant under full super-Weyl group $U(1)_\SW$.  
Furthermore, unlike the analogous situation in the chiral formalism, 
the superconformal theory in the linear formalism also directly inherits 
the full local $U(1)'_\FI$ gauge symmetry of the
original theory  thanks to the fact 
 --- already discussed below Eq.~(\ref{u1fishift}) ---
that $U(1)'_\FI$ gauge transformations shift the Lagrangian
by total derivatives and thus leave the action invariant.
Indeed, by suitably covariantizing the partial derivatives appearing
in this Lagrangian, it can be shown that
all of these symmetries can be gauged without further modification of  
the action~\cite{West,Lessons}.
Finally, as discussed in Sect.~5.1, the Lagrangian in  
Eq.~(\ref{eq:LCompLag})
also possesses an additional local $U(1)_L$ symmetry.
However, it should be emphasized that the $U(1)_L$ symmetry is not a {\it gauge}\/ symmetry, 
since it has no corresponding gauge bosons.
It is nevertheless a local symmetry of the action 
because --- just as with $U(1)'_\FI$  ---
an arbitrary local $U(1)_L$ transformation causes the 
Lagrangian of the theory to shift by a total derivative, thereby leaving the action invariant.

In  summary, then,
we see that 
our compensated superconformal covariantized theory in the linear formalism
will have a local symmetry of the form
$U(1)'_\FI \times U(1)_\SW \times U(1)_L$.
It is, however, to recast this symmetry structure into a slightly different
form.

In order to do this, we first observe that
the local $U(1)_L$ symmetry
has the net effect of rendering the  
additional degrees
of freedom in $\SigL$ and $\SigbL$ unphysical.
Thus, the presence of the $U(1)_L$ symmetry  allows these 
degrees of freedom to be  gauged away.
This can be seen most readily when the Lagrangian is expressed in  
terms of the original
matter fields $\Phi_i$, for we then find that $\SigL$ and $\SigbL$ are
the only fields which transform non-trivially under $U(1)_L$.  

This implies that any $U(1)$ charges which are assigned to $\SigL$ and $\SigbL$
are ultimately unphysical, since their effects under $U(1)$ gauge transformations
can be eliminated through 
a compensating $U(1)_L$ transformation. 
Given this, we are free to define a new FI symmetry 
--- to be denoted $U(1)_\FI$ ---
under which $\SigL$  and $\SigbL$
are charged.  
For reasons to become clear shortly, we shall choose
$\Sigma_L,\Sigmabar_L$ to carry the same $U(1)_\FI$ charges as the corresponding
$\Sigma,\Sigmabar$ fields carry in the chiral formalism, namely
\beq
        Q_{\Sigma_L,\Sigmabar_L} ~=~ \pm {2\xi\over 3 M_P^2}~.
\label{qchoice}
\eeq
Thus, while our original matter fields $\Phi_i$ have 
$U(1)_\FI$ charges which coincide with their $U(1)'_\FI$ charges,
the rescaled matter fields $\widetilde \Phi_{Li}$ will not.

One way to think of this shift from the $U(1)'_\FI$ description to the 
$U(1)_\FI$ description is as a basis change realized through linear combination
\begin{equation}
         U(1)_\FI ~\equiv ~ U(1)'_\FI - \frac{2\xi}{3M_P^2} U(1)_L~.
\label{eq:InterrelateLambdasFIL}
\end{equation}
In this way, 
we see that $U(1)'_\FI$ and $U(1)_\FI$ can be freely substituted for each
other in the presence of the $U(1)_L$ symmetry.  However,
it is important to stress that this replacement of $U(1)'_\FI$ by $U(1)_\FI$
is not a true change of basis because we should not think of the $U(1)'_\FI$ gauge boson
as somehow picking up a $U(1)_L$ charge when being repackaged as a $U(1)_\FI$ gauge boson.
Rather, we must think of the $U(1)'_\FI$ gauge boson and the 
$U(1)_\FI$ gauge boson as being one and the same object.
As discussed above, this is because the change from $U(1)'_\FI$ to $U(1)_\FI$ has
no physical import, and does not alter our theory in any way.

There is, however, one important difference between $U(1)_\FI$ and $U(1)'_\FI$,
and this is what ultimately motivates the specific charge choice in Eq.~(\ref{qchoice}).
We have already remarked that $U(1)'_\FI$ is a symmetry of the action of our superconformal,
compensated theory in the linear formalism;
specifically, $U(1)'_\FI$ transformations cause the Lagrangian of the superconformal theory to
shift by total derivatives.
However, we expect that a canonical gauge symmetry 
should not merely leave the {\it action}\/ invariant;
it should leave the {\it Lagrangian}\/ invariant as well. 
Unfortunately, $U(1)'_\FI$ does not do this in any manifest way.
However, by charging the chiral compensator fields $\Sigma_L,\Sigmabar_L$ under $U(1)_\FI$
as in Eq.~(\ref{qchoice}),
we see that $U(1)_\FI$ is now a full manifest gauge symmetry of the theory, 
one which leaves the Lagrangian as well as the action invariant. 
This happens because variations of
$V$, $\SigL$, and $\SigbL$ under $U(1)_\FI$ transformations now explicitly
cancel against each other.  

Given these observations, we shall therefore recast the symmetries of our 
covariantized superconformal 
theory in the linear formalism as $U(1)_\FI\times U(1)_\SW \times U(1)_L$.
All of these symmetries are local;  they are also ``big'' in the sense 
defined below Eq.~(\ref{gaugetransforms}).
Likewise, only $U(1)_\SW$ is an $R$-type symmetry.
Finally, we again stress that only $U(1)_\FI$ and $U(1)_\SW$ are gauge symmetries;
indeed $U(1)_L$, despite being local, lacks a corresponding gauge boson.

Having now redefined the symmetry content of the theory in terms of
$U(1)_\FI$, we recast the theory in terms of the rescaled matter fields  
$\wtPhi_{Li}$ defined
in Eq.~(\ref{eq:RedefinePhiLinear}).  Since $\SigL$ is charged under  
both $U(1)_\FI$
and $U(1)_L$, the $\wtPhi_{Li}$ fields acquire charges under both symmetries  
due to rescaling.
Specifically, we find that the $U(1)_L$, $U(1)_\FI$, and $U(1)_\SW$ 
charges of the $\wtPhi_{Li}$ matter fields
are given by
\begin{equation}
    Q^L_{\wtPhi_{Li}} = w_i~, ~~~~~~
    Q^{\FI}_{\wtPhi_{Li}}  =  Q^\FI_{\Phi_i} - \frac{2\xi}{3M_P^2} w_i~, ~~~~~~
    Q^{\SW}_{\wtPhi_{Li}}  =  0~,
\label{eq:ChgsOfRescaled}
\end{equation}
while the corresponding charges of the $\Sigma_L$ compensator field are
given by
\begin{equation}
   Q^{L}_{\Sigma_L}  =  -1~, ~~~~~~
   Q^{\FI}_{\Sigma_L}  =  \frac{2\xi}{3M_P^2}~, ~~~~~~
   Q^{\SW}_{\Sigma_L}  =  \frac{2}{3}~.
\label{chgsforSigma}
\end{equation}
However,
when we compare these charge assignments to the corresponding ones in the
chiral-formalism description of the same theory,
we find that they agree exactly.
In other words, the  charges of $\Sigma_L$ and $\wtPhi_{Li}$ under $U(1)_\FI$ in
the linear formalism are precisely those of $\Sigma$ and $\Phitilde_i$ under 
$U(1)_\FI$ in the chiral formalism.  Furthermore, the
global $U(1)'_\FI$ symmetry of the chiral formalism is identical to
the $U(1)'_\FI$ symmetry of the linear formalism, which is simply a linear
combination of $U(1)_L$ and $U(1)_\FI$, as discussed above.  

We thus see that 
the superconformal compensated theories in the chiral and linear formalisms have an almost
identical symmetry structure.  Both theories are invariant under a gauged
$U(1)_\FI \times U(1)_\SW$ symmetry as well as  under an additional
global symmetry which
can be viewed, depending of the basis of $U(1)$ factors chosen,
as either $U(1)_L$ or $U(1)_\FI$.  
Indeed, the only difference between the two
formalisms is whether this additional symmetry may also be considered
local (despite not being gauged) in a particular basis.
This, then, is another explicit verification of the duality relation between
the chiral and linear formalisms.

As discussed above, this duality relation is only expected to hold at the
level of the superconformal theories prior to freezing our compensator fields.
Our next step, therefore, is to investigate what happens in the linear formalism
once the compensator fields are frozen.  In the linear formalism, there are two
sets of compensator fields that must be frozen:  the chiral compensators $\Sigma_L,\Sigmabar_L$,
and the linear compensator $L$.
Freezing each of these has a distinct effect.

Let us first consider freezing the chiral compensators $\Sigma_L,\Sigmabar_L$.
Given the charge assignments in Eq.~(\ref{chgsforSigma}),
we find that there are always two independent symmetries which survive this freezing
(\ie, two independent symmetries under which $\Sigma_L,\Sigmabar_L$ are uncharged).
It is particularly convenient to choose a specific basis for these symmetries: 
\beqn
         U(1)'_\FI &\equiv& U(1)_\FI + {2\xi \over 3 M_P^2} U(1)_L~,\nonumber\\
         R_G &\equiv&  {2\over 3} U(1)_L + U(1)_\SW ~.
\label{bestbasis}
\eeqn
Indeed, if we formally ``solve'' for $U(1)_L$ from the first equation in Eq.~(\ref{bestbasis})  
and substitute this into the second equation, we obtain precisely the definition~(\ref{eq:RG})
for $R_G$ in the chiral formalism. 
Note that $U(1)'_\FI$ and $R_G$ are both local $U(1)$ symmetries which are ``big'', operating
at the superfield level.

The next step is to implement the freezing of the linear compensator $L$.
However, $L$ only carries a Weyl charge.
Thus, the net effect of freezing $L$ is simply to break the Weyl symmetry of the theory, \ie,
to demote the ``big'' $U(1)_\SW$ symmetry to its ``little'' (Wess-Zumino-gauge) remnant $R_5$.
This in turn leaves the $U(1)'_\FI$ symmetry of Eq.~(\ref{bestbasis})
intact, but demotes $R_G$ from ``big'' to ``little'' status.
Both symmetries remain local, however.

The upshot, then, is that coupling our original globally $R_5$-invariant theory to supergravity
in the linear formalism 
results in a frozen theory with two local symmetries:  the $U(1)'_\FI$  
gauge symmetry of our original theory, and a new ``little'' local $R_G$ symmetry defined as
\beq
         R_G ~\equiv~  {2\over 3} U(1)_L + R_5~.
\eeq
However, $R_G$ is essentially equivalent to $R_5$ as far as 
the physics of our theory is concerned;  indeed, if we work in terms of the original
unscaled matter fields $\Phi_i$, these symmetries are identical.
We thus see that the linear formalism indeed preserves both the original FI gauge symmetry
 {\it and}\/ the original $R_5$-symmetry of the theory, as advertised.  
 
We conclude, then, that the process of coupling an $R_5$-invariant theory to supergravity
in the linear formalism does not alter the original symmetry structure of the
theory.  Indeed, the only change is that $R_5$ is promoted from a global symmetry to a local one
whose gauge field is the auxiliary field $b_\mu$ of the gravitational  
multiplet.

\subsubsection{An explicit example}

For a concrete illustration of these ideas,
let us revisit the $R_5$-invariant toy model  with non-zero FI term
that was presented
in Sect.~4.3.2.  We shall now trace how its symmetry structure changes through  
the various
steps in the conformal-compensator calculus, this time in the linear formalism.

We begin by considering the compensated superconformal theory.
As discussed above, the linear formalism allows us the option of writing
the action for our conformally-compensated
theory in terms of either the original matter fields  
$\Phi_i$ of the uncompensated theory, or the rescaled matter fields  
$\wtPhi_i$.
In the former notation, the Lagrangian for the conformally-compensated
theory is given by
\begin{equation}
   \mathcal{L}~=~ \int d^4\theta\left[L\ln\left(\frac{L}{\SigL\SigbL}\right)+
      \frac{2\xi}{3M_P^2}LV
      + K'\right] +\left[\int d^2\theta \, W +\mathrm{h.c.}\right]~
\label{eq:LinLagEx2}
\end{equation}
where $W$ and $K'$ are defined in Eqs.~(\ref{eq:Wex})
and (\ref{eq:Kex}), respectively.  
Alternatively, when written in terms of the compensated fields 
$\wtPhi_{Li}$, this  
Lagrangian becomes
\begin{equation}
   \mathcal{L}~=~ \int d^4\theta\left[L\ln\left(\frac{L}{\SigL\SigbL}\right)
      + \frac{2\xi}{3M_P^2}LV
      + \frac{\SigL\SigbL}{3M_P^2} \widetilde{K}'_L\right]
      +\left[\int d^2\theta\left(\frac{\SigL}{\sqrt{3}M_P}\right)^3\widetilde{W}_L
      +\mathrm{h.c.}\right]~
\label{eq:LinLagEx2Rescaled}
\end{equation}
where $\widetilde{K}'_L$ and 
$\widetilde{W}_L$ take the same form as  the corresponding
expressions for $K'$ and $W$ in 
Eqs.~(\ref{eq:Kex}) and (\ref{eq:Wex}),  
but with $\Phi_i$ and $\Phi^\dagger_i$ replaced everywhere by $\wtPhi_{Li}$  
and $\wtPhid_{Li}$.
This action is invariant under not only $U(1)'_\FI$ gauge  
transformations and local
superconformal transformations (including local $R_5$-rotations), but  
also under an
additional, local $U(1)_L$ symmetry (which is a symmetry of the  
action, but not of
the Lagrangian).  As discussed above, this $U(1)_L$ symmetry
allows us complete freedom to redefine $\SigL$ and $\SigbL$.  Consequently we
may assign charges to these fields under the $U(1)_\FI$ gauge symmetry  
defined in
Eq.~(\ref{eq:InterrelateLambdasFIL}) without affecting the physics.

\begin{table}[b!]
\begin{center}
\begin{tabular}{||c||c|ccc|c||}
   \hline 
   \hline 
   ~~ Field ~~ & ~$U(1)_{\FI}$~ &~ $U(1)_\SW$:~ & $R_5$~ &~ Weyl~& ~ $U(1)_L$ ~ \\
     \hline
     \hline
$\Phi_1$ &   +1  &2/3 & 2/3 & 1 & 0   \\
$\Phi_2$ & $-1$  &2/3 & 2/3 & 1 & 0  \\
$\Phi_3$ &    0  &2/3 & 2/3 & 1 & 0  \\
     \hline
$\wtPhi_{L1}$ & $\phantom{-}1-2\xi/3M_P^2$   & 0& 0 & 0 & +1   \\
$\wtPhi_{L2}$ & ~$-1-2\xi/3M_P^2$~  & 0& 0 & 0 & +1  \\
$\wtPhi_{L3}$ &  $-2\xi/3M_P^2$   & 0& 0 & 0 & +1  \\
\hline
$\SigL$             &  $\phantom{-}2\xi/3M_P^2$ & 2/3 & 2/3 &  1  & $-1$  \\
$L$                 &      0         &  $\ast$  &  0  &  2  & 0    \\
     \hline
$\lambda_\alpha$    &      0         &  $\ast$  & 1  & 3/2 & 0    \\
$\psi_{\mu \alpha}$ &      0         &  $\ast$    & 1  & 3/2 & 0    \\
     \hline
     \hline
\end{tabular}
\end{center}
\caption{The symmetry structure of the linearly-compensated version
     of the toy model presented in Sect.~4.3.2. 
    This table may be compared with Table~\ref{tab:chgs2}, which shows 
    the corresponding symmetry structure within the chiral formalism. 
    Charges are listed for the original matter fields $\Phi_i$, their  
    rescaled counterparts $\wtPhi_{Li}$, the compensators $\SigL$ and $L$, the  
     $U(1)_\FI$ gaugino $\lambda_\alpha$, and the gravitino $\psi_{\mu\alpha}$.  
    An entry `$\ast$' indicates that the corresponding field is not a chiral
    superfield, and therefore its $R_5$- and Weyl-charges cannot 
    be packaged as a chiral charge under $U(1)_\SW$. 
    Note that $R_G$ is an independent symmetry only when the theory is written
    in terms of the rescaled fields $\wtPhi_{Li}$.
\label{tab:chgs2lin}}
\end{table}

We therefore find that our superconformal theory has 
the symmetry structure shown in Table~\ref{tab:chgs2lin}.
Moreover, comparing the symmetries and charge assignments 
listed in this table to those 
given in Table~\ref{tab:chgs2} for
the chirally-compensated theory, we see that the $U(1)_\FI$ symmetries of the
two theories are identical;  that is, the field charges in the linear-formalism
description of the theory, written in terms of the rescaled fields  
$\wtPhi_{Li}$,
coincide with the field charges of the chiral-formalism description.
The $R_5$-charges and Weyl-weights in the two
theories are likewise identical.  
Thus, this explicit $R_5$-invariant example nicely illustrates
the general result of Sect.~6.1, namely that the chiral- and linear-formalism 
descriptions
of such a theory
are dual to each other at the level of their  
conformally-compensated Lagrangians.

\begin{table}[t!]
\begin{center}
\begin{tabular}{||c||c|c||}
   \hline 
   \hline 
   ~~ Field ~~ & ~$U(1)'_{\FI}$~~ &~ $R_G$~ \\
     \hline
   \hline 
$\Phi_1$ &   +1  & 2/3  \\
$\Phi_2$ &   $-1$  & 2/3  \\
$\Phi_3$ &    0  & 2/3  \\
     \hline
$\wtPhi_{L1}$ & +1  & 2/3  \\
$\wtPhi_{L2}$ & $-1$  & 2/3  \\
$\wtPhi_{L3}$ &  0  & 2/3  \\
     \hline
$\lambda_\alpha$    &   0   &  1  \\
$\psi_{\mu \alpha}$ &   0   &  1  \\
     \hline
     \hline 
\end{tabular}
\end{center}
\caption{The symmetry structure of the final version of our toy model in Table~\ref{tab:chgs2lin},
     after the compensator fields are ``frozen'' to their fixed values in the linear
    formalism.
    This table may be compared with Table~\ref{tab:chgs2after}, which shows 
    the corresponding symmetry structure within the chiral formalism. 
    Once again, charges are listed for both the original matter fields $\Phi_i$ and their  
    rescaled counterparts
    $\wtPhi_{Li}$, as well as for the $U(1)_\FI$ gaugino
    $\lambda_\alpha$ and gravitino  $\psi_{\mu\alpha}$.
\label{tab:chgs2linafter}}
\end{table}

We now turn to examine the effects that emerge upon freezing the compensator fields
$\SigL$, $\SigbL$, and $L$.  As discussed
in Sect.~6.3.1, freezing these compensators 
breaks the local $U(1)_L\times U(1)_\FI \times U(1)_\SW$  
invariance of the compensated superconformal theory 
down to a subgroup comprising a $U(1)'_\FI$ gauge symmetry  
and a local $R_G$ symmetry.  The charge assignments for the various fields of the frozen  
theory under
these symmetries are shown in Table~\ref{tab:chgs2linafter}.

We see, then, that 
the FI gauge symmetry
of the frozen theory is identical to that of the original,  
uncompensated theory.
This stands in stark contrast to the situation in the chiral formalism in Sect.~4.3.2.
But even more importantly, we see that no
additional global  
symmetries are
present.
Indeed, what remains in addition to the $U(1)'_\FI$ gauge symmetry is merely a local
version of our original $R_5$-symmetry, this time with $R_G$ playing the role of $R_5$.

\section{Discussion}
\setcounter{footnote}{0}

Although supersymmetric theories with non-zero Fayet-Iliopoulos terms 
have played a vital role in many areas of particle physics over the past
twenty years,
history has shown that it is 
surprisingly difficult to couple such theories to supergravity.
While there exists a relatively extensive literature dealing with
this and related issues~[2--24], 
recent discussions~\cite{Seiberg} have sparked renewed interest 
in this question and added urgency to the task of understanding its resolution.

In this work, we have sought to clarify the status of Fayet-Iliopoulos
terms in supergravity theories.  Following the lead
of Ref.~\cite{Seiberg}, we have focused on issues pertaining to the
supercurrent supermultiplets of globally supersymmetric theories with non-zero FI terms. 
However, we have also broadened our
investigation by studying the
overall symmetry properties of such theories,
particularly insofar as questions of gauge invariance and $R_5$-symmetry conservation 
are concerned.
Towards this end, we have reviewed both the chiral (or ``old
minimal'') and linear (or ``new minimal'') compensator formalisms of
supergravity.  This has enabled us 
to analyze, within each of these formalisms,
the salient properties of the actions and supercurrent supermultiplets corresponding to
theories containing non-zero FI terms.
This also has enabled us to ascertain which features of these theories
are intrinsic to FI terms, and which tend to be more closely associated
with the formalisms themselves and might therefore be viewed as artifacts of 
those formalisms.
We have also examined the extent to which the 
well-established duality~\cite{RelateChiralLinearSUGRA}
between the conformally-compensated theories in these two formulations is disturbed
after their respective compensator fields are ``frozen'' in order to break
the extraneous symmetries of the superconformal group.  

The primary results of our analysis may be summarized as follows.

\begin{itemize}

\item First, within the context of a generic globally supersymmetric theory
      with supercurrent supermultiplet $J_{\alpha\alphadot}$, 
      we have studied the specific contribution $\Xi_{\alpha\alphadot}$ within
      $J_{\alpha\alphadot}$ that arises due to the possible existence of 
      a non-zero FI term.   
      Specifically, $\Xi_{\alpha\alphadot}$ may be identified
      as that part of the overall supercurrent $J_{\alpha\alphadot}$
      which depends explicitly on the Fayet-Iliopoulos coefficient $\xi$  ---
      and which therefore has a linear dependence on the fields associated with
      the $U(1)_{\rm FI}$ vector superfield --- without use of the equations of motion.
      Our conclusion is that no self-consistent solutions for 
      $\Xi_{\alpha\alphadot}$ exist for theories with unbroken $R_5$-symmetry.
      (Such theories are natural candidates for study, since FI terms by themselves 
      intrinsically preserve $R_5$-symmetry.)
      In the chiral formalism, this happens because one cannot construct a
      consistent supercurrent supermultiplet $J_{\alpha\alphadot}^{(C)}$ from
      the three Noether currents $j^{(5)}_\mu$, $j_{\mu\alpha}$, and $T_{\mu\nu}$ of
      the flat-space theory.  Instead, as we demonstrated in Sect.~4.2,
      explicit contributions from the compensator fields of this formalism 
      must be included in order to obtain a consistent multiplet, and 
      we have shown in Sect.~4.4 that there 
      is no way in which these contributions can preserve $R_5$-symmetry. 
      This peculiar property of FI terms is
      indicative of the fundamental difficulties which arise in coupling $R_5$-invariant
      theories with non-zero FI terms to supergravity using the chiral formalism.
      Moreover, in the linear formalism, we have shown in Sect.~5.2 
      that $\Xi_{\alpha\alphadot}$ must vanish outright.   

\item  On the other hand, for theories that break $R_5$-symmetry, we find
      that the FI supercurrent contribution $\Xi_{\alpha\alphadot}$ 
      exists, but necessarily breaks the $U(1)_\FI$ gauge symmetry of the original
      theory.  This result is in complete agreement with the results of Ref.~\cite{Seiberg}.
      However, as we discuss in Sect.~2, the FI term by itself does not 
      break this gauge symmetry.  Instead, as we demonstrate in Sect.~4, this breaking
      of the $U(1)_\FI$ symmetry in the presence of a non-zero FI term
      is entirely an artifact of the chiral-compensator formalism,
      as originally pointed out in Ref.~\cite{StelleAndWestEToTheV}.
      Indeed, in the linear formalism, the $U(1)_\FI$ gauge symmetry 
      is preserved, even in the presence of a non-zero FI term.  
      However, the linear formalism is not applicable 
      for theories which break $R_5$-symmetry.
      
\item  Taking a bird's-eye view, these observations suggest that there is
      a fundamental connection between $R_5$-symmetry and $U(1)_\FI$ gauge invariance
      in the presence of a non-zero FI term:  either both symmetries are simultaneously
      preserved (as in the linear formalism), or both are broken (as in the chiral
      formalism~\cite{BarbieriEtAlEToTheV}).  
      Using our results from Sect.~5.2, this implies 
      at the level of supercurrent supermultiplets 
      that the only way to have 
      a non-zero FI contribution $\Xi_{\alpha\alphadot}$ is to break $U(1)_\FI$ gauge
      invariance.  This is, of course, precisely what occurs in Ref.~\cite{Seiberg}, 
      but we now see that this is indeed an expected and generic property, forced
      upon us by the minimal compensator formalisms.

\item As we have shown in Sects.~4.4 and 5.4,
      there is one situation in which we may construct a non-zero 
      FI contribution $\Xi_{\alpha\alphadot}$ to the supercurrent even when
      $R_5$-symmetry is conserved:  this occurs if our $U(1)_\FI$ gauge boson
      has an explicit supersymmetric mass.  Such a situation might arise, \eg,
      for effective FI theories at lower energy scales.  This mass breaks the 
      $U(1)_\FI$ gauge symmetry explicitly, but demonstrates that non-zero solutions
      for $\Xi_{\alpha\alphadot}$ can exist in such situations.
      Moreover, to the best of our knowledge, 
      this is the only known situation in which the minimal supergravity
      formalisms permit $R_5$ symmetry to be manifestly preserved while 
      $U(1)_\FI$ gauge symmetry is broken in the presence of a non-zero FI term.

\item In the chiral formalism, the existence of a non-zero FI term results 
     in the presence of an exact, global symmetry in the conformally-compensated, 
     locally supersymmetric theory at the classical level.  
     This is discussed explicitly in Sect.~4.3.~
     Even after the compensator fields of the theory are ``frozen'' 
     in order to break superconformal invariance down to
     Poincar\'e supersymmetry, this exact global symmetry remains unbroken.  
     This result is in complete agreement with the results of Ref.~\cite{Seiberg}.
     Moreover, the symmetries of the frozen theory also include a gauged $R$-symmetry,
     under which the gravitino and all gauginos in the theory are charged.

\item In the linear formalism, the conformally-compensated theory in the 
       presence of a non-zero FI term is dual to the corresponding theory in the 
      chiral formalism.  (Indeed, the analogue of the additional $U(1)'_\FI$ symmetry 
      in the latter is the built-in $U(1)_L$ symmetry in the former.)  Upon freezing the
      compensators of the linear formalism, however, the symmetry content of the
      theory reduces to that of the uncompensated theory, 
      the one difference being that $R_5$ has become a local symmetry.  
      Consequently, as discussed in Sect.~6.3,
      all of the symmetries of the final, frozen theory are local.

\end{itemize}

Several comments about these results are in order.

First, it should be noted that there are many 
subtleties associated with the process of freezing the conformal
compensator fields.
Of course, this process is intended to eliminate the extraneous degrees of
gauge freedom associated with transformations under the superconformal group
(scale invariance, $S$-supersymmetry, {\it etc}\/.) which were not symmetries
of the original theory.  However, the gauge-fixing conditions imposed in order
to break these symmetries often also break other
symmetries of that theory
which ought to be preserved in the frozen theory.  
For example, the gauge-fixing conditions break supersymmetry.
Likewise, the $R_5$-symmetry of an $R_5$-invariant theory will unfortunately be
broken in the chiral formalism.
In such cases, a set of analogous
symmetries can be defined for this theory, the generators of which
correspond to linear combinations of the generators of the symmetries of the
compensated theory (for an excellent review, see Ref.~\cite{KugoUehara}).
Nevertheless, issues of this sort do not directly impact the status of FI terms
in supergravity theories, and we have therefore refrained from
discussing them in detail in the present work.

Second, it should be emphasized that the analysis presented here
has been entirely classical.  At the quantum level, the results outlined
above are still valid;  however, one must be slightly more careful in choosing
the compensator formalism one uses in order to couple a given theory to supergravity.
In particular, theories which admit a consistent description in the linear
formalism at the classical level often do not admit such a description at
the quantum level;  in models in which the $R_5$-symmetry is anomalous,
higher-order corrections will induce superconformal-anomaly contributions
which are not of the form $L_\alpha$, thereby rendering the
linear-formalism description invalid~\cite{SupercurrentAnomaliesRenorm}.
Indeed, this kind of issue is known to affect the anomaly structures
of supersymmetric QED and QCD~\cite{SQED,SQCD}.
This poses additional constraints on the applicability of the linear formalism,
but these constraints may be overcome.  The massless Wess-Zumino
model, for example, satisfies such constraints and may be consistently coupled to
supergravity using the linear formalism, even at the quantum level.

Third, it is worth mentioning that other supergravity formulations
exist in addition to the chiral and linear formalisms discussed here.
These include the non-minimal formalism of Ref.~\cite{Breitenlohner},
as well as a variety of additional possibilities which utilize
reducible multiplets as compensators (for a discussion of some of
these possibilities, see Ref.~\cite{KugoNonminimal}).
One interesting non-minimal alternative, especially as far as FI terms
are concerned, is the vectorial formalism
of Ref.~\cite{deWitAuxiliary}, in which the compensator fields include
a vector multiplet $V$, a chiral superfield $\Sigma_V$, and its
hermitian conjugate $\Sigmabar_V$.  This formalism can be
viewed as being analogous to the chiral formalism,
but with an FI term, whose corresponding gauge field is the vector
compensator $V$, already incorporated~\cite{KugoUehara}.  We will not
discuss these possibilities further here, except to note that use of such 
non-minimal formalisms 
could offer valuable additional perspectives on many the issues surrounding
FI terms in supergravity.

Finally, even within the minimal formalisms we have focused on here, 
the question of whether theories with primordial non-zero FI terms can
be consistently coupled to supergravity is likely to depend on whether 
$R_5$-symmetry is ultimately broken in nature (\ie, on whether the $R_5$-symmetry breaking
scale is at or near the SUSY-breaking scale).  If so, and if the
chiral formalism is applicable, then such a coupling to supergravity 
is likely to be impossible
as a result of the extra global symmetry that remains in this formalism,
as discussed in Sect.~4.3  and  originally pointed out in Ref.~\cite{Seiberg}. 
In such a case, if our underlying primordial theory at the highest energy scales is coupled
to supergravity, then this theory cannot contain FI terms.
As a result, if  supersymmetry is realized in nature and broken dynamically at lower energy
scales, the mechanism responsible for that breaking will turn
out to be predominately $F$-term in nature.
However, if $R_5$-symmetry is ultimately preserved in nature, then the chiral
formalism will no longer be appropriate and the  
the linear formalism will be more relevant as a description of the
corresponding supergravity. 
In such cases, as discussed in Sect.~6.3, 
the resulting symmetry structure may well be different.

We conclude, then, that 
the issue of coupling theories with FI 
terms to supergravity is indeed a subject with many subtleties and 
unexpected consequences.
Incorporating the FI term into supergravity appears to be 
uniquely and inherently wedded to the specific formalism employed,
which motivates the comparative treatment we have provided here.
Moreover, we see that the $R_5$-symmetry properties of a theory
appear to be closely linked to its FI properties, and both together
appear to determine the degree of difficulty with which 
such a theory can ultimately be
coupled to supergravity.  Finally, we have seen that
the FI contribution $\Xi_{\alpha\alphadot}$ to the supercurrent 
$J_{\alpha\alphadot}$ is a 
very strange beast:  in one formalism it breaks gauge invariance, and in 
another formalism it doesn't even exist.
All of these features together explain why the 
issue of coupling theories with non-zero FI terms to supergravity has had a long
and tortu(r)ous history, and suggest that this topic is likely to have a
long and tortu(r)ous future as well.

\bigskip

\section*{Acknowledgments}

We are happy to thank Z.~Chacko, E.~Dudas, S.J.~Gates, and especially 
N.~Seiberg for discussions.
This work was supported in part by the
U.S. Department of Energy under Grant~DE-FG02-04ER-41298.
The opinions and conclusions expressed here are those of the authors, 
and do not represent either the Department of Energy or the National Science Foundation.

\bigskip

\bibliographystyle{unsrt}

\end{document}